\documentclass[aps,prc,twocolumn,preprintnumbers, floatfix]{revtex4-2}
\makeatletter

\renewcommand*{\p@subsection}{}

\renewcommand*{\p@subsubsection}{}
\makeatother

\usepackage{subfigure}
\usepackage{multirow}
 \usepackage{xcolor}
\usepackage{fancyhdr}
\usepackage{longtable}
\usepackage{parskip}
\usepackage[T1]{fontenc}
\usepackage{dcolumn}   
\usepackage{float}
\usepackage{bm}        
\usepackage{amsfonts}  
\usepackage{placeins}
\usepackage[]{graphicx}
\usepackage{wrapfig}
\usepackage{amsmath}
\usepackage{nameref}

\newcommand{\pwisein}{\left\{ \begin{array}{ll}}
\newcommand{\pwiseout}{\end{array}\right.}

\usepackage[colorlinks,citecolor=blue,urlcolor=blue,linkcolor=blue]{hyperref}
\begin{document}
\title{\Large Time Projection Chamber for GADGET II}


\author{Ruchi Mahajan$^{1}$}
\email{mahajan@frib.msu.edu}
\author{T. Wheeler$^{1,2,3}$}
\author{E. Pollacco$^{4}$}
\author{C. Wrede$^{2,1}$}
\email{wrede@frib.msu.edu}
\author{A. Adams$^{1,2}$}
\author{H. Alvarez-Pol$^{5}$}
\author{A. Andalib$^{1}$}
\author{A. Anthony$^{6}$}
\author{Y. Ayyad$^{5}$}
\author{D. Bazin$^{1,2}$}
\author{T. Budner$^{1,2}$}
\author{M. Cortesi$^{1}$}
\author{J. Dopfer$^{1,2}$}
\author{M. Friedman$^{7}$}
\author{A. Jaros$^{1}$}
\author{D. Perez-Loureiro$^{8}$}
\author{B. Mehl$^{9}$}
\author{R. De Oliveira$^{9}$}
\author{L. J. Sun$^{1}$}
\author{J. Surbrook$^{1,2}$}

\affiliation{$^1$Facility for Rare Isotope Beams, Michigan State University, East Lansing, Michigan 48824, USA}
\affiliation{$^2$Department of Physics and Astronomy, Michigan State University, East Lansing, Michigan 48824, USA}
\affiliation{$^3$Department of Computational Mathematics, Science, and Engineering, Michigan State University, 
East Lansing, Michigan 48824, USA}
\affiliation{$^{4}$IRFU / DEDIP, CEA Saclay, F91191 Gif-sur-Yvette, France}
\affiliation{$^5$IGFAE, Universidade de Santiago de Compostela, E-15782, Santiago de Compostela, Spain.}
\affiliation{$^{6}$Department of Physics and Astronomy, High Point University, High Point, NC, 27268, USA} 
\affiliation{$^7$The RACAH institute of Physics, Hebrew University of Jerusalem, Israel.}
\affiliation{$^{8}$Canadian Nuclear Laboratories, Canada.}
\affiliation{$^9$CERN Esplanade des Particules 1 P.O. Box 1211 Geneva 23, Switzerland}
\date{\today}
\begin{abstract}
\textbf{Background:} The established GAseous Detector with GErmanium Tagging (GADGET) detection system is used to measure weak, low-energy $\beta$-delayed proton decays. It consists of the gaseous Proton Detector equipped with a MICROMEGAS (MM) readout to detect protons and other charged particles calorimetrically, surrounded by the Segmented Germanium Array (SeGA) for high-resolution detection of prompt $\gamma$-rays.
\newline \textbf{Purpose:} To upgrade GADGET's Proton Detector to operate as a compact Time Projection Chamber (TPC) for the detection, 3D imaging and identification of low-energy $\beta$-delayed single- and multi-particle emissions mainly of interest to astrophysical studies. 
\newline\textbf{Method:} A new high granularity MM board with 1024 pads has been designed, fabricated, installed and tested. A high-density data acquisition system based on Generic Electronics for TPCs (GET) has been installed and optimized to record and process the gas avalanche signals collected on the readout pads. The TPC's performance has been tested using a $^{220}$Rn $\alpha$-particle source and cosmic-ray muons. In addition, decay events in the TPC have been simulated by adapting the ATTPCROOT data analysis framework. Further, a novel application of 2D convolutional neural networks for GADGET II event classification is introduced.  The optimization of data throughput is also addressed.\newline\textbf{Results:} The GADGET II TPC is capable of detecting and identifying $\alpha$-particles, as well as measuring their track direction, range, and energy. The extracted energy resolution of the GADGET II TPC using P10 gas is about 5.4\% at 6.288 MeV ($^{220}$Rn $\alpha$-events), computed using charge integration. Based on a systematic simulation study, we estimated the detection efficiency of the GADGET II TPC for protons and $\alpha$-particles, respectively. It has also been demonstrated that the GADGET II TPC is capable of tracking minimum ionizing particles (i.e. cosmic-ray muons). From these measurements, the electron drift velocity was measured under typical operating conditions. In addition to being one of the first generation of micro pattern gaseous detectors (MPGDs) to utilize a resistive anode applied to low-energy nuclear physics,  the GADGET II TPC will also be the first TPC surrounded by a high-efficiency array of high-purity germanium $\gamma$-ray detectors. \newline\textbf{Conclusions:} The TPC of GADGET II has been designed, fabricated and tested, and is ready for operation at the Facility for Rare Isotope Beams (FRIB) for radioactive beam-line experiments. 
\end{abstract}
\maketitle 
\section{Introduction:Astrophysical Motivation}
Type I X-ray bursts are short-lived bursts of X-rays detected by space-based X-ray 
telescopes that can last from a few seconds to a few minutes. The mechanism behind these bursts is not fully understood, but they are 
thought to be caused by thermonuclear runaways occurring on the surfaces of neutron stars accreting hydrogen and helium from close binary 
companions. The occurrence of these X-ray bursts is dependent on the mass, temperature, accretion rate, and magnetic field of the neutron 
star \cite{Rolfs:1988}. These factors along with nuclear reaction and decay rates play a crucial role in influencing the behavior and 
characteristics of X-ray bursts. The nuclear reactions in models of Type I X-ray bursts take place at temperatures that peak in the range 
of 1-2 GK.  For these temperatures, the range of center-of-mass energies that contribute most to the reaction rates (the Gamow window) is 
between 100 and 400 keV for the lower masses (A < 20) and up to 1 MeV or higher for the higher masses (A > 20). The cross sections of 
these reactions are very difficult to measure directly at astrophysical energies because they involve radioactive beams limited by 
intensity. Therefore, indirect methods including $\beta$-delayed charged particle emission by implanting radioactivity in Si detectors 
have been introduced to probe such reaction rates \cite{Borge_2013, PhysRevC.83.045808, MCCLESKEY2013124, WALLACE201259}.  This approach 
is not ideal for studying these astrophysically interesting decays with energies below 400 keV due to the presence of a large $\beta$-
background and summing. To further reduce the $\beta$-background, novel gaseous detectors such as AstroBox and the GAseous Detector with 
GErmanium Tagging (GADGET) were developed \cite{POLLACCO2013102,2016NIMPB.376..357S,Friedman:2019}.  The position-sensitive readouts of 
these detectors were based on a micro pattern gaseous amplifier detector (MPGD) technology.  The precursor nuclei are produced by nuclear 
reactions and thermalized in the gas volume of the detector; the subsequent emission of charged particles ionizes the gas. The ionization 
electrons created in the event are drifted toward the position-sensitive gas avalanche readout by a uniform electric field, and amplified 
by a strong avalanche electric field within the MPGD gas gap. Calorimetric measurements with these systems were successfully proven to 
detect weak and low energy $\beta$-delayed charged particle branches (without particle identification) 
\cite{POLLACCO2013102,2016NIMPB.376..357S,Friedman:2019}.
\newline GADGET is an operational detection system at the National Superconducting Cyclotron Laboratory (NSCL)/FRIB consisting of the 
gaseous Proton Detector surrounded by the Segmented Germanium Array (SeGA) of High Purity Germanium (HPGe) $\gamma$-ray detectors. It has 
been used at the NSCL for calorimetric measurements of low-energy $\beta$-delayed protons in cases where particle identification is not 
needed \cite{PhysRevLett.128.182701, PhysRevC.101.052802, PhysRevC.103.014322, Surbrook2022}. For example, this system has been used by 
Budner {\em et al.} \cite{PhysRevLett.128.182701} to study the $\beta$-decay of $^{31}$Cl to probe the $^{30}P$(p, $\gamma$)$^{31}$S 
reaction leading to the measurement of a 260-keV resonance which represents the weakest $\beta$-delayed, charged-particle intensity ever 
measured below 400 keV. Friedman {\em et al.} \cite{PhysRevC.101.052802} have reported a measurement of the branching ratios of the 
$^{23}$Al $\beta$-delayed protons as a probe of the key (204-keV center-of-mass) $^{22}$Na(p,$\gamma$)$^{23}$Mg resonance strength using 
GADGET. Sun {\em et al.} \cite{PhysRevC.103.014322} have used GADGET to study the proton-$\gamma$ coincidences from the $\beta$-decay of 
$^{25}$Si. This system has also been used to study the $\beta$-decay of $^{11}$Be to search for exotic decay modes including 
$\beta$$^{-}$-delayed proton emission and dark decay of the neutron \cite{Surbrook2022}. \newline The present work details an expansion of GADGET's scientific capabilities through a technical upgrade of GADGET's Proton 
Detector to operate as a Time Projection Chamber (TPC), enabling 3D track imaging, particle identification and the detection of multiple 
particle emissions. The first experiment with GADGET II seeks to determine the thermonuclear reaction rate of the 
$^{15}$O($\alpha$,$\gamma$)$^{19}$Ne reaction, which has long been regarded as one of the most important thermonuclear reactions in Type 
I X-ray bursts \cite{Wiescher_1999, PhysRevC.96.032801,Cyburt_2016,Fisker_2006}. At hot CNO cycle break-out temperatures, the rate of 
this reaction is strongly dominated by a single resonance with a center of mass energy of 506 keV corresponding to a $^{19}$Ne state 
having an excitation energy of 4034 keV \cite{Davids_2011} . It is technically infeasible to measure the strength of this resonance 
directly with current facilities due to the need for a very intense and low-energy beam of radioactive $^{15}$O. However, it is possible 
to determine the resonance strength indirectly from the 
 spin, lifetime and branching ratio ($\Gamma$$_{\alpha}$/$\Gamma$) of the 4034 keV state. Since the spin and lifetime are well known the 
 resonance strength can be constructed from $\Gamma$$_{\alpha}$/$\Gamma$ \cite{PhysRevC.77.035803,PhysRevC.72.041302,PhysRevC.74.045803}. 
 Attempts have been made to measure $\Gamma$$_{\alpha}$/$\Gamma$ using direct reactions to populate the 4034 keV state and search for the 
 $\alpha$-particle emission, resulting in an evaluated strong upper limit of 4$\times$10$^{-4}$ \cite{Davids_2011}. The goal with GADGET 
 II is to measure a finite value for 
 $\Gamma$$_{\alpha}$/$\Gamma$ and hence the reaction rate at FRIB. These measurements proceed via the $^{20}$Mg($\beta$p$\alpha$)$^{15}$O 
 decay sequence using $\beta$-delayed proton emission to populate the 4034 keV $^{19}$Ne state followed by a rare $\alpha$ emission 
 \cite{PhysRevC.96.032801, PhysRevC.99.065801}. We are also planning to investigate the second and third most important reactions in Type 
 I X-ray bursts \cite{Cyburt_2016}: $^{59}$Cu(p, $\gamma$)$^{60}$Zn and $^{59}$Cu(p, $\alpha$)$^{56}$Ni, both of which will be studied 
 simultaneously using the $\beta$-decay of $^{60}$Ga to $^{60}$Zn. In addition to the program on Type I X-ray bursts, the unique 
 sensitivity of GADGET II can be used to 
 investigate important reactions in other astrophysical scenarios and to search for exotic decays. 
\newline A newly designed position-sensitive MICROMEGAS (MM) board has been implemented. The high-granularity readout plane consists of 1024 square 
pads, 1016 of which are used for tracking reconstruction, while 8 are used as vetos. A high-density Generic Electronics for TPCs (GET) data 
acquisition system \cite{POLLACCO201881} has been installed to cover an electronic channel per pad \cite{Mahajan, Wheeler}. This increases the 
number of measurement pads by over a factor of 200 as compared to the Proton Detector which had only 5 measurement pads. The higher granularity of 
the upgraded readout allows for better 3D reconstruction of charged particles emitted from radioactive decays inside the detector volume. In this 
work the performance of the TPC has been investigated using an $\alpha$-particle source ($^{228}$Th) and cosmic-ray muons. A dedicated ATTPCROOT 
data analysis framework based on the FairRoot package was used to compare experimental data with simulations\cite{AYYAD2020161341, Ayyad_2017, 
ATTPCROOT_git, anthony_2023_10027879}.
\newline The paper is organized as follows: A brief overview of the TPC and operating principles is followed by a detailed description of individual 
components in Sect.~\ref{sec-2}. The performance of the GADGET II TPC using cosmic ray muons and radioactive sources is demonstrated in 
Sect.~\ref{sec-3}. The optimization of the TPC data throughput is discussed in Sect.~\ref{sec-4}. The ATTPCROOT framework used to simulate the 
events of interest, and the VGG16 Convolutional Neural Network (ConvNet) used to facilitate the identification of real events in the data is 
described briefly in Sect.~\ref{sec-5}. The summary and outlook for this initial experimental campaign using the GADGET II TPC  is provided in 
Sect.~\ref{sec-6}.
\section{Detector description} 
\label{sec-2}
\subsection{\textit{Overview}}
\begin{figure}[htbp]
\includegraphics[width=2.7 in]{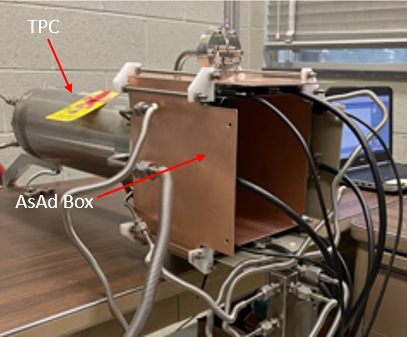}
\caption{\label{fig1} A photograph of the GADGET II TPC setup in a test space off the beam line.}
\end{figure}

The GADGET II TPC, shown in Figure~\ref{fig1}, is a cylindrical gaseous detector, operated in a mode in which the radioactive ions of interest are thermalized inside the gas volume and allowed to decay. 
Ionization electrons created by electrically charged decay products are drifted by a uniform electric field towards the position-sensitive MM-based 
gaseous amplifier \cite{POLLACCO2013102}. Online experiments at FRIB are operated in implant-decay cycles. The delayed charged particles are 
typically detected during the ‘‘beam off'' mode. During the ‘‘beam on'' mode an electrostatic gating grid \cite{Friedman:2019} is activated to 
remove the large ionization originating from the beam. TABLE~\ref{tab}  lists some of the nominal operating parameters of the GADGET II TPC. The 
following subsection describes the individual components in detail.
\begin{table}[ht]
\centering
\caption{\label{tab}: Nominal operating parameters of GADGET II TPC.}
\centering
\begin{tabular}{|p{4.5cm}|p{3.5cm}|}
\hline
\multicolumn{2}{|c|}{TPC parameters} \\
\hline
No: of measurement pads   & 1016 \\
Pad size &  2.2 $\times$ 2.2 mm$^{2}$\\
Pad plane area & 50.24 cm$^{2}$ \\
No: of veto pads   & 8 \\
Total veto pad area& 28.26 cm$^{2}$ \\
Length of drift region  & 400 mm\\
Amplification gap & 128 $\mu$m\\
 \hline
 \multicolumn{2}{|c|}{Gating grid parameters} \\
 \hline
 No: of gold plated copper wires & 60 \\
 Diameter & 20 $\mu$m \\
 Wire separation & 2 mm \\
 \hline
 \multicolumn{2}{|c|}{GET parameters} \\
 \hline
 Electronic sampling frequency & 50 MHz\\
 Signal shaping time & 502 nsec\\
 Event rate & 1 kHz \\
 GET gain & 1 pC \\
 \hline
 \multicolumn{2}{|c|}{Gas amplifier parameters} \\
 \hline
 Typical gas composition & P10(90\%Ar+10\% CH$_{4}$) \\
 Gas pressure & 800 Torr  \\
 Gas gain & 40 for $^{220}$Rn $\alpha$-particles   \\
 Drift field & 150 V/cm \\
 Amplification field & 30 kV/cm \\
 Drift velocity & 5.44 $\pm$0.03 cm/$\mu$sec \\
 Temperature & 25$^{\circ}$C \\
 \hline
 \multicolumn{2}{|c|}{Micromesh parameters} \\
 \hline
 Resistance & 10 M$\Omega$/square \\
 Capacitance (calculated) & 287 nF \\
 Mesh - Anode separation & 128 $\mu$m \\
\hline
\end{tabular}
\end{table}

\subsection{\textit{Resistive MICROMEGAS}}
\label{subsec:2_2}
\begin{figure}[htbp]
\centering
\includegraphics[width=3.4 in]{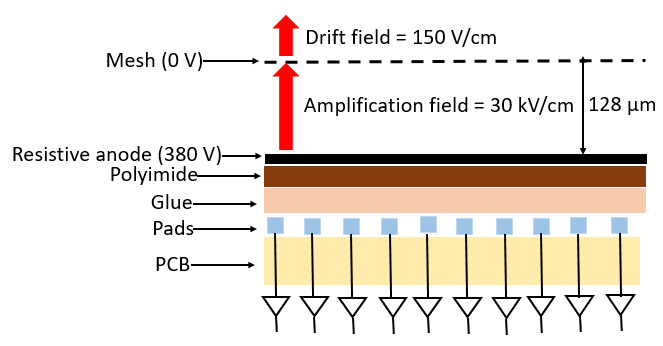}
\caption{\label{fig2}Schematic of the resistive MM structure.}
\end{figure}

The GADGET II TPC readout board is based on a resistive-anode MM, which also serves as an end cap for the detector’s gas volume. While MPGDs with resistive electrodes are widely employed in high-energy physics and other fields, resistive-anode MM are novel in low-energy nuclear physics \cite{ATTIE2022166109}. The GADGET II TPC resistive-anode MM board was custom-designed and manufactured at CERN. The anode is charaterized by a resistivity of 10 M$\Omega$ per square, which protects the front-end electronics \cite{CHEFDEVILLE2021165268}  from sporadic disrupting discharges. The MM comprises a stainless steel micromesh, with 18 $\mu$m wire diameter and 45 $\mu$m
micromesh opening. This micromesh underwent calendaring, leading to a 30 $\mu$m thickness and 45\% optical transparency. The micromesh is held by insulating pillars at 128 $\mu$m above the anode plane and is kept at ground voltage.  Avalanche amplification of the electrons occurs at typical electric field strength of approximately 30 kV/cm established aross the 128 $\mu$m amplification gap between the micromesh and the resistive anode. The resistive anode lies on top of a thin layer of 50 $\mu$m Polyimide, which is fastened to the segmented readout plane by a thin layer of glue, as shown in Figure~\ref{fig2}. The Polyimide and glue act as dielectrics in a two dimensional RC network. The resistance within this network is determined by the surface resistivity of the anode, achieved by evaporating a thin and uniform layer of diamond like carbon (DLC). The capacitance, on the other hand, is determined by the relative permittivity of the glue and Polyimide, as well as the separation between the anode and readout plane. Effectively, the resistive anode evenly spreads a deposited charge cluster \cite{DIXIT2006281}.  In the specified geometry, charge dispersion is minimal and is confined to a single pad, covering an area of 4.84 mm$^{2}$.

\begin{figure}[htbp]
\centering
\includegraphics[width=3.4 in]{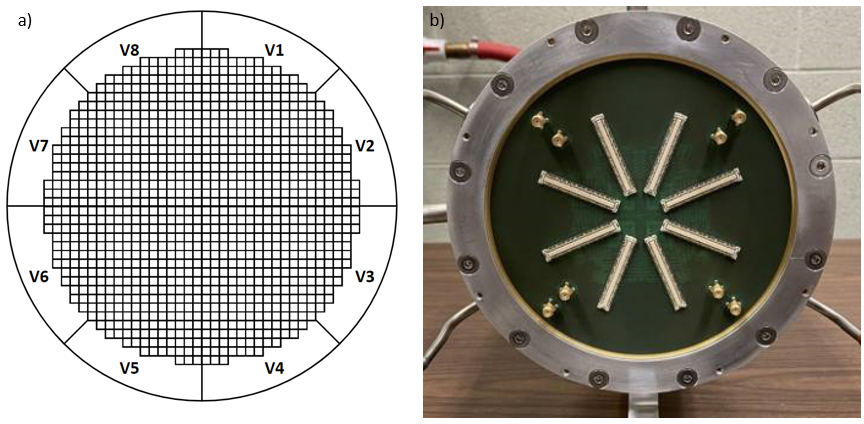}
\caption{\label{fig3}a) Schematic front view of MM board depicting the 1016 measurement pads in the central region and the 8 veto pads (marked as V1 to V8). b) Rear end cap view of the installed MM board with 8 multi - pin connectors of 144 channels which is further attached to T-Zap boards (not shown).}
\end{figure}

The anode is divided into 1024 sections, forming a circular area with a diameter of 10 cm on a PCB frame that runs parallel to the micromesh. To enhance the adhesion of the Polyimide layer supporting the resistive layer, the pads are deliberately oxidized. There are 1016 measurement pads of 2.2 $\times$ 2.2 mm$^{2}$ in the circular array, each with a small square shape and a total area of 50.24 cm$^{2}$. The array is also encircled by 8 larger veto pads, having an area of 28.26 cm$^{2}$. This arrangement allows for vetoing of charged particles that might escape the active volume and deposit only part of their energy in the active volume. Figure~\ref{fig3} a) and Figure~\ref{fig3} b) show the front and rear view, respectively, of the installed MM at one end of the TPC. 

\subsection{\textit{GET System}}
\begin{figure*}[htbp]
\centering
\includegraphics[width=4.8in]{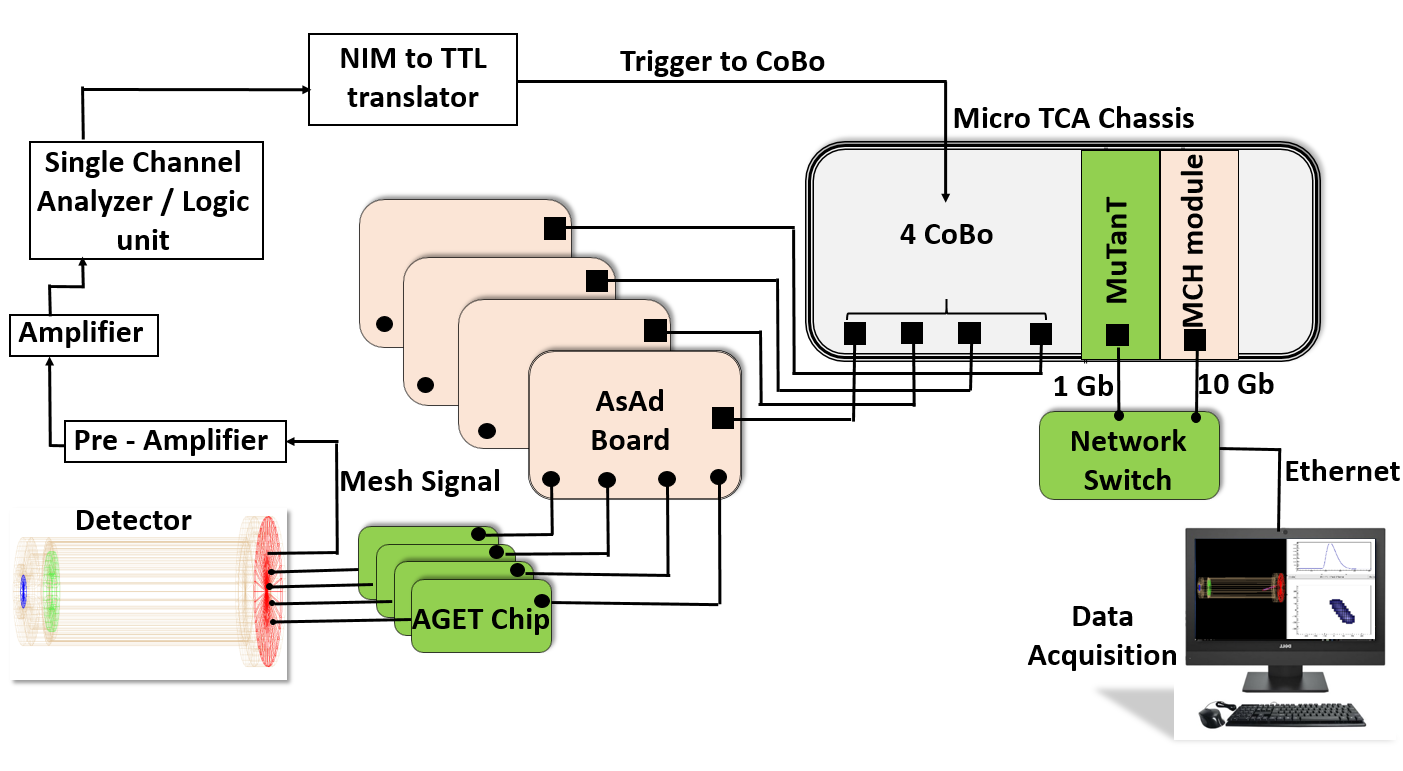}
\caption{\label{fig4} Schematic overview of the GET system with a single $\mu$TCA (Micro Telecommunications Computing Architecture) chassis, 4 CoBos (Concentration Boards) and a MuTanT (Multiplicity Trigger and Time).  }
\end{figure*}

Due to the large number of pads on the segmented readout board, a high-density data acquisition system is required. The GET system was chosen for this purpose, which is a scalable and generic electronics system that was originally designed for gas-filled detector applications in nuclear physics including TPCs. The GET system has an electronics architecture based on a versatile Application Specific Integrated Circuit (ASIC) design with several modes of acquisition. The 256 channel ASIC and ADC (AsAd) front-end card hold 4 AGET (ASICS) chips and ADCs that perform the first concentration of the data from 64 input channels to one analog output (see Figure~\ref{fig4}). 
A small Field Programmable Gate Array (FPGA) is also implemented to handle the slow control of the chips and monitor the currents, voltages and temperature. A concentration board (CoBo) originally designed for 4 AsAd boards holds an FPGA to handle the clock distribution, removal of data below threshold, and the data flow. An internal clock of the system is in the Multiplicity Trigger and Time (MuTanT) module. The MuTanT also provides triggering which may be generated by the total multiplicity and/or by an external trigger. More details on the GET data acquisition system are found in Ref. \cite{POLLACCO201881}. Figure~\ref{fig4} shows the schematic overview of the GET system highlighting all of the main components and the designed mesh trigger (Sect.~\ref{subsec:2_6}) used for the data acquisition system.  For the first experiment with GADGET II at FRIB,  4 CoBos are used to read out all 1024 channels of the MM. While a single CoBo can effectively read out all 1024 channels of MM, this 4 CoBo configuration was chosen to address a data throughput limitation, which is thoroughly explained in Sect.~\ref{sec-4}. 

\subsection{\textit{AsAd Box Design and T-Zap Boards}}
In order to integrate the front-end electronics for the GET system with the TPC, the AsAd Box was designed. One side of the AsAd Box consists of 4 triangular PCB boards, called T-Zap boards, which were custom designed and fabricated at CERN. These 4 T-Zap boards connect directly to the MM to transmit signals from the TPC to the AsAds. The 4 AsAd boards connect perpendicularly to the T-Zap boards, with each one accommodating a total of 256 signals from MM pads. All of these components are housed in a box made of copper plates that act as a Faraday cage to reduce the pickup of external electromagnetic noise. Figure~\ref{fig1} shows the assembled AsAd Box configuration including the T-Zaps. This configuration was chosen to place the AsAd boards as close as possible to the MM.
\subsection{\textit{Mesh Trigger}}
\label{subsec:2_6}
 A data acquisition trigger for the GADGET II TPC (see Figure~\ref{fig4}) signals the arrival of the charge at the MM mesh. The ionization electrons drift toward the MM mesh. These electrons cross the mesh and amplifier in the gas gap between the mesh and the resistive anode. The movement of the avalanche charges (electrons toward the resistive anode and the ions toward the mesh) generates signals on the resistive anode (transferred to the segmented pad plane by capacitive coupling) and on the mesh. Both signals (on the mesh and on the anode) have an identical pulse height but possess opposite polarity - negative on the resistive anode and positive on the mesh. By grounding the mesh (see Figure~\ref{fig2})  through a low impedance charge amplifier (modified Canberra model 2006) a trigger logic  signal is generated via a fast amplifier and leading-edge discriminator. All particle tracks above threshold entering the active volume thus generate a mesh trigger. The signal-to-noise ratio (S/N) for the mesh signals using an $\alpha$-particle source (see ~\ref{subsec:3_1}) was measured to be 16. Further, as described in Sect.~\ref{sec-4} an anti-coincidence circuit was introduced to reduce the mesh trigger rate using the veto pad signals.
 
\section{GADGET II TPC performance}
\label{sec-3}
\subsection{\texorpdfstring{\textit{$\alpha$-Particle Source Test}}{Alpha-Particle Source Test}}
\label{subsec:3_1}
\begin{figure}[htbp]
\includegraphics[width=3.3in]{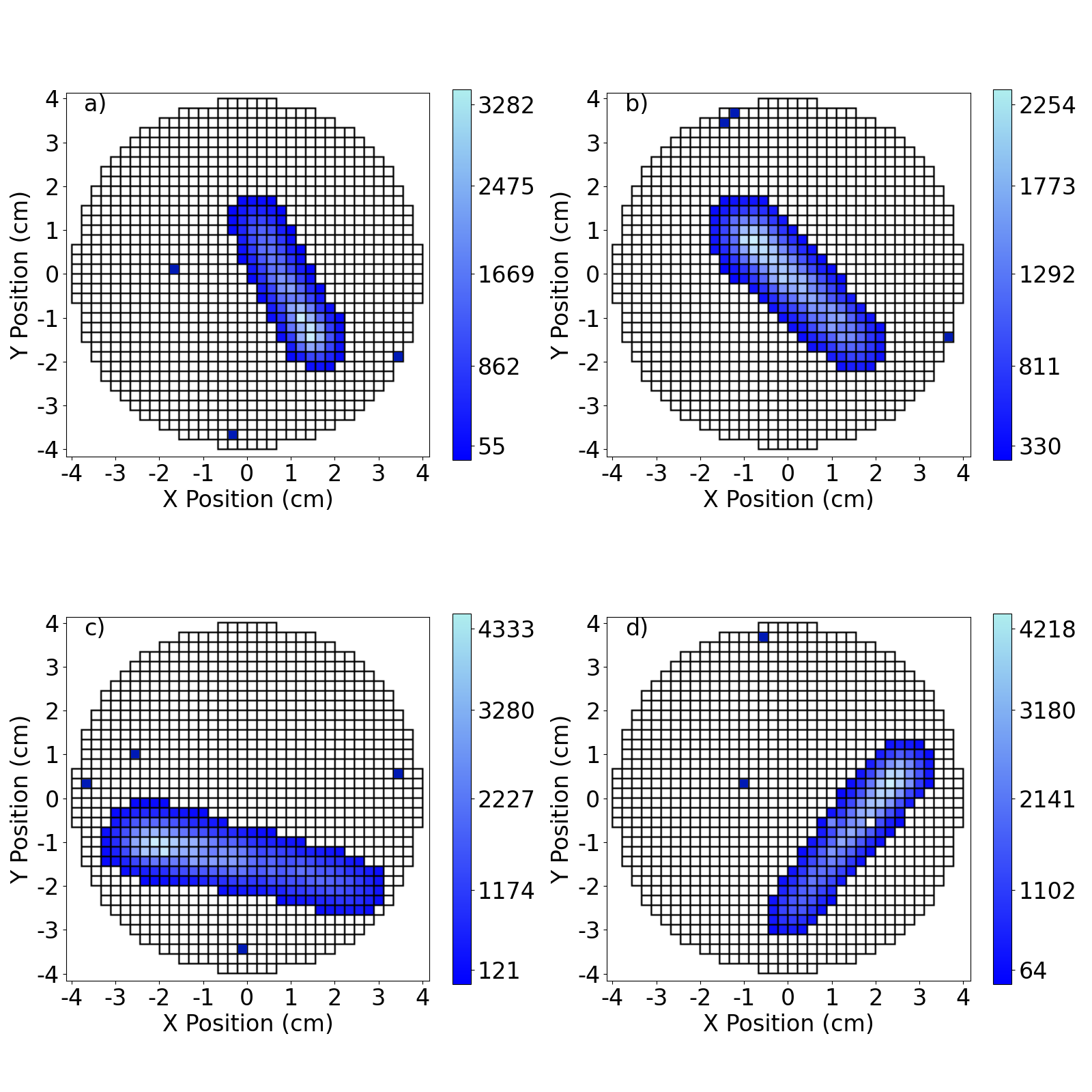}
\caption{\label{fig5} Label a) and b): Projected 2D MM pad plane images of $^{220}$Rn alpha tracks, while c) and d) depict $^{216}$Po alpha tracks 
within the GADGET II TPC. The diffusely illuminated pads refer to the outlier points.}
\end{figure}

\begin{figure}[htbp]
\includegraphics[width=3.3in]{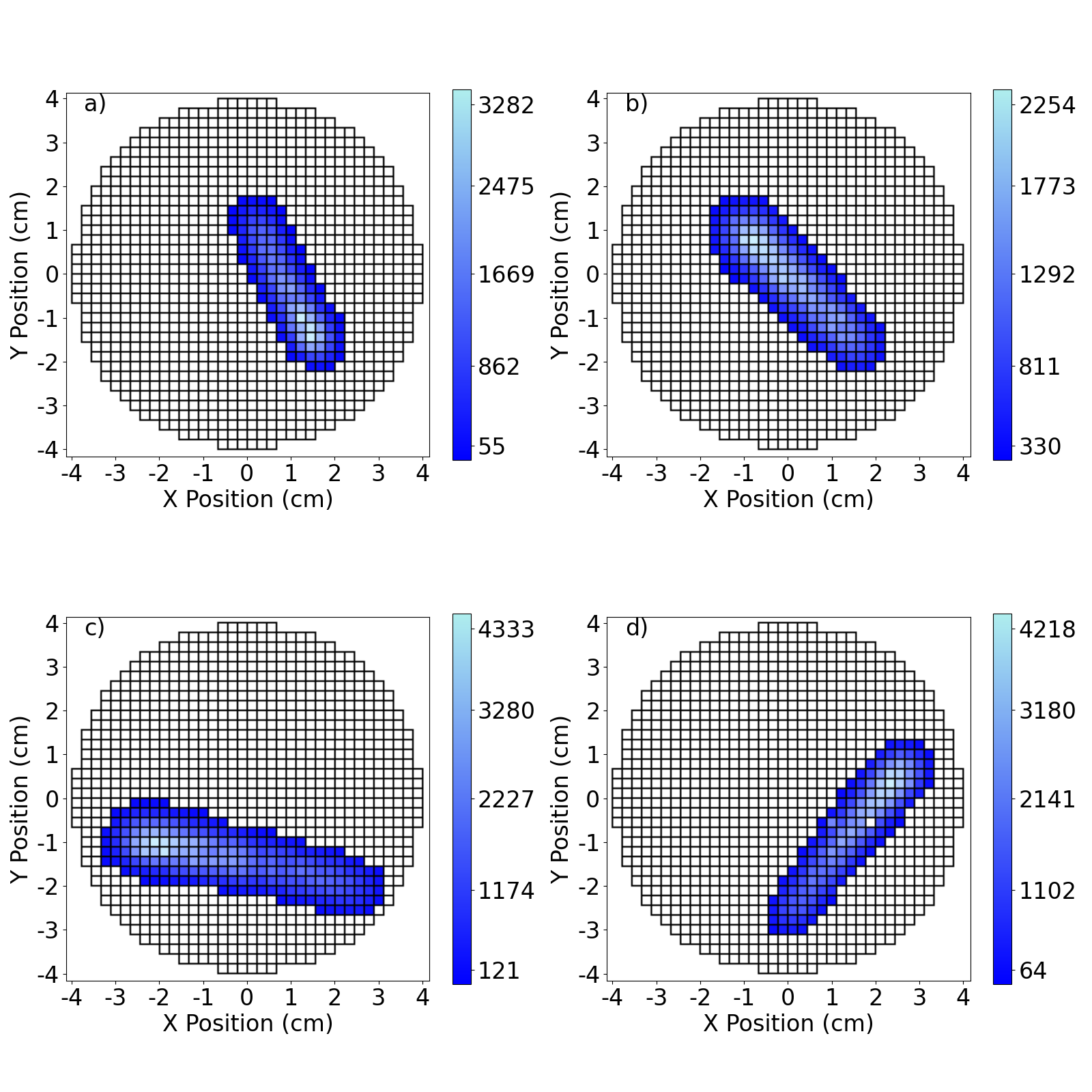}
\caption{\label{fig6} Label a) and b): Projected 2D MM pad plane images of $^{220}$Rn alpha tracks, while c) and d) 
depict $^{216}$Po alpha tracks within the GADGET II TPC after outlier removal..}
\end{figure}

\begin{figure}[htbp]
\includegraphics[width=3.3in]{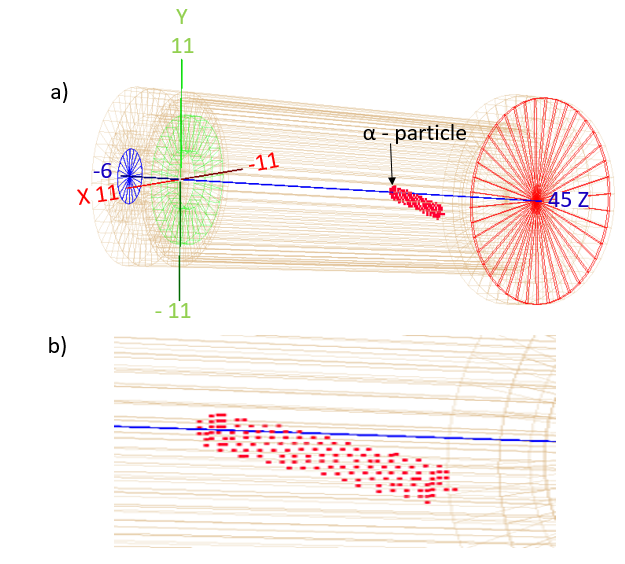}
\caption{\label{fig7} a) 3D hit pattern representation of a $^{220}$Rn alpha track inside the GADGET II TPC. b) Magnified in image of $^{220}$Rn alpha track inside the GADGET II TPC. Here, the z-coordinate is chosen arbitrarily as the absolute z position of the decay event inside the TPC is unknown. The numbers (11, -11, 45 and -6) on the axis label are the lengths of the X, Y  and Z axes in cm. }
\end{figure}

A first performance evaluation of the TPC was carried out by irradiating the drift volume from within using $\alpha$-particles, while the detector volume was filled with P10 gas mixture at 800 Torr.  A 1 $\mu$Ci  $^{228}$Th source was installed in the gas inlet line leading to the generation of $^{220}$Rn (T$_{1/2}$ = 55.4 s), which occasionally escapes the thin window of the sealed source due to recoil from alpha emission, mixes with P10 gas in the inlet line and then flows into the detector. The $^{220}$Rn decays to $^{216}$Po emitting $\alpha$-particles having an energy of 6.288 MeV (with 99.886\% branching ratio \cite{WU20071057}). $^{216}$Po subsequently decays by emission of a 6.778 MeV $\alpha$-particle (with 99.9981\% branching ratio \cite{AURANEN2020117}).  
\newline To establish a clean track analysis any signals on pads outside the locality of the main track in a particular event are removed. This is achieved by using two different outlier detection algorithms (Hotelling and Squared Prediction Error (SPE)) in a pipeline fashion \cite{article}. Hotelling calculates distances between data points and their mean based on variance, marking points farther from the mean as potential outliers. SPE computes squared prediction errors for each point using a statistical model, flagging points with high SPE values as potential outliers. Charge collection in a pad (point) is removed only when both algorithms agree that a point is an outlier. Figure~\ref{fig5} and Figure~\ref{fig6} shows the projected 2D images of the $^{220}$Rn alpha tracks on the pad plane, before and after eliminating outliers, respectively. Figure~\ref{fig7} shows a 3D reconstruction of an alpha track inside the TPC.

\begin{figure}[htbp]
\includegraphics[width=3.3 in]{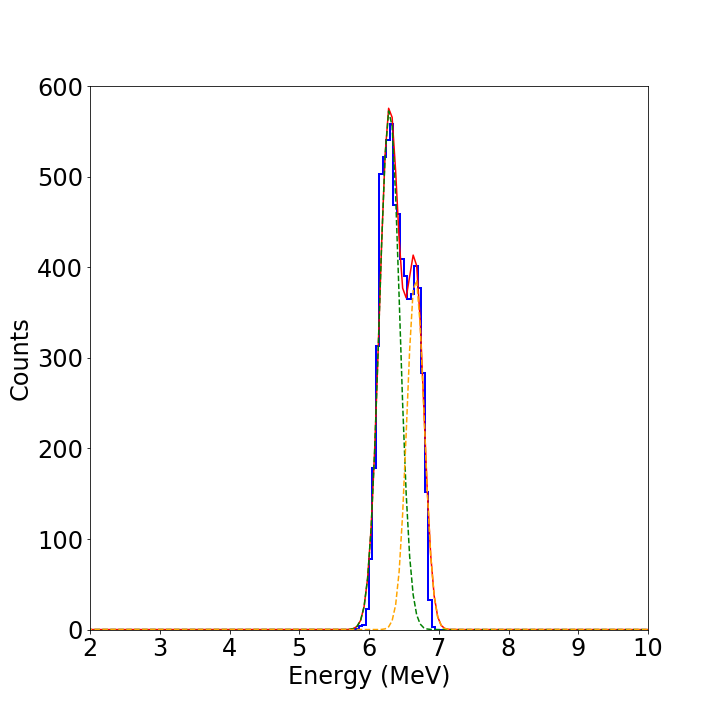}
\caption{\label{fig8} Aggregate $^{220}$Rn energy spectrum (blue), with a fit (red) demonstrating 5.4\% energy resolution at 6.288 MeV. The higher energy peak is from the $^{216}$Po 6.778 MeV $\alpha$-particle.}
\end{figure}

\begin{figure}[htbp]
\includegraphics[width= 3 in]{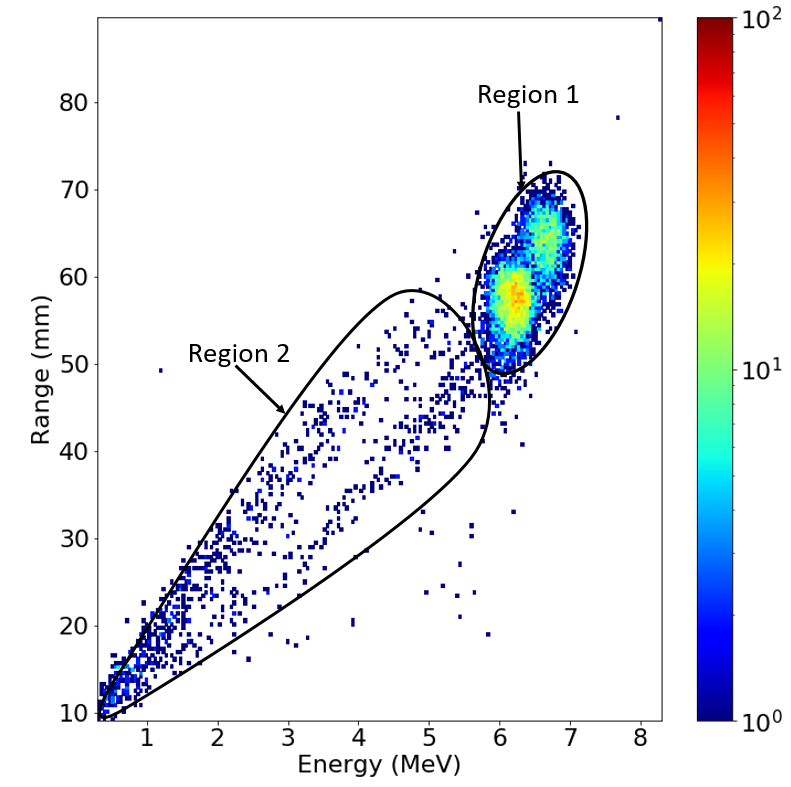}
\caption{\label{fig9} Aggregate range versus energy histogram of $^{220}$Rn and $^{216}$Po $\alpha$-particles for events with an angle range between 0$^{\circ}$ to 70$^{\circ}$ with respect to the pad plane.}
\end{figure}

Principal Component Analysis (PCA) is used to fit tracks by identifying lines and planes that best approximate the data through least squares optimization. The first principal component maximizes the variance of data points along it, while the second principal component, orthogonal to the first, minimizes variance \cite{WOLD198737}. In this analysis, the first and second principal components correspond to the length and width of a track, respectively. After extracting the length of a given track, the charge is integrated over all pads and then converted to energy. The resulting energy spectrum for $^{220}$Rn alphas can be seen in Figure~\ref{fig8}. After gain matching pads using signals induced from pulses on the mesh, an energy resolution of 5.4\% was achieved for events with an angle range between 0$^{0}$ to 70$^{0}$ with respect to the pad plane. This is consistent with other quoted TPC energy resolutions \cite{ACTAR2018,ACTAR2020,FissionTPC,KONCZYKOWSKI2019125}. However, the energy resolution of TPCs in general is variable depending on the different gain matching procedures and operating parameters (gas, pressure, drift voltage etc.) \cite{ACTAR2020,FissionTPC}. Additionally, the quoted values in Ref. \cite{ACTAR2020} correspond to a smaller angular range with respect to the pad plane, which improves the energy resolution at the cost of statistics.
The 6.778 MeV $\alpha$-peak appears weaker compared to the 6.288 MeV $\alpha$-peak because the $^{216}$Po may be positively charged from $^{220}$Rn $\alpha$ decay and drift to the cathode, where $\alpha$-particles emitted into the inactive cathode from $^{216}$Po decay are lost. After extracting the energy for an ensemble of events, a range versus energy plot has been generated (see Figure~\ref{fig9}) which is crucial for particle identification. 

There are two predominant features that appear in the range versus energy plot. The dense region where the 6.288 MeV and 6.778 MeV $\alpha$-particles reside is marked as Region 1. For most of the events inside Region 1 $\alpha$-particles deposit their full energy in the active volume and have expected ranges for 6.288 MeV and 6.778 MeV $\alpha$-particles. The events lying in the region marked as Region 2 are signatures of the so called ‘‘wall effect'' where decays happen near the upstream or downstream end of the TPC and the $\alpha$-particle tracks terminate in the inactive anode or cathode after partial energy deposition in the active region. Events with tracks traversing the volume projected by the veto pads have been eliminated by using an anti-coincidence condition (as discussed in Sect.~\ref{sec-4}).

\subsection{\textit{Cosmic-ray muon events}}
The GADGET II TPC's detection capabilities have also been tested via cosmic-ray muon measurements. For the muons with tracks perpendicular to the MM, there are only 27 electron-ion pairs per pad formed from the resulting ionization in the drift zone based on the cosmic muon stopping power in P10 \cite{2017APSG}. Thus, detection of minimum-ionizing cosmic-ray muons is a clear signature of a good signal to noise ratio on the pads. 
\newline To tag and locate a cosmic-ray muon, two plastic scintillators (BC408), each attached to a photo multiplier tube  (PMT), were 
placed above and below the drift chamber of the TPC \cite{spirit}. 
\begin{figure}[htbp]
\centering
\includegraphics[width=3.2 in]{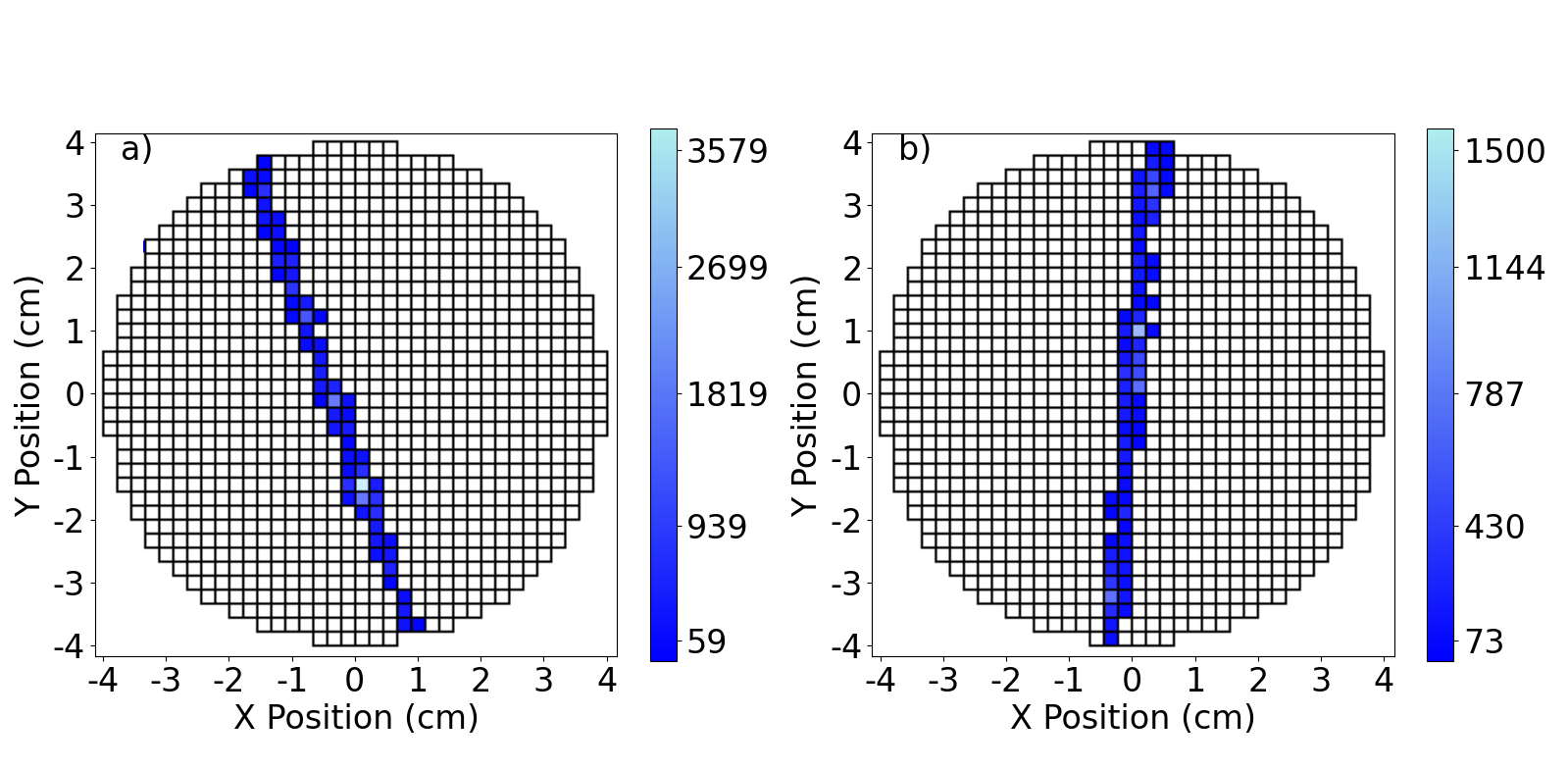}
\caption{\label{fig10} Label a)-b): Two examples of tracks of cosmic muons detected in the GADGET II TPC.}
\end{figure}
The scintillators are 2.5 cm wide and 40 cm long and have a thickness of 2 mm. Logic signals from the PMTs were generated to produce a  coincidence signal when a cosmic muon was detected in both scintillators. This coincidence signal was sent to the CoBo as a trigger for the data acquisition system  (instead of using the mesh trigger described in subsection~\ref{subsec:2_6}). For these measurements the the amplification field was increased to 44 kV/cm to achieve sufficient gain. A sample of the tracks recorded by this method is shown in Figure~\ref{fig10}. 
These cosmic-ray muon measurements are useful to study the effect of the diffusion of the ionization electrons in the gas. During beam-line experiments using the TPC, there is no global external trigger that yields information on where a decay happened inside the TPC. So, by extracting the width of the cosmic muon tracks from the timing distribution as a function of distance, the absolute position of the charged particle event ($\alpha$ and p) versus width can be calibrated. For this purpose the whole drift length of detector was scanned in steps of 5 cm to extract the widths of the cosmic muon tracks as a function of distance. The widths (W) of the cosmic muon tracks as a function of drift length is modeled using following function \cite{diff}:
\begin{equation}
 W = A + B \cdot C^{x}
\end{equation}
where A is a constant representing the track width, B is a constant representing the amount of diffusion that occurs, C is a constant representing the rate of diffusion, and $x$ is the drift distance in the gas. The width of cosmic muon tracks increases with greater drift distance due to electron diffusion. Figure~\ref{fig11} illustrates this relationship, showing that tracks near the upstream end (cathode) appear wider compared to those near the downstream end (anode).
\begin{figure}[htbp]
\centering
\includegraphics[width=3.4 in]{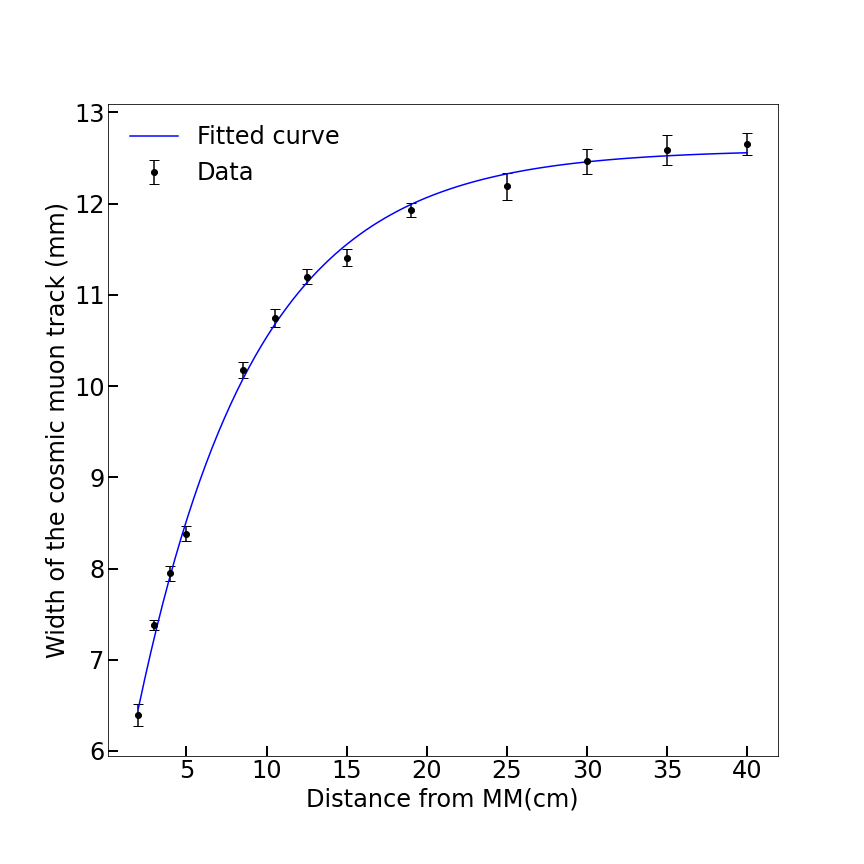}
\caption{\label{fig11} Average width of the cosmic muon tracks as a function of distance from the MM.}
\end{figure}
\newline Further, the drift velocity was extracted from the slope of the drift time versus distance from the MM by fitting the data using a straight line. Figure~\ref{fig12} shows that this fit yields a drift velocity of 5.44 $\pm$ 0.03 cm/$\mu$sec. The maximum possible drift time for primary ionization electrons in a 40-cm drift region is therefore 7.352 $\pm$ 0.041 $\mu$sec. These values are consistent with the previous measurements performed with the original GADGET detection system using a particle-gamma coincidence method \cite{Friedman:2019}.
\begin{figure}[htbp]
\centering
\includegraphics[width=3.in]{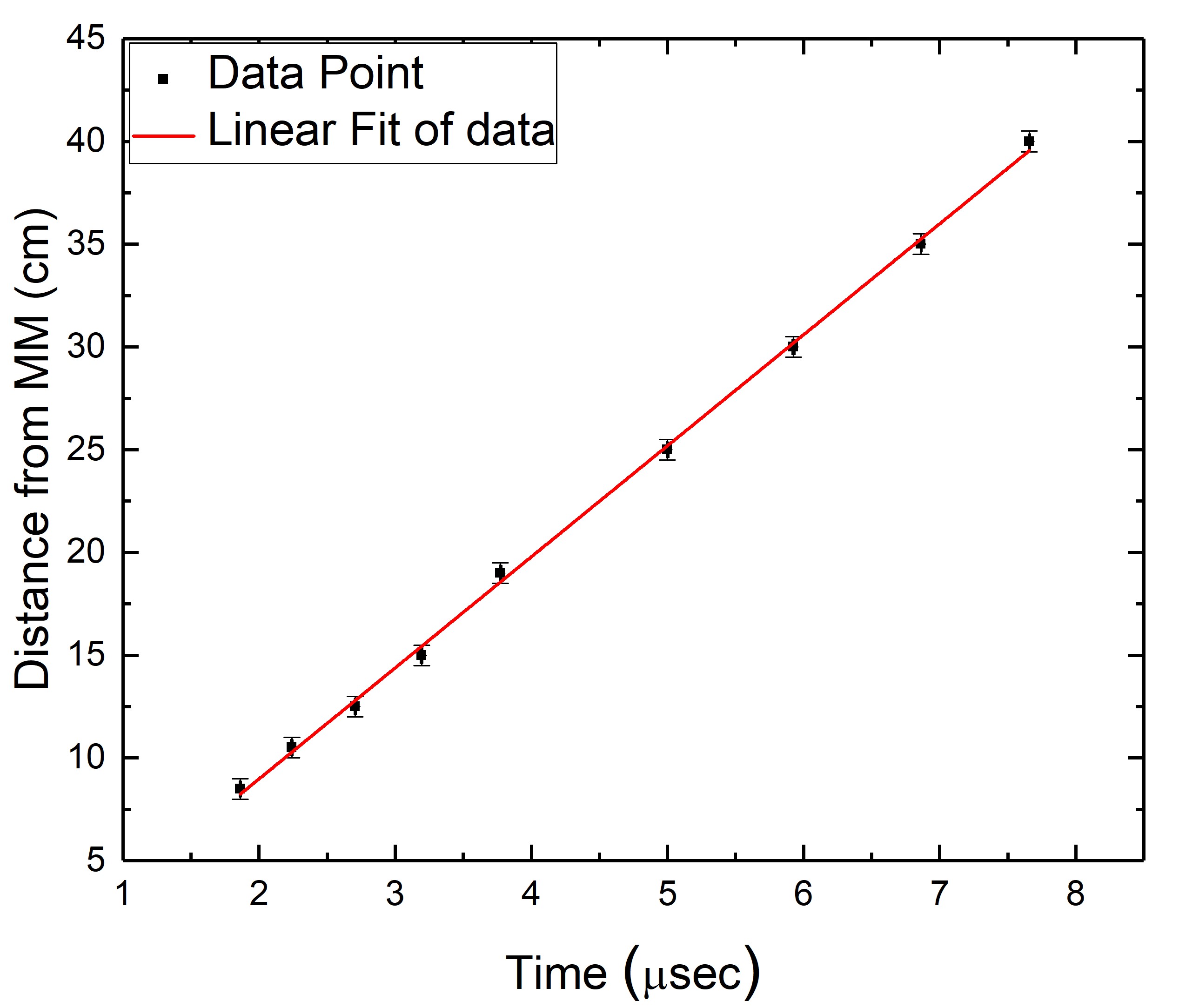}
\caption{\label{fig12} Linear fit (red) of the drift time versus distance measurements for cosmic muons. The drift velocity of 5.44 $\pm$ 0.03 cm/$\mu$sec is extracted from the slope of this fit.}
\end{figure}
\section{TPC data throughput optimization}
\label{sec-4}
\begin{figure}[htbp]
\centering
\includegraphics[width=3.2in]{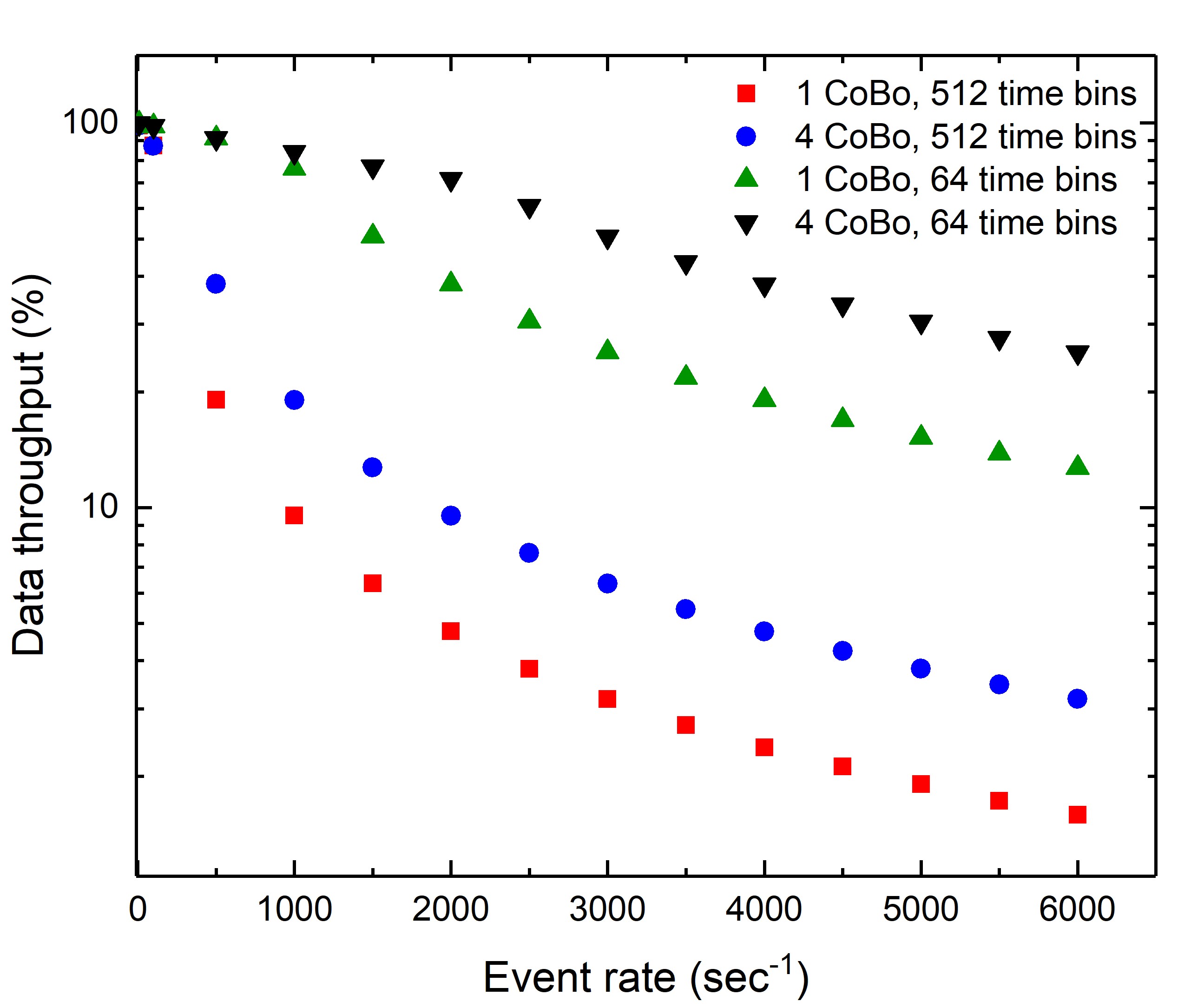}
\caption{\label{fig13} Data throughput as a function of event rate for different time binnings and CoBo configurations.}
\end{figure}

A limitation of the GET-based data acquisition system is the data throughput, namely the amount of data it can acquire and process per unit time. This affects the overall detection rate capability of the TPC. The first experiment with GADGET II at FRIB seeks to determine the thermonuclear reaction rate of the $^{15}$O($\alpha$,$\gamma$)$^{19}$Ne reaction, requiring the  acquisition of $\beta$-delayed particles (protons and alpha particles from $^{20}$Mg and $^{20}$Na daughter decays, respectively) at an event rate of thousands of particles per second. The GET system originally purchased for GADGET II to operate at NSCL with lower beam delivery rates is not able to handle such high event rates. This limitation is principally due to the bottleneck in the single CoBo that is used to concentrate and transfer data to the computer. This bottleneck results in a significant dead time for event rates above 1 kHz. The data throughput limitation can be improved by four methods:  1) increasing the number of CoBos, 2) fine tuning the parameter settings (channel hit readout, threshold etc.) and in particular reducing the number of time bins recorded in each trace, 3) implementing an active veto trigger, and 4) implementing upgraded CoBos.
\newline To address this bottleneck issue, it has been tested and established (using a random pulse generator and $\alpha$ - decay events from $^{220}$Rn decay) that increasing the number of CoBos leads to a roughly linear increase in data throughput. For high event rate experiments, data throughput can also be improved by fine tuning the parameter settings (channel hit readout, threshold etc.) and in particular reducing the number of time bins recorded in each trace from a default readout depth of 512 to 64 time bins \cite{GF}. Figure~\ref{fig13} shows the data throughput as a function of event rate for different time binnings and CoBo configurations.
Using a four-CoBo configuration coupled with a reduced readout depth of 64 time bins, demonstrates a notable enhancement in data throughput. These results are in agreement with the findings presented in Ref. \cite{POLLACCO201881}. The data throughput can be further improved by implementing an active veto trigger where any events with tracks that hit the veto pads are discarded before being stored. To effectively manage a high event rate involving thousands of particles per second, the implementation of an active veto trigger system proves to be extremely beneficial, enabling beam rates up to an order of magnitude higher, or more, but depending on the case. The design of an upgraded version of the original CoBo is in progress and this will increase the throughput per CoBo by improving the GET bandwidth by a factor of 2. For the first beam-line experiment with GADGET II, the four-CoBo configuration has been successfully implemented along with the active veto trigger. This method will reduce the total data rate by about an order of magnitude which allows full use of the available beam for this particular case.

\section{GADGET II Simulation}
\label{sec-5}
To facilitate data analysis, the ATTPCROOT \cite{AYYAD2020161341, Ayyad_2017, ATTPCROOT_git,anthony_2023_10027879} framework was adapted to work with the GADGET II geometry. Monte Carlo simulations using the ATTPCROOT framework have been performed to determine the behavior of charged particles in the TPC using a P10 gas mixture at 800 Torr. The response of the full GADGET II TPC is simulated using $\alpha$-events (from $^{220}$Rn decay) and proton-$\alpha$ events (from $^{20}$Mg decay) of different energies. These simulations will be useful in the analysis of the experimental data from the GADGET II setup. This simulated data will also be used to train a ConvNet (see subsection ~\ref{subsec:5_3})  to identify the proton-alpha coincidence events of interest.
\subsection{\texorpdfstring{\textit{Simulated $^{20}$Mg and $^{220}$Rn Decay Events}}{Simulated 20Mg and 220Rn Decay Events}}
\label{subsec:5_1}
ATTPCROOT is a ROOT based framework which requires external libraries (FairSoft and Fair Root) and a set of physics generators originally developed to simulate and analyze data from the Active Target Time Projection Chamber (AT-TPC) and its prototype detectors \cite{AYYAD2020161341, Ayyad_2017, ATTPCROOT_git,anthony_2023_10027879}. Using this framework a  user can unpack and analyze data, as well as create a customized geometry of interest for performing realistic simulations on an event-by-event basis using a virtual Monte Carlo package. Each generated event can correspond to a particular decay sequence. These simulations are performed using the Geant4 toolkit \cite{geant_paper}. The identical format of real and simulated data makes this useful not only for simulating data but also for analyzing experimental data on equal footing. 
\begin{figure}[htbp]
\centering
\includegraphics[width=3.4in, height = 2.5 in]{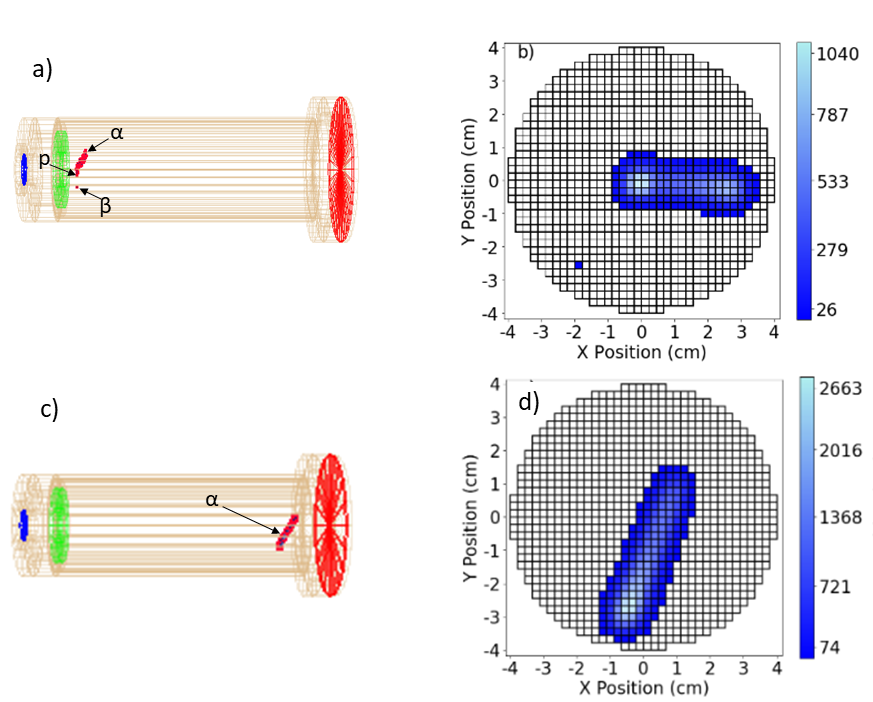}
\caption{\label{fig14} Panel (a) and (b)-ATTPCROOT simulation of the GADGET II TPC for $^{20}$Mg($\beta$p$\alpha$)$^{15}$O, 3D render and 2D projection. Panel (c) and (d)-ATTPCROOT simulation of the GADGET II TPC for $^{220}$Rn $\alpha$-decay, 3D render and 2D projection.}
\end{figure}
Once the simulated data is created, pulse shape analysis is used to process the readout from the pad planes. Pattern recognition algorithms can be used to evaluate each event, and the trajectories of the particles in the detector can then be tracked. More information about this simulation package and digitization can be found in Ref. \cite{AYYAD2020161341, Ayyad_2017, ATTPCROOT_git, Anyhony2023,anthony_2023_10027879}. 
\newline For the first experiment with the GADGET II system, $^{20}$Mg decay events have been simulated. These simulations relied on the GADGET II TPC geometry and utilized $^{20}$Mg decay probabilities as input data. These simulations also consider the effects of electron diffusion and charge dispersion. Diffusion arises from the movement of electrons through the gaseous medium, while charge dispersion is a result of the presence of a thin DLC coating on our MM pads (as discussed in sub section~\ref{subsec:2_2}). The $^{20}$Mg($\beta$p$\alpha$)$^{15}$O events of interest will have a unique 3D topology in the TPC. The proton energies will be roughly 1.2 MeV based on Doppler broadening analysis of the 4034-keV $\gamma$-ray peak \cite{PhysRevC.99.065801}. For these simulations, a P10 gas mixture at atmospheric pressure was considered. The protons will be identified by their characteristic Bragg curves and have a range of 3.7 cm. The $\alpha$ particles will deposit 506 keV of total energy in very short (<4mm), dense tracks at the point of proton emission. Figure~\ref{fig14} (a) shows a simulated event of interest inside the active volume of the TPC and the projection of a proton-$\alpha$ track on the pad plane is depicted in Figure~\ref{fig14} (b). $^{220}$Rn $\alpha$-decay events were also simulated as shown in Figure~\ref{fig14} (c) and Figure~\ref{fig14} (d), respectively.
\subsection{\textit{Simulated detection efficiency of GADGET II TPC}}
\label{subsec:5_2}
\begin{figure}[htbp]
\includegraphics[width=3.3 in]{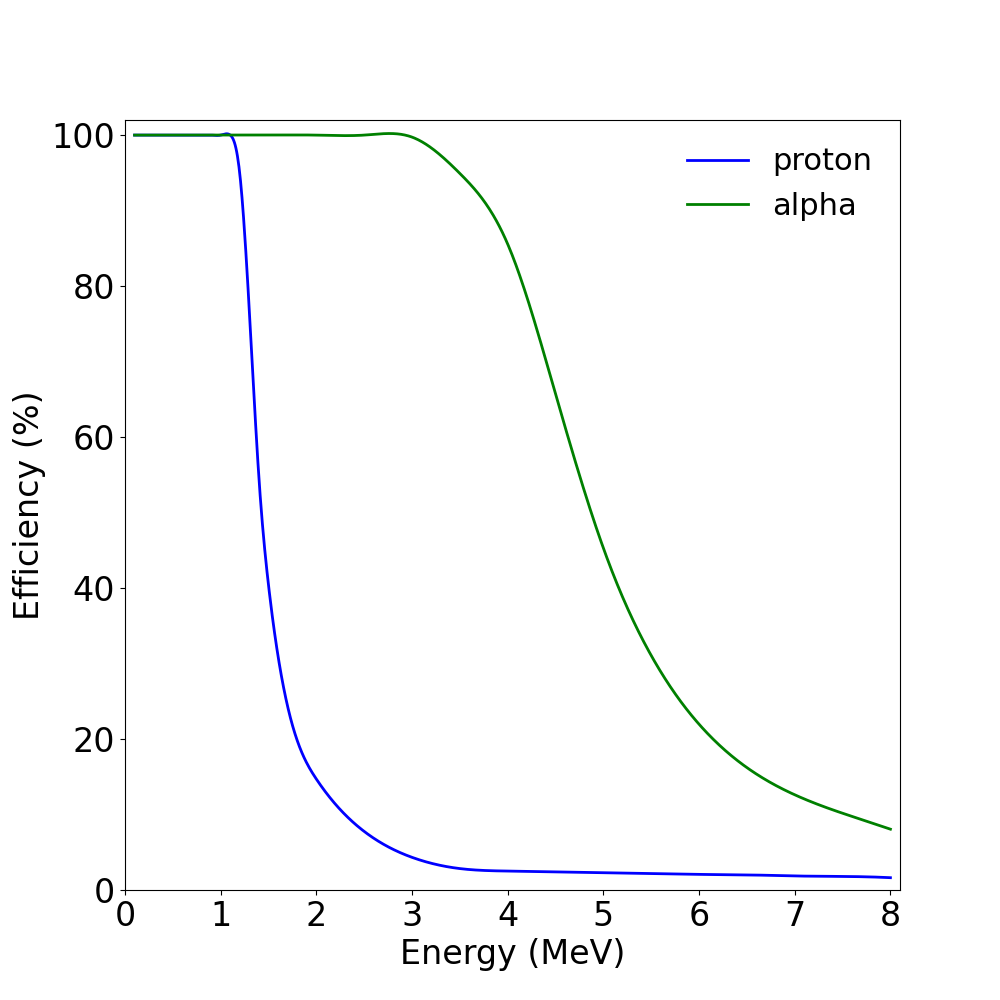}
\caption{\label{fig15} Simulated efficiency of the GADGET II TPC filled with P10 gas at 800 Torr to detect protons and 
$\alpha$-particles for the case of $^{60}$Ga decay with FRIB beam. }
\end{figure}

The detection efficiency of the GADGET II TPC filled with P10 gas mixture at 800 Torr for protons and $\alpha$ particles has been projected through ATTPCROOT simulations, utilizing a realistic experiment scenario: the decay of $^{60}$Ga. 
For the $^{60}$Ga decay, LISE++ simulations \cite{TARASOV2016185} were employed to generate a 3D thermalized beam distribution using a primary beam of $^{60}$Ge on a $^{12}$C target at FRIB facility, the Advanced Rare Isotope Separator (ARIS) \cite{HAUSMANN2013349}, and a beam-energy degrader.  Due to the short half life and relatively long timescales for Brownian motion, this distribution was used as a source distribution of protons and $\alpha$-particles. In these simulations $\beta$-particles were not included. The efficiency was then calculated by determining the ratio of events with tracks that hit the measurement pads excluding the veto pads (discussed in sub section~\ref{subsec:2_2}) and the wall effect events to the total number of simulated events.  Figure~\ref{fig15} illustrates the simulated efficiency curve for protons and $\alpha$-particles across an energy range from 0.1 to 8 MeV.  Upon acquiring experimental beam data,  the measured three-dimensional beam distribution will be utilized to evaluate efficiency for each experiment. The precise distribution of x-y hits on the pad plane yields the convolution of the beam distribution and diffusion. Furthermore, the drift time can be determined through the timing of particle-$\gamma$ coincidences, to measure the z-distribution \cite{Friedman:2019}.
\subsection{\textit{VGG16 Convolutional neural network}}
\label{subsec:5_3}
Convolutional neural networks (ConvNets) are deep learning algorithms that are ideal for image classification. One such ConvNet is the VGG16 model \cite{tamira, Simonyan2014VeryDC, KUCHERA2019156}. The VGG16 has 16 weighted layers, over 130 million parameters, and has been trained on millions of images. 
\begin{figure}[htbp]
\centering
\includegraphics[width=3.2in]{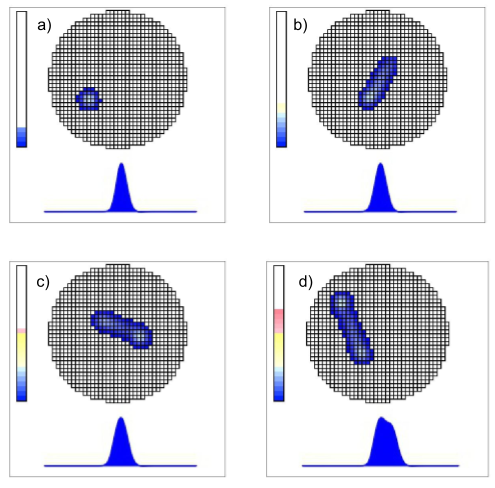}
\caption{\label{fig16} Label a)-d): Images generated for the VGG16 ConvNet using simulated data from ATTPCROOT. Each image contains a 2D projection of a track on the pad plane with relative charge density, the integrated charge as a function of time, and the total integrated charge in the form of an energy bar. The images include: a) 500 keV alpha,  b) 1200 keV proton c) 1700 keV proton-$\alpha$ event, and d) 2000 keV proton-$\alpha$-event from the decay of $^{20}$Mg. }
\end{figure}
To leverage this powerful algorithm, fine-tuning of this model is being conducted.  This involves freezing the weights of the early layers in the 
network, and developing a custom classifier that is appended to the existing architecture. It is only the custom classifier weights that are 
updated during the training/validation stage. The full 3D topology of the events of interest need to be considered for proper classification. 
However, the use of 3D ConvNets is computationally expensive. To ameliorate this, a process known as early data fusion is implemented. This allows 
condensation of the relevant 3D track information into a single 2D image with strong features for a 2D ConvNet. This involves appending the feature 
space of 2D image with three separate data modalities, namely, the 2D projection of the track with the relative charge density, the time 
projection, and the integrated charge in the form of an ``energy bar'' (Figure~\ref{fig16}). Subsequently, the ConvNet is entrusted with 
determining the relationship between these features during training. This model has undergone training and testing using simulated data from 
$^{20}$Mg events (protons, alphas, and proton-alphas), achieving a training and testing accuracy of 100$\%$. The ConvNet will be trained on a 
combination of simulated and manually labeled real data for identifying events of interest in unseen experimental data. The operation of GADGET II 
and the success of its first science experiment do not necessarily depend on this tool. Instead, this tool is specifically designed to reduce the 
labor required in rare event searches. More details on this will be communicated in a subsequent publication. 
\section{Summary and future outlook}
\label{sec-6}
A compact TPC employing a MPGD has been built  to measure low energy charged particles from $\beta$-delayed charged particle emitters of interest to nuclear astrophysics and searches for rare decays. This system is a substantial upgrade to the Proton Detector of the original GADGET system, incorporating a new high granularity resistive MM board with 1024 pads. The resistive part of this technology has been introduced for the first time in low-energy nuclear physics experiments with the primary benefit of providing added protection of the electronics from the large ionization generated by heavy ion beams. To accommodate the large number of electronics channels, the high-density GET electronics data acquisition system was implemented. For the GADGET II detection system, the number of measurement pads has increased by a factor of over 200 providing the needed spatial resolution to construct 2D images of multi -particle decay events. The third spatial dimension is obtained by measuring the relative electron drift time in the chamber. These combined measurements provide a 3D reconstruction of the decay events. This enables measurement of the range and total charge providing total kinetic energy, particle ID and detection of multi-particle emissions. The GET electronics data acquisition system was tested using a random pulse generator and $\alpha$-decay events from $^{220}$Rn taking into account the data throughput modifications. The functionality of the GADGET II TPC has been tested using a $^{228}$Th ($^{220}$Rn) source and the aggregate energy resolution is found to be 5.4 \% at 6.288 MeV. In addition, the detection capabilities of the TPC have been tested with cosmic-ray muon measurements, and the effect of diffusion as a function of distance from the MM has been studied. With these measurements, the electron drift velocity has also been extracted. The TPC has been simulated using the ATTPCROOT data analysis framework based on the FairRoot package for $^{20}$Mg and $^{220}$Rn decay events. Additionally, a novel method has been introduced for leveraging a 2D ConvNet for event classification with GADGET II. ATTPCROOT simulations have been used to produce efficiency curves for protons and $\alpha$-particles.
\newline For beam-line experiments, the high efficiency and high resolution HPGe arrays like SeGA \cite{MUELLER2001492} and FRIB Decay Station's DeGAi \cite{dega} will be integrated with the GADGET II TPC for detailed studies of the coincidences between gamma rays and charged particle emissions. Both germanium arrays use a 250 MHz XIA Pixie-16 Digital Data Acquisition System (DDAS) module \cite{PROKOP2014163}. The data streams from the GET electronics for the TPC and DDAS for the germanium detectors will be combined using time stamps. In order to improve the data throughput in future, the next generation of CoBos will be installed offering improved data throughput by a factor 2.  For experiments focusing on calorimetric measurements of low-energy $\beta$-delayed protons, particularly in cases where particle identification is not needed, one can also revert to the original GADGET detection system with the benefit of simplicity. Additionally, the TPC will be tested with different gas mixtures and gas pressures. The GADGET II scientific campaign is currently in progress including research on nuclear astrophysics and searches for exotic decays.
\FloatBarrier
\section*{Acknowledgements}
This work has been supported by the U. S. Department of Energy under award no: DE-SC0016052, DE-SC0024587 and the U. S. National Science Foundation under award no: 1565546, 1913554 and 2209429.  We gratefully acknowledge the support of NSCL/FRIB technical staff.
\medskip
\providecommand{\noopsort}[1]{}\providecommand{\singleletter}[1]{#1}%


\providecommand{\noopsort}[1]{}\providecommand{\singleletter}[1]{#1}%
\begin{thebibliography}{53}%
\makeatletter
\providecommand \@ifxundefined [1]{%
 \@ifx{#1\undefined}
}%
\providecommand \@ifnum [1]{%
 \ifnum #1\expandafter \@firstoftwo
 \else \expandafter \@secondoftwo
 \fi
}%
\providecommand \@ifx [1]{%
 \ifx #1\expandafter \@firstoftwo
 \else \expandafter \@secondoftwo
 \fi
}%
\providecommand \natexlab [1]{#1}%
\providecommand \enquote  [1]{``#1''}%
\providecommand \bibnamefont  [1]{#1}%
\providecommand \bibfnamefont [1]{#1}%
\providecommand \citenamefont [1]{#1}%
\providecommand \href@noop [0]{\@secondoftwo}%
\providecommand \href [0]{\begingroup \@sanitize@url \@href}%
\providecommand \@href[1]{\@@startlink{#1}\@@href}%
\providecommand \@@href[1]{\endgroup#1\@@endlink}%
\providecommand \@sanitize@url [0]{\catcode `\\12\catcode `\$12\catcode `\&12\catcode `\#12\catcode `\^12\catcode `\_12\catcode `\%12\relax}%
\providecommand \@@startlink[1]{}%
\providecommand \@@endlink[0]{}%
\providecommand \url  [0]{\begingroup\@sanitize@url \@url }%
\providecommand \@url [1]{\endgroup\@href {#1}{\urlprefix }}%
\providecommand \urlprefix  [0]{URL }%
\providecommand \Eprint [0]{\href }%
\providecommand \doibase [0]{https://doi.org/}%
\providecommand \selectlanguage [0]{\@gobble}%
\providecommand \bibinfo  [0]{\@secondoftwo}%
\providecommand \bibfield  [0]{\@secondoftwo}%
\providecommand \translation [1]{[#1]}%
\providecommand \BibitemOpen [0]{}%
\providecommand \bibitemStop [0]{}%
\providecommand \bibitemNoStop [0]{.\EOS\space}%
\providecommand \EOS [0]{\spacefactor3000\relax}%
\providecommand \BibitemShut  [1]{\csname bibitem#1\endcsname}%
\let\auto@bib@innerbib\@empty
\bibitem [{\citenamefont {Rolfs}(1988)}]{Rolfs:1988}%
  \BibitemOpen
  \bibfield  {author} {\bibinfo {author} {\bibfnamefont {C.~E.}\ \bibnamefont {Rolfs}},\ }\href@noop {} {\emph {\bibinfo {title} {Cauldrons in the cosmos: Nuclear astrophysics}}}\ (\bibinfo  {publisher} {University of Chicago Press},\ \bibinfo {year} {1988})\BibitemShut {NoStop}%
\bibitem [{\citenamefont {Borge}(2013)}]{Borge_2013}%
  \BibitemOpen
  \bibfield  {author} {\bibinfo {author} {\bibfnamefont {M.~J.~G.}\ \bibnamefont {Borge}},\ }\bibfield  {title} {\bibinfo {title} {Beta-delayed particle emission},\ }\href {https://doi.org/10.1088/0031-8949/2013/t152/014013} {\bibfield  {journal} {\bibinfo  {journal} {Physica Scripta}\ }\textbf {\bibinfo {volume} {T152}},\ \bibinfo {pages} {014013} (\bibinfo {year} {2013})}\BibitemShut {NoStop}%
\bibitem [{\citenamefont {Saastamoinen}\ \emph {et~al.}(2011)\citenamefont {Saastamoinen}, \citenamefont {Trache}, \citenamefont {Banu}, \citenamefont {Bentley}, \citenamefont {Davinson}, \citenamefont {Hardy}, \citenamefont {Iacob}, \citenamefont {McCleskey}, \citenamefont {Roeder}, \citenamefont {Simmons}, \citenamefont {Tabacaru}, \citenamefont {Tribble}, \citenamefont {Woods},\ and\ \citenamefont {\"Ayst\"o}}]{PhysRevC.83.045808}%
  \BibitemOpen
  \bibfield  {author} {\bibinfo {author} {\bibfnamefont {A.}~\bibnamefont {Saastamoinen}}, \bibinfo {author} {\bibfnamefont {L.}~\bibnamefont {Trache}}, \bibinfo {author} {\bibfnamefont {A.}~\bibnamefont {Banu}}, \bibinfo {author} {\bibfnamefont {M.~A.}\ \bibnamefont {Bentley}}, \bibinfo {author} {\bibfnamefont {T.}~\bibnamefont {Davinson}}, \bibinfo {author} {\bibfnamefont {J.~C.}\ \bibnamefont {Hardy}}, \bibinfo {author} {\bibfnamefont {V.~E.}\ \bibnamefont {Iacob}}, \bibinfo {author} {\bibfnamefont {M.}~\bibnamefont {McCleskey}}, \bibinfo {author} {\bibfnamefont {B.~T.}\ \bibnamefont {Roeder}}, \bibinfo {author} {\bibfnamefont {E.}~\bibnamefont {Simmons}}, \bibinfo {author} {\bibfnamefont {G.}~\bibnamefont {Tabacaru}}, \bibinfo {author} {\bibfnamefont {R.~E.}\ \bibnamefont {Tribble}}, \bibinfo {author} {\bibfnamefont {P.~J.}\ \bibnamefont {Woods}},\ and\ \bibinfo {author} {\bibfnamefont {J.}~\bibnamefont {\"Ayst\"o}},\ }\bibfield  {title} {\bibinfo {title} {Experimental study of
  $\ensuremath{\beta}$-delayed proton decay of $^{23}\mathrm{Al}$ for nucleosynthesis in novae},\ }\href {https://doi.org/10.1103/PhysRevC.83.045808} {\bibfield  {journal} {\bibinfo  {journal} {Phys. Rev. C}\ }\textbf {\bibinfo {volume} {83}},\ \bibinfo {pages} {045808} (\bibinfo {year} {2011})}\BibitemShut {NoStop}%
\bibitem [{\citenamefont {McCleskey}\ \emph {et~al.}(2013)\citenamefont {McCleskey}, \citenamefont {Trache}, \citenamefont {Saastamoinen}, \citenamefont {Banu}, \citenamefont {Simmons}, \citenamefont {Roeder}, \citenamefont {Tabacaru}, \citenamefont {Tribble}, \citenamefont {Davinson}, \citenamefont {Woods},\ and\ \citenamefont {Äystö}}]{MCCLESKEY2013124}%
  \BibitemOpen
  \bibfield  {author} {\bibinfo {author} {\bibfnamefont {M.}~\bibnamefont {McCleskey}}, \bibinfo {author} {\bibfnamefont {L.}~\bibnamefont {Trache}}, \bibinfo {author} {\bibfnamefont {A.}~\bibnamefont {Saastamoinen}}, \bibinfo {author} {\bibfnamefont {A.}~\bibnamefont {Banu}}, \bibinfo {author} {\bibfnamefont {E.}~\bibnamefont {Simmons}}, \bibinfo {author} {\bibfnamefont {B.}~\bibnamefont {Roeder}}, \bibinfo {author} {\bibfnamefont {G.}~\bibnamefont {Tabacaru}}, \bibinfo {author} {\bibfnamefont {R.}~\bibnamefont {Tribble}}, \bibinfo {author} {\bibfnamefont {T.}~\bibnamefont {Davinson}}, \bibinfo {author} {\bibfnamefont {P.}~\bibnamefont {Woods}},\ and\ \bibinfo {author} {\bibfnamefont {J.}~\bibnamefont {Äystö}},\ }\bibfield  {title} {\bibinfo {title} {Implantation-decay station for low-energy $\beta$-delayed proton measurements},\ }\href {https://doi.org/https://doi.org/10.1016/j.nima.2012.10.069} {\bibfield  {journal} {\bibinfo  {journal} {Nuclear Instruments and Methods in Physics Research Section A:
  Accelerators, Spectrometers, Detectors and Associated Equipment}\ }\textbf {\bibinfo {volume} {700}},\ \bibinfo {pages} {124} (\bibinfo {year} {2013})}\BibitemShut {NoStop}%
\bibitem [{\citenamefont {Wallace}\ \emph {et~al.}(2012)\citenamefont {Wallace}, \citenamefont {Woods}, \citenamefont {Lotay}, \citenamefont {Alharbi}, \citenamefont {Banu}, \citenamefont {David}, \citenamefont {Davinson}, \citenamefont {McCleskey}, \citenamefont {Roeder}, \citenamefont {Simmons}, \citenamefont {Spiridon}, \citenamefont {Trache},\ and\ \citenamefont {Tribble}}]{WALLACE201259}%
  \BibitemOpen
  \bibfield  {author} {\bibinfo {author} {\bibfnamefont {J.}~\bibnamefont {Wallace}}, \bibinfo {author} {\bibfnamefont {P.}~\bibnamefont {Woods}}, \bibinfo {author} {\bibfnamefont {G.}~\bibnamefont {Lotay}}, \bibinfo {author} {\bibfnamefont {A.}~\bibnamefont {Alharbi}}, \bibinfo {author} {\bibfnamefont {A.}~\bibnamefont {Banu}}, \bibinfo {author} {\bibfnamefont {H.}~\bibnamefont {David}}, \bibinfo {author} {\bibfnamefont {T.}~\bibnamefont {Davinson}}, \bibinfo {author} {\bibfnamefont {M.}~\bibnamefont {McCleskey}}, \bibinfo {author} {\bibfnamefont {B.}~\bibnamefont {Roeder}}, \bibinfo {author} {\bibfnamefont {E.}~\bibnamefont {Simmons}}, \bibinfo {author} {\bibfnamefont {A.}~\bibnamefont {Spiridon}}, \bibinfo {author} {\bibfnamefont {L.}~\bibnamefont {Trache}},\ and\ \bibinfo {author} {\bibfnamefont {R.}~\bibnamefont {Tribble}},\ }\bibfield  {title} {\bibinfo {title} {{$\beta$-Delayed proton-decay study of $^{20}$Mg and its implications for the $^{19}$Ne(p,$\gamma$)$^{20}$Na breakout reaction in X-ray
  bursts}},\ }\href {https://doi.org/https://doi.org/10.1016/j.physletb.2012.04.046} {\bibfield  {journal} {\bibinfo  {journal} {Physics Letters B}\ }\textbf {\bibinfo {volume} {712}},\ \bibinfo {pages} {59} (\bibinfo {year} {2012})}\BibitemShut {NoStop}%
\bibitem [{\citenamefont {Pollacco}\ \emph {et~al.}(2013)\citenamefont {Pollacco}, \citenamefont {Trache}, \citenamefont {Simmons}, \citenamefont {Spiridon}, \citenamefont {McCleskey}, \citenamefont {Roeder}, \citenamefont {Saastamoinen}, \citenamefont {Tribble}, \citenamefont {Pascovici}, \citenamefont {Kebbiri}, \citenamefont {Mols},\ and\ \citenamefont {Raillot}}]{POLLACCO2013102}%
  \BibitemOpen
  \bibfield  {author} {\bibinfo {author} {\bibfnamefont {E.}~\bibnamefont {Pollacco}}, \bibinfo {author} {\bibfnamefont {L.}~\bibnamefont {Trache}}, \bibinfo {author} {\bibfnamefont {E.}~\bibnamefont {Simmons}}, \bibinfo {author} {\bibfnamefont {A.}~\bibnamefont {Spiridon}}, \bibinfo {author} {\bibfnamefont {M.}~\bibnamefont {McCleskey}}, \bibinfo {author} {\bibfnamefont {B.}~\bibnamefont {Roeder}}, \bibinfo {author} {\bibfnamefont {A.}~\bibnamefont {Saastamoinen}}, \bibinfo {author} {\bibfnamefont {R.}~\bibnamefont {Tribble}}, \bibinfo {author} {\bibfnamefont {G.}~\bibnamefont {Pascovici}}, \bibinfo {author} {\bibfnamefont {M.}~\bibnamefont {Kebbiri}}, \bibinfo {author} {\bibfnamefont {J.}~\bibnamefont {Mols}},\ and\ \bibinfo {author} {\bibfnamefont {M.}~\bibnamefont {Raillot}},\ }\bibfield  {title} {\bibinfo {title} {{AstroBox: A novel detection system for very low-energy protons from $\beta$-delayed proton decay}},\ }\href {https://doi.org/https://doi.org/10.1016/j.nima.2013.04.084} {\bibfield  {journal}
  {\bibinfo  {journal} {Nuclear Instruments and Methods in Physics Research Section A: Accelerators, Spectrometers, Detectors and Associated Equipment}\ }\textbf {\bibinfo {volume} {723}},\ \bibinfo {pages} {102} (\bibinfo {year} {2013})}\BibitemShut {NoStop}%
\bibitem [{\citenamefont {{Saastamoinen}}\ \emph {et~al.}(2016)\citenamefont {{Saastamoinen}}, \citenamefont {{Pollacco}}, \citenamefont {{Roeder}}, \citenamefont {{Spiridon}}, \citenamefont {{Daq}}, \citenamefont {{Trache}}, \citenamefont {{Pascovici}}, \citenamefont {{De Oliveira}}, \citenamefont {{Rodrigues}},\ and\ \citenamefont {{Tribble}}}]{2016NIMPB.376..357S}%
  \BibitemOpen
  \bibfield  {author} {\bibinfo {author} {\bibfnamefont {A.}~\bibnamefont {{Saastamoinen}}}, \bibinfo {author} {\bibfnamefont {E.}~\bibnamefont {{Pollacco}}}, \bibinfo {author} {\bibfnamefont {B.~T.}\ \bibnamefont {{Roeder}}}, \bibinfo {author} {\bibfnamefont {A.}~\bibnamefont {{Spiridon}}}, \bibinfo {author} {\bibfnamefont {M.}~\bibnamefont {{Daq}}}, \bibinfo {author} {\bibfnamefont {L.}~\bibnamefont {{Trache}}}, \bibinfo {author} {\bibfnamefont {G.}~\bibnamefont {{Pascovici}}}, \bibinfo {author} {\bibfnamefont {R.}~\bibnamefont {{De Oliveira}}}, \bibinfo {author} {\bibfnamefont {M.~R.~D.}\ \bibnamefont {{Rodrigues}}},\ and\ \bibinfo {author} {\bibfnamefont {R.~E.}\ \bibnamefont {{Tribble}}},\ }\bibfield  {title} {\bibinfo {title} {{AstroBox2 - Detector for low-energy {\ensuremath{\beta}}-delayed particle detection}},\ }\href {https://doi.org/10.1016/j.nimb.2016.02.020} {\bibfield  {journal} {\bibinfo  {journal} {Nuclear Instruments and Methods in Physics Research B}\ }\textbf {\bibinfo {volume} {376}},\
  \bibinfo {pages} {357} (\bibinfo {year} {2016})}\BibitemShut {NoStop}%
\bibitem [{\citenamefont {Friedman}(2019)}]{Friedman:2019}%
  \BibitemOpen
  \bibfield  {author} {\bibinfo {author} {\bibfnamefont {M.}~\bibnamefont {Friedman}},\ }\bibfield  {title} {\bibinfo {title} {Gadget: a gaseous detector with germanium tagging},\ }\bibfield  {journal} {\bibinfo  {journal} {Nuclear Instruments and Methods in Physics Research Section A: Accelerators, Spectrometers, Detectors and Associated Equipment}\ }\textbf {\bibinfo {volume} {940}},\ \href {https://doi.org/10.1016/j.nima.2019.05.100} {10.1016/j.nima.2019.05.100} (\bibinfo {year} {2019})\BibitemShut {NoStop}%
\bibitem [{\citenamefont {Budner}\ \emph {et~al.}(2022)\citenamefont {Budner}, \citenamefont {Friedman}, \citenamefont {Wrede}, \citenamefont {Brown}, \citenamefont {Jos\'e}, \citenamefont {P\'erez-Loureiro}, \citenamefont {Sun}, \citenamefont {Surbrook}, \citenamefont {Ayyad}, \citenamefont {Bardayan}, \citenamefont {Chae}, \citenamefont {Chen}, \citenamefont {Chipps}, \citenamefont {Cortesi}, \citenamefont {Glassman}, \citenamefont {Hall}, \citenamefont {Janasik}, \citenamefont {Liang}, \citenamefont {O'Malley}, \citenamefont {Pollacco}, \citenamefont {Psaltis}, \citenamefont {Stomps},\ and\ \citenamefont {Wheeler}}]{PhysRevLett.128.182701}%
  \BibitemOpen
  \bibfield  {author} {\bibinfo {author} {\bibfnamefont {T.}~\bibnamefont {Budner}}, \bibinfo {author} {\bibfnamefont {M.}~\bibnamefont {Friedman}}, \bibinfo {author} {\bibfnamefont {C.}~\bibnamefont {Wrede}}, \bibinfo {author} {\bibfnamefont {B.~A.}\ \bibnamefont {Brown}}, \bibinfo {author} {\bibfnamefont {J.}~\bibnamefont {Jos\'e}}, \bibinfo {author} {\bibfnamefont {D.}~\bibnamefont {P\'erez-Loureiro}}, \bibinfo {author} {\bibfnamefont {L.~J.}\ \bibnamefont {Sun}}, \bibinfo {author} {\bibfnamefont {J.}~\bibnamefont {Surbrook}}, \bibinfo {author} {\bibfnamefont {Y.}~\bibnamefont {Ayyad}}, \bibinfo {author} {\bibfnamefont {D.~W.}\ \bibnamefont {Bardayan}}, \bibinfo {author} {\bibfnamefont {K.}~\bibnamefont {Chae}}, \bibinfo {author} {\bibfnamefont {A.~A.}\ \bibnamefont {Chen}}, \bibinfo {author} {\bibfnamefont {K.~A.}\ \bibnamefont {Chipps}}, \bibinfo {author} {\bibfnamefont {M.}~\bibnamefont {Cortesi}}, \bibinfo {author} {\bibfnamefont {B.}~\bibnamefont {Glassman}}, \bibinfo {author} {\bibfnamefont {M.~R.}\
  \bibnamefont {Hall}}, \bibinfo {author} {\bibfnamefont {M.}~\bibnamefont {Janasik}}, \bibinfo {author} {\bibfnamefont {J.}~\bibnamefont {Liang}}, \bibinfo {author} {\bibfnamefont {P.}~\bibnamefont {O'Malley}}, \bibinfo {author} {\bibfnamefont {E.}~\bibnamefont {Pollacco}}, \bibinfo {author} {\bibfnamefont {A.}~\bibnamefont {Psaltis}}, \bibinfo {author} {\bibfnamefont {J.}~\bibnamefont {Stomps}},\ and\ \bibinfo {author} {\bibfnamefont {T.}~\bibnamefont {Wheeler}},\ }\bibfield  {title} {\bibinfo {title} {{Constraining the $^{30}\mathrm{P}(p,\text{ }\ensuremath{\gamma})^{31}\mathrm{S}$ Reaction Rate in ONe Novae via the Weak, Low-Energy, $\ensuremath{\beta}$-Delayed Proton Decay of $^{31}\mathrm{Cl}$}},\ }\href {https://doi.org/10.1103/PhysRevLett.128.182701} {\bibfield  {journal} {\bibinfo  {journal} {Phys. Rev. Lett.}\ }\textbf {\bibinfo {volume} {128}},\ \bibinfo {pages} {182701} (\bibinfo {year} {2022})}\BibitemShut {NoStop}%
\bibitem [{\citenamefont {Friedman}\ \emph {et~al.}(2020)\citenamefont {Friedman}, \citenamefont {Budner}, \citenamefont {P\'erez-Loureiro}, \citenamefont {Pollacco}, \citenamefont {Wrede}, \citenamefont {Jos\'e}, \citenamefont {Brown}, \citenamefont {Cortesi}, \citenamefont {Fry}, \citenamefont {Glassman}, \citenamefont {Heideman}, \citenamefont {Janasik}, \citenamefont {Roosa}, \citenamefont {Stomps}, \citenamefont {Surbrook},\ and\ \citenamefont {Tiwari}}]{PhysRevC.101.052802}%
  \BibitemOpen
  \bibfield  {author} {\bibinfo {author} {\bibfnamefont {M.}~\bibnamefont {Friedman}}, \bibinfo {author} {\bibfnamefont {T.}~\bibnamefont {Budner}}, \bibinfo {author} {\bibfnamefont {D.}~\bibnamefont {P\'erez-Loureiro}}, \bibinfo {author} {\bibfnamefont {E.}~\bibnamefont {Pollacco}}, \bibinfo {author} {\bibfnamefont {C.}~\bibnamefont {Wrede}}, \bibinfo {author} {\bibfnamefont {J.}~\bibnamefont {Jos\'e}}, \bibinfo {author} {\bibfnamefont {B.~A.}\ \bibnamefont {Brown}}, \bibinfo {author} {\bibfnamefont {M.}~\bibnamefont {Cortesi}}, \bibinfo {author} {\bibfnamefont {C.}~\bibnamefont {Fry}}, \bibinfo {author} {\bibfnamefont {B.}~\bibnamefont {Glassman}}, \bibinfo {author} {\bibfnamefont {J.}~\bibnamefont {Heideman}}, \bibinfo {author} {\bibfnamefont {M.}~\bibnamefont {Janasik}}, \bibinfo {author} {\bibfnamefont {M.}~\bibnamefont {Roosa}}, \bibinfo {author} {\bibfnamefont {J.}~\bibnamefont {Stomps}}, \bibinfo {author} {\bibfnamefont {J.}~\bibnamefont {Surbrook}},\ and\ \bibinfo {author} {\bibfnamefont
  {P.}~\bibnamefont {Tiwari}},\ }\bibfield  {title} {\bibinfo {title} {Low-energy $^{23}\mathrm{Al}\phantom{\rule{0.28em}{0ex}}\ensuremath{\beta}$-delayed proton decay and $^{22}\mathrm{Na}$ destruction in novae},\ }\href {https://doi.org/10.1103/PhysRevC.101.052802} {\bibfield  {journal} {\bibinfo  {journal} {Phys. Rev. C}\ }\textbf {\bibinfo {volume} {101}},\ \bibinfo {pages} {052802} (\bibinfo {year} {2020})}\BibitemShut {NoStop}%
\bibitem [{\citenamefont {Sun}\ \emph {et~al.}(2021)\citenamefont {Sun}, \citenamefont {Friedman}, \citenamefont {Budner}, \citenamefont {P\'erez-Loureiro}, \citenamefont {Pollacco}, \citenamefont {Wrede}, \citenamefont {Brown}, \citenamefont {Cortesi}, \citenamefont {Fry}, \citenamefont {Glassman}, \citenamefont {Heideman}, \citenamefont {Janasik}, \citenamefont {Kruskie}, \citenamefont {Magilligan}, \citenamefont {Roosa}, \citenamefont {Stomps}, \citenamefont {Surbrook},\ and\ \citenamefont {Tiwari}}]{PhysRevC.103.014322}%
  \BibitemOpen
  \bibfield  {author} {\bibinfo {author} {\bibfnamefont {L.~J.}\ \bibnamefont {Sun}}, \bibinfo {author} {\bibfnamefont {M.}~\bibnamefont {Friedman}}, \bibinfo {author} {\bibfnamefont {T.}~\bibnamefont {Budner}}, \bibinfo {author} {\bibfnamefont {D.}~\bibnamefont {P\'erez-Loureiro}}, \bibinfo {author} {\bibfnamefont {E.}~\bibnamefont {Pollacco}}, \bibinfo {author} {\bibfnamefont {C.}~\bibnamefont {Wrede}}, \bibinfo {author} {\bibfnamefont {B.~A.}\ \bibnamefont {Brown}}, \bibinfo {author} {\bibfnamefont {M.}~\bibnamefont {Cortesi}}, \bibinfo {author} {\bibfnamefont {C.}~\bibnamefont {Fry}}, \bibinfo {author} {\bibfnamefont {B.~E.}\ \bibnamefont {Glassman}}, \bibinfo {author} {\bibfnamefont {J.}~\bibnamefont {Heideman}}, \bibinfo {author} {\bibfnamefont {M.}~\bibnamefont {Janasik}}, \bibinfo {author} {\bibfnamefont {A.}~\bibnamefont {Kruskie}}, \bibinfo {author} {\bibfnamefont {A.}~\bibnamefont {Magilligan}}, \bibinfo {author} {\bibfnamefont {M.}~\bibnamefont {Roosa}}, \bibinfo {author} {\bibfnamefont
  {J.}~\bibnamefont {Stomps}}, \bibinfo {author} {\bibfnamefont {J.}~\bibnamefont {Surbrook}},\ and\ \bibinfo {author} {\bibfnamefont {P.}~\bibnamefont {Tiwari}},\ }\bibfield  {title} {\bibinfo {title} {$^{25}\mathrm{Si}\phantom{\rule{4pt}{0ex}}{\ensuremath{\beta}}^{+}$-decay spectroscopy},\ }\href {https://doi.org/10.1103/PhysRevC.103.014322} {\bibfield  {journal} {\bibinfo  {journal} {Phys. Rev. C}\ }\textbf {\bibinfo {volume} {103}},\ \bibinfo {pages} {014322} (\bibinfo {year} {2021})}\BibitemShut {NoStop}%
\bibitem [{\citenamefont {Surbrook}(2023)}]{Surbrook2022}%
  \BibitemOpen
  \bibfield  {author} {\bibinfo {author} {\bibfnamefont {J.}~\bibnamefont {Surbrook}},\ }\emph {\bibinfo {title} {A SEARCH FOR NOVEL DECAY MODES IN 11-BERYLLIUM}},\ \href@noop {} {Ph.D. thesis},\ \bibinfo  {school} {MSU} (\bibinfo {year} {2023}),\ \bibinfo {note} {published thesis}\BibitemShut {NoStop}%
\bibitem [{\citenamefont {Wiescher}\ \emph {et~al.}(1999)\citenamefont {Wiescher}, \citenamefont {Görres},\ and\ \citenamefont {Schatz}}]{Wiescher_1999}%
  \BibitemOpen
  \bibfield  {author} {\bibinfo {author} {\bibfnamefont {M.}~\bibnamefont {Wiescher}}, \bibinfo {author} {\bibfnamefont {J.}~\bibnamefont {Görres}},\ and\ \bibinfo {author} {\bibfnamefont {H.}~\bibnamefont {Schatz}},\ }\bibfield  {title} {\bibinfo {title} {Break-out reactions from the cno cycles},\ }\href {https://doi.org/10.1088/0954-3899/25/6/201} {\bibfield  {journal} {\bibinfo  {journal} {Journal of Physics G: Nuclear and Particle Physics}\ }\textbf {\bibinfo {volume} {25}},\ \bibinfo {pages} {R133} (\bibinfo {year} {1999})}\BibitemShut {NoStop}%
\bibitem [{\citenamefont {Wrede}\ \emph {et~al.}(2017)\citenamefont {Wrede}, \citenamefont {Glassman}, \citenamefont {P\'erez-Loureiro}, \citenamefont {Allen}, \citenamefont {Bardayan}, \citenamefont {Bennett}, \citenamefont {Brown}, \citenamefont {Chipps}, \citenamefont {Febbraro}, \citenamefont {Fry}, \citenamefont {Hall}, \citenamefont {Hall}, \citenamefont {Liddick}, \citenamefont {O'Malley}, \citenamefont {Ong}, \citenamefont {Pain}, \citenamefont {Schwartz}, \citenamefont {Shidling}, \citenamefont {Sims}, \citenamefont {Thompson},\ and\ \citenamefont {Zhang}}]{PhysRevC.96.032801}%
  \BibitemOpen
  \bibfield  {author} {\bibinfo {author} {\bibfnamefont {C.}~\bibnamefont {Wrede}}, \bibinfo {author} {\bibfnamefont {B.~E.}\ \bibnamefont {Glassman}}, \bibinfo {author} {\bibfnamefont {D.}~\bibnamefont {P\'erez-Loureiro}}, \bibinfo {author} {\bibfnamefont {J.~M.}\ \bibnamefont {Allen}}, \bibinfo {author} {\bibfnamefont {D.~W.}\ \bibnamefont {Bardayan}}, \bibinfo {author} {\bibfnamefont {M.~B.}\ \bibnamefont {Bennett}}, \bibinfo {author} {\bibfnamefont {B.~A.}\ \bibnamefont {Brown}}, \bibinfo {author} {\bibfnamefont {K.~A.}\ \bibnamefont {Chipps}}, \bibinfo {author} {\bibfnamefont {M.}~\bibnamefont {Febbraro}}, \bibinfo {author} {\bibfnamefont {C.}~\bibnamefont {Fry}}, \bibinfo {author} {\bibfnamefont {M.~R.}\ \bibnamefont {Hall}}, \bibinfo {author} {\bibfnamefont {O.}~\bibnamefont {Hall}}, \bibinfo {author} {\bibfnamefont {S.~N.}\ \bibnamefont {Liddick}}, \bibinfo {author} {\bibfnamefont {P.}~\bibnamefont {O'Malley}}, \bibinfo {author} {\bibfnamefont {W.-J.}\ \bibnamefont {Ong}}, \bibinfo {author}
  {\bibfnamefont {S.~D.}\ \bibnamefont {Pain}}, \bibinfo {author} {\bibfnamefont {S.~B.}\ \bibnamefont {Schwartz}}, \bibinfo {author} {\bibfnamefont {P.}~\bibnamefont {Shidling}}, \bibinfo {author} {\bibfnamefont {H.}~\bibnamefont {Sims}}, \bibinfo {author} {\bibfnamefont {P.}~\bibnamefont {Thompson}},\ and\ \bibinfo {author} {\bibfnamefont {H.}~\bibnamefont {Zhang}},\ }\bibfield  {title} {\bibinfo {title} {New portal to the $^{15}\mathbf{O}(\ensuremath{\alpha},\ensuremath{\gamma})^{19}\mathbf{Ne}$ resonance triggering cno-cycle breakout},\ }\href {https://doi.org/10.1103/PhysRevC.96.032801} {\bibfield  {journal} {\bibinfo  {journal} {Phys. Rev. C}\ }\textbf {\bibinfo {volume} {96}},\ \bibinfo {pages} {032801} (\bibinfo {year} {2017})}\BibitemShut {NoStop}%
\bibitem [{\citenamefont {Cyburt}\ \emph {et~al.}(2016)\citenamefont {Cyburt}, \citenamefont {Amthor}, \citenamefont {Heger}, \citenamefont {Johnson}, \citenamefont {Keek}, \citenamefont {Meisel}, \citenamefont {Schatz},\ and\ \citenamefont {Smith}}]{Cyburt_2016}%
  \BibitemOpen
  \bibfield  {author} {\bibinfo {author} {\bibfnamefont {R.~H.}\ \bibnamefont {Cyburt}}, \bibinfo {author} {\bibfnamefont {A.~M.}\ \bibnamefont {Amthor}}, \bibinfo {author} {\bibfnamefont {A.}~\bibnamefont {Heger}}, \bibinfo {author} {\bibfnamefont {E.}~\bibnamefont {Johnson}}, \bibinfo {author} {\bibfnamefont {L.}~\bibnamefont {Keek}}, \bibinfo {author} {\bibfnamefont {Z.}~\bibnamefont {Meisel}}, \bibinfo {author} {\bibfnamefont {H.}~\bibnamefont {Schatz}},\ and\ \bibinfo {author} {\bibfnamefont {K.}~\bibnamefont {Smith}},\ }\bibfield  {title} {\bibinfo {title} {{DEPENDENCE} {OF} x-{RAY} {BURST} {MODELS} {ON} {NUCLEAR} {REACTION} {RATES}},\ }\href {https://doi.org/10.3847/0004-637x/830/2/55} {\bibfield  {journal} {\bibinfo  {journal} {The Astrophysical Journal}\ }\textbf {\bibinfo {volume} {830}},\ \bibinfo {pages} {55} (\bibinfo {year} {2016})}\BibitemShut {NoStop}%
\bibitem [{\citenamefont {Fisker}\ \emph {et~al.}(2006)\citenamefont {Fisker}, \citenamefont {Gorres}, \citenamefont {Wiescher},\ and\ \citenamefont {Davids}}]{Fisker_2006}%
  \BibitemOpen
  \bibfield  {author} {\bibinfo {author} {\bibfnamefont {J.~L.}\ \bibnamefont {Fisker}}, \bibinfo {author} {\bibfnamefont {J.}~\bibnamefont {Gorres}}, \bibinfo {author} {\bibfnamefont {M.}~\bibnamefont {Wiescher}},\ and\ \bibinfo {author} {\bibfnamefont {B.}~\bibnamefont {Davids}},\ }\bibfield  {title} {\bibinfo {title} {{The Importance of $^{15}$O($\alpha$, $\gamma$)$^{19}$Ne to X-Ray Bursts and Superbursts}},\ }\href {https://doi.org/10.1086/507083} {\bibfield  {journal} {\bibinfo  {journal} {The Astrophysical Journal}\ }\textbf {\bibinfo {volume} {650}},\ \bibinfo {pages} {332} (\bibinfo {year} {2006})}\BibitemShut {NoStop}%
\bibitem [{\citenamefont {Davids}\ \emph {et~al.}(2011)\citenamefont {Davids}, \citenamefont {Cyburt}, \citenamefont {Jos{\'{e}}},\ and\ \citenamefont {Mythili}}]{Davids_2011}%
  \BibitemOpen
  \bibfield  {author} {\bibinfo {author} {\bibfnamefont {B.}~\bibnamefont {Davids}}, \bibinfo {author} {\bibfnamefont {R.~H.}\ \bibnamefont {Cyburt}}, \bibinfo {author} {\bibfnamefont {J.}~\bibnamefont {Jos{\'{e}}}},\ and\ \bibinfo {author} {\bibfnamefont {S.}~\bibnamefont {Mythili}},\ }\bibfield  {title} {\bibinfo {title} {{The influence of uncertainties in the $^{15}$O($\alpha$, $\gamma$)$^{19}$Ne reaction rate on the models of Type I X-ray bursts}},\ }\href {https://doi.org/10.1088/0004-637x/735/1/40} {\bibfield  {journal} {\bibinfo  {journal} {The Astrophysical Journal}\ }\textbf {\bibinfo {volume} {735}},\ \bibinfo {pages} {40} (\bibinfo {year} {2011})}\BibitemShut {NoStop}%
\bibitem [{\citenamefont {Mythili}\ \emph {et~al.}(2008)\citenamefont {Mythili}, \citenamefont {Davids}, \citenamefont {Alexander}, \citenamefont {Ball}, \citenamefont {Chicoine}, \citenamefont {Chakrawarthy}, \citenamefont {Churchman}, \citenamefont {Forster}, \citenamefont {Gujrathi}, \citenamefont {Hackman}, \citenamefont {Howell}, \citenamefont {Kanungo}, \citenamefont {Leslie}, \citenamefont {Padilla}, \citenamefont {Pearson}, \citenamefont {Ruiz}, \citenamefont {Ruprecht}, \citenamefont {Schumaker}, \citenamefont {Tanihata}, \citenamefont {Vockenhuber}, \citenamefont {Walden},\ and\ \citenamefont {Yen}}]{PhysRevC.77.035803}%
  \BibitemOpen
  \bibfield  {author} {\bibinfo {author} {\bibfnamefont {S.}~\bibnamefont {Mythili}}, \bibinfo {author} {\bibfnamefont {B.}~\bibnamefont {Davids}}, \bibinfo {author} {\bibfnamefont {T.~K.}\ \bibnamefont {Alexander}}, \bibinfo {author} {\bibfnamefont {G.~C.}\ \bibnamefont {Ball}}, \bibinfo {author} {\bibfnamefont {M.}~\bibnamefont {Chicoine}}, \bibinfo {author} {\bibfnamefont {R.~S.}\ \bibnamefont {Chakrawarthy}}, \bibinfo {author} {\bibfnamefont {R.}~\bibnamefont {Churchman}}, \bibinfo {author} {\bibfnamefont {J.~S.}\ \bibnamefont {Forster}}, \bibinfo {author} {\bibfnamefont {S.}~\bibnamefont {Gujrathi}}, \bibinfo {author} {\bibfnamefont {G.}~\bibnamefont {Hackman}}, \bibinfo {author} {\bibfnamefont {D.}~\bibnamefont {Howell}}, \bibinfo {author} {\bibfnamefont {R.}~\bibnamefont {Kanungo}}, \bibinfo {author} {\bibfnamefont {J.~R.}\ \bibnamefont {Leslie}}, \bibinfo {author} {\bibfnamefont {E.}~\bibnamefont {Padilla}}, \bibinfo {author} {\bibfnamefont {C.~J.}\ \bibnamefont {Pearson}}, \bibinfo {author}
  {\bibfnamefont {C.}~\bibnamefont {Ruiz}}, \bibinfo {author} {\bibfnamefont {G.}~\bibnamefont {Ruprecht}}, \bibinfo {author} {\bibfnamefont {M.~A.}\ \bibnamefont {Schumaker}}, \bibinfo {author} {\bibfnamefont {I.}~\bibnamefont {Tanihata}}, \bibinfo {author} {\bibfnamefont {C.}~\bibnamefont {Vockenhuber}}, \bibinfo {author} {\bibfnamefont {P.}~\bibnamefont {Walden}},\ and\ \bibinfo {author} {\bibfnamefont {S.}~\bibnamefont {Yen}},\ }\bibfield  {title} {\bibinfo {title} {Lifetimes of states in $^{19}\mathrm{Ne}$ above the $^{15}\mathrm{O}$$+\ensuremath{\alpha}$ breakup threshold},\ }\href {https://doi.org/10.1103/PhysRevC.77.035803} {\bibfield  {journal} {\bibinfo  {journal} {Phys. Rev. C}\ }\textbf {\bibinfo {volume} {77}},\ \bibinfo {pages} {035803} (\bibinfo {year} {2008})}\BibitemShut {NoStop}%
\bibitem [{\citenamefont {Tan}\ \emph {et~al.}(2005)\citenamefont {Tan}, \citenamefont {G\"orres}, \citenamefont {Daly}, \citenamefont {Couder}, \citenamefont {Couture}, \citenamefont {Lee}, \citenamefont {Stech}, \citenamefont {Strandberg}, \citenamefont {Ugalde},\ and\ \citenamefont {Wiescher}}]{PhysRevC.72.041302}%
  \BibitemOpen
  \bibfield  {author} {\bibinfo {author} {\bibfnamefont {W.~P.}\ \bibnamefont {Tan}}, \bibinfo {author} {\bibfnamefont {J.}~\bibnamefont {G\"orres}}, \bibinfo {author} {\bibfnamefont {J.}~\bibnamefont {Daly}}, \bibinfo {author} {\bibfnamefont {M.}~\bibnamefont {Couder}}, \bibinfo {author} {\bibfnamefont {A.}~\bibnamefont {Couture}}, \bibinfo {author} {\bibfnamefont {H.~Y.}\ \bibnamefont {Lee}}, \bibinfo {author} {\bibfnamefont {E.}~\bibnamefont {Stech}}, \bibinfo {author} {\bibfnamefont {E.}~\bibnamefont {Strandberg}}, \bibinfo {author} {\bibfnamefont {C.}~\bibnamefont {Ugalde}},\ and\ \bibinfo {author} {\bibfnamefont {M.}~\bibnamefont {Wiescher}},\ }\bibfield  {title} {\bibinfo {title} {{Lifetime of the astrophysically important 4.03-MeV state in $^{19}\mathrm{Ne}$}},\ }\href {https://doi.org/10.1103/PhysRevC.72.041302} {\bibfield  {journal} {\bibinfo  {journal} {Phys. Rev. C}\ }\textbf {\bibinfo {volume} {72}},\ \bibinfo {pages} {041302} (\bibinfo {year} {2005})}\BibitemShut {NoStop}%
\bibitem [{\citenamefont {Kanungo}\ \emph {et~al.}(2006)\citenamefont {Kanungo}, \citenamefont {Alexander}, \citenamefont {Andreyev}, \citenamefont {Ball}, \citenamefont {Chakrawarthy}, \citenamefont {Chicoine}, \citenamefont {Churchman}, \citenamefont {Davids}, \citenamefont {Forster}, \citenamefont {Gujrathi}, \citenamefont {Hackman}, \citenamefont {Howell}, \citenamefont {Leslie}, \citenamefont {Morton}, \citenamefont {Mythili}, \citenamefont {Pearson}, \citenamefont {Ressler}, \citenamefont {Ruiz}, \citenamefont {Savajols}, \citenamefont {Schumaker}, \citenamefont {Tanihata}, \citenamefont {Walden},\ and\ \citenamefont {Yen}}]{PhysRevC.74.045803}%
  \BibitemOpen
  \bibfield  {author} {\bibinfo {author} {\bibfnamefont {R.}~\bibnamefont {Kanungo}}, \bibinfo {author} {\bibfnamefont {T.~K.}\ \bibnamefont {Alexander}}, \bibinfo {author} {\bibfnamefont {A.~N.}\ \bibnamefont {Andreyev}}, \bibinfo {author} {\bibfnamefont {G.~C.}\ \bibnamefont {Ball}}, \bibinfo {author} {\bibfnamefont {R.~S.}\ \bibnamefont {Chakrawarthy}}, \bibinfo {author} {\bibfnamefont {M.}~\bibnamefont {Chicoine}}, \bibinfo {author} {\bibfnamefont {R.}~\bibnamefont {Churchman}}, \bibinfo {author} {\bibfnamefont {B.}~\bibnamefont {Davids}}, \bibinfo {author} {\bibfnamefont {J.~S.}\ \bibnamefont {Forster}}, \bibinfo {author} {\bibfnamefont {S.}~\bibnamefont {Gujrathi}}, \bibinfo {author} {\bibfnamefont {G.}~\bibnamefont {Hackman}}, \bibinfo {author} {\bibfnamefont {D.}~\bibnamefont {Howell}}, \bibinfo {author} {\bibfnamefont {J.~R.}\ \bibnamefont {Leslie}}, \bibinfo {author} {\bibfnamefont {A.~C.}\ \bibnamefont {Morton}}, \bibinfo {author} {\bibfnamefont {S.}~\bibnamefont {Mythili}}, \bibinfo {author}
  {\bibfnamefont {C.~J.}\ \bibnamefont {Pearson}}, \bibinfo {author} {\bibfnamefont {J.~J.}\ \bibnamefont {Ressler}}, \bibinfo {author} {\bibfnamefont {C.}~\bibnamefont {Ruiz}}, \bibinfo {author} {\bibfnamefont {H.}~\bibnamefont {Savajols}}, \bibinfo {author} {\bibfnamefont {M.~A.}\ \bibnamefont {Schumaker}}, \bibinfo {author} {\bibfnamefont {I.}~\bibnamefont {Tanihata}}, \bibinfo {author} {\bibfnamefont {P.}~\bibnamefont {Walden}},\ and\ \bibinfo {author} {\bibfnamefont {S.}~\bibnamefont {Yen}},\ }\bibfield  {title} {\bibinfo {title} {{Lifetime of ${}^{19}{\mathrm{Ne}}^{*}$(4.03 MeV)}},\ }\href {https://doi.org/10.1103/PhysRevC.74.045803} {\bibfield  {journal} {\bibinfo  {journal} {Phys. Rev. C}\ }\textbf {\bibinfo {volume} {74}},\ \bibinfo {pages} {045803} (\bibinfo {year} {2006})}\BibitemShut {NoStop}%
\bibitem [{\citenamefont {Glassman}\ \emph {et~al.}(2019)\citenamefont {Glassman}, \citenamefont {P\'erez-Loureiro}, \citenamefont {Wrede}, \citenamefont {Allen}, \citenamefont {Bardayan}, \citenamefont {Bennett}, \citenamefont {Chipps}, \citenamefont {Febbraro}, \citenamefont {Friedman}, \citenamefont {Fry}, \citenamefont {Hall}, \citenamefont {Hall}, \citenamefont {Liddick}, \citenamefont {O'Malley}, \citenamefont {Ong}, \citenamefont {Pain}, \citenamefont {Schwartz}, \citenamefont {Shidling}, \citenamefont {Sims}, \citenamefont {Sun}, \citenamefont {Thompson},\ and\ \citenamefont {Zhang}}]{PhysRevC.99.065801}%
  \BibitemOpen
  \bibfield  {author} {\bibinfo {author} {\bibfnamefont {B.~E.}\ \bibnamefont {Glassman}}, \bibinfo {author} {\bibfnamefont {D.}~\bibnamefont {P\'erez-Loureiro}}, \bibinfo {author} {\bibfnamefont {C.}~\bibnamefont {Wrede}}, \bibinfo {author} {\bibfnamefont {J.}~\bibnamefont {Allen}}, \bibinfo {author} {\bibfnamefont {D.~W.}\ \bibnamefont {Bardayan}}, \bibinfo {author} {\bibfnamefont {M.~B.}\ \bibnamefont {Bennett}}, \bibinfo {author} {\bibfnamefont {K.~A.}\ \bibnamefont {Chipps}}, \bibinfo {author} {\bibfnamefont {M.}~\bibnamefont {Febbraro}}, \bibinfo {author} {\bibfnamefont {M.}~\bibnamefont {Friedman}}, \bibinfo {author} {\bibfnamefont {C.}~\bibnamefont {Fry}}, \bibinfo {author} {\bibfnamefont {M.~R.}\ \bibnamefont {Hall}}, \bibinfo {author} {\bibfnamefont {O.}~\bibnamefont {Hall}}, \bibinfo {author} {\bibfnamefont {S.~N.}\ \bibnamefont {Liddick}}, \bibinfo {author} {\bibfnamefont {P.}~\bibnamefont {O'Malley}}, \bibinfo {author} {\bibfnamefont {W.-J.}\ \bibnamefont {Ong}}, \bibinfo {author} {\bibfnamefont
  {S.~D.}\ \bibnamefont {Pain}}, \bibinfo {author} {\bibfnamefont {S.~B.}\ \bibnamefont {Schwartz}}, \bibinfo {author} {\bibfnamefont {P.}~\bibnamefont {Shidling}}, \bibinfo {author} {\bibfnamefont {H.}~\bibnamefont {Sims}}, \bibinfo {author} {\bibfnamefont {L.~J.}\ \bibnamefont {Sun}}, \bibinfo {author} {\bibfnamefont {P.}~\bibnamefont {Thompson}},\ and\ \bibinfo {author} {\bibfnamefont {H.}~\bibnamefont {Zhang}},\ }\bibfield  {title} {\bibinfo {title} {Doppler broadening in $^{20}\mathrm{Mg}(\ensuremath{\beta}p\ensuremath{\gamma})^{19}\mathrm{Ne}$ decay},\ }\href {https://doi.org/10.1103/PhysRevC.99.065801} {\bibfield  {journal} {\bibinfo  {journal} {Phys. Rev. C}\ }\textbf {\bibinfo {volume} {99}},\ \bibinfo {pages} {065801} (\bibinfo {year} {2019})}\BibitemShut {NoStop}%
\bibitem [{\citenamefont {Pollacco}\ \emph {et~al.}(2018)\citenamefont {Pollacco}, \citenamefont {Grinyer}, \citenamefont {Abu-Nimeh}, \citenamefont {Ahn}, \citenamefont {Anvar}, \citenamefont {Arokiaraj}, \citenamefont {Ayyad}, \citenamefont {Baba}, \citenamefont {Babo}, \citenamefont {Baron}, \citenamefont {Bazin}, \citenamefont {Beceiro-Novo}, \citenamefont {Belkhiria}, \citenamefont {Blaizot}, \citenamefont {Blank}, \citenamefont {Bradt}, \citenamefont {Cardella}, \citenamefont {Carpenter}, \citenamefont {Ceruti}, \citenamefont {{De Filippo}}, \citenamefont {Delagnes}, \citenamefont {{De Luca}}, \citenamefont {{De Witte}}, \citenamefont {Druillole}, \citenamefont {Duclos}, \citenamefont {Favela}, \citenamefont {Fritsch}, \citenamefont {Giovinazzo}, \citenamefont {Gueye}, \citenamefont {Isobe}, \citenamefont {Hellmuth}, \citenamefont {Huss}, \citenamefont {Lachacinski}, \citenamefont {Laffoley}, \citenamefont {Lebertre}, \citenamefont {Legeard}, \citenamefont {Lynch}, \citenamefont {Marchi}, \citenamefont
  {Martina}, \citenamefont {Maugeais}, \citenamefont {Mittig}, \citenamefont {Nalpas}, \citenamefont {Pagano}, \citenamefont {Pancin}, \citenamefont {Poleshchuk}, \citenamefont {Pedroza}, \citenamefont {Pibernat}, \citenamefont {Primault}, \citenamefont {Raabe}, \citenamefont {Raine}, \citenamefont {Rebii}, \citenamefont {Renaud}, \citenamefont {Roger}, \citenamefont {Roussel-Chomaz}, \citenamefont {Russotto}, \citenamefont {Saccà}, \citenamefont {Saillant}, \citenamefont {Sizun}, \citenamefont {Suzuki}, \citenamefont {Swartz}, \citenamefont {Tizon}, \citenamefont {Trifiró}, \citenamefont {Usher}, \citenamefont {Wittwer},\ and\ \citenamefont {Yang}}]{POLLACCO201881}%
  \BibitemOpen
  \bibfield  {author} {\bibinfo {author} {\bibfnamefont {E.}~\bibnamefont {Pollacco}}, \bibinfo {author} {\bibfnamefont {G.}~\bibnamefont {Grinyer}}, \bibinfo {author} {\bibfnamefont {F.}~\bibnamefont {Abu-Nimeh}}, \bibinfo {author} {\bibfnamefont {T.}~\bibnamefont {Ahn}}, \bibinfo {author} {\bibfnamefont {S.}~\bibnamefont {Anvar}}, \bibinfo {author} {\bibfnamefont {A.}~\bibnamefont {Arokiaraj}}, \bibinfo {author} {\bibfnamefont {Y.}~\bibnamefont {Ayyad}}, \bibinfo {author} {\bibfnamefont {H.}~\bibnamefont {Baba}}, \bibinfo {author} {\bibfnamefont {M.}~\bibnamefont {Babo}}, \bibinfo {author} {\bibfnamefont {P.}~\bibnamefont {Baron}}, \bibinfo {author} {\bibfnamefont {D.}~\bibnamefont {Bazin}}, \bibinfo {author} {\bibfnamefont {S.}~\bibnamefont {Beceiro-Novo}}, \bibinfo {author} {\bibfnamefont {C.}~\bibnamefont {Belkhiria}}, \bibinfo {author} {\bibfnamefont {M.}~\bibnamefont {Blaizot}}, \bibinfo {author} {\bibfnamefont {B.}~\bibnamefont {Blank}}, \bibinfo {author} {\bibfnamefont {J.}~\bibnamefont {Bradt}},
  \bibinfo {author} {\bibfnamefont {G.}~\bibnamefont {Cardella}}, \bibinfo {author} {\bibfnamefont {L.}~\bibnamefont {Carpenter}}, \bibinfo {author} {\bibfnamefont {S.}~\bibnamefont {Ceruti}}, \bibinfo {author} {\bibfnamefont {E.}~\bibnamefont {{De Filippo}}}, \bibinfo {author} {\bibfnamefont {E.}~\bibnamefont {Delagnes}}, \bibinfo {author} {\bibfnamefont {S.}~\bibnamefont {{De Luca}}}, \bibinfo {author} {\bibfnamefont {H.}~\bibnamefont {{De Witte}}}, \bibinfo {author} {\bibfnamefont {F.}~\bibnamefont {Druillole}}, \bibinfo {author} {\bibfnamefont {B.}~\bibnamefont {Duclos}}, \bibinfo {author} {\bibfnamefont {F.}~\bibnamefont {Favela}}, \bibinfo {author} {\bibfnamefont {A.}~\bibnamefont {Fritsch}}, \bibinfo {author} {\bibfnamefont {J.}~\bibnamefont {Giovinazzo}}, \bibinfo {author} {\bibfnamefont {C.}~\bibnamefont {Gueye}}, \bibinfo {author} {\bibfnamefont {T.}~\bibnamefont {Isobe}}, \bibinfo {author} {\bibfnamefont {P.}~\bibnamefont {Hellmuth}}, \bibinfo {author} {\bibfnamefont {C.}~\bibnamefont {Huss}},
  \bibinfo {author} {\bibfnamefont {B.}~\bibnamefont {Lachacinski}}, \bibinfo {author} {\bibfnamefont {A.}~\bibnamefont {Laffoley}}, \bibinfo {author} {\bibfnamefont {G.}~\bibnamefont {Lebertre}}, \bibinfo {author} {\bibfnamefont {L.}~\bibnamefont {Legeard}}, \bibinfo {author} {\bibfnamefont {W.}~\bibnamefont {Lynch}}, \bibinfo {author} {\bibfnamefont {T.}~\bibnamefont {Marchi}}, \bibinfo {author} {\bibfnamefont {L.}~\bibnamefont {Martina}}, \bibinfo {author} {\bibfnamefont {C.}~\bibnamefont {Maugeais}}, \bibinfo {author} {\bibfnamefont {W.}~\bibnamefont {Mittig}}, \bibinfo {author} {\bibfnamefont {L.}~\bibnamefont {Nalpas}}, \bibinfo {author} {\bibfnamefont {E.}~\bibnamefont {Pagano}}, \bibinfo {author} {\bibfnamefont {J.}~\bibnamefont {Pancin}}, \bibinfo {author} {\bibfnamefont {O.}~\bibnamefont {Poleshchuk}}, \bibinfo {author} {\bibfnamefont {J.}~\bibnamefont {Pedroza}}, \bibinfo {author} {\bibfnamefont {J.}~\bibnamefont {Pibernat}}, \bibinfo {author} {\bibfnamefont {S.}~\bibnamefont {Primault}}, \bibinfo
  {author} {\bibfnamefont {R.}~\bibnamefont {Raabe}}, \bibinfo {author} {\bibfnamefont {B.}~\bibnamefont {Raine}}, \bibinfo {author} {\bibfnamefont {A.}~\bibnamefont {Rebii}}, \bibinfo {author} {\bibfnamefont {M.}~\bibnamefont {Renaud}}, \bibinfo {author} {\bibfnamefont {T.}~\bibnamefont {Roger}}, \bibinfo {author} {\bibfnamefont {P.}~\bibnamefont {Roussel-Chomaz}}, \bibinfo {author} {\bibfnamefont {P.}~\bibnamefont {Russotto}}, \bibinfo {author} {\bibfnamefont {G.}~\bibnamefont {Saccà}}, \bibinfo {author} {\bibfnamefont {F.}~\bibnamefont {Saillant}}, \bibinfo {author} {\bibfnamefont {P.}~\bibnamefont {Sizun}}, \bibinfo {author} {\bibfnamefont {D.}~\bibnamefont {Suzuki}}, \bibinfo {author} {\bibfnamefont {J.}~\bibnamefont {Swartz}}, \bibinfo {author} {\bibfnamefont {A.}~\bibnamefont {Tizon}}, \bibinfo {author} {\bibfnamefont {A.}~\bibnamefont {Trifiró}}, \bibinfo {author} {\bibfnamefont {N.}~\bibnamefont {Usher}}, \bibinfo {author} {\bibfnamefont {G.}~\bibnamefont {Wittwer}},\ and\ \bibinfo {author}
  {\bibfnamefont {J.}~\bibnamefont {Yang}},\ }\bibfield  {title} {\bibinfo {title} {{GET: A generic electronics system for TPCs and nuclear physics instrumentation}},\ }\href {https://doi.org/https://doi.org/10.1016/j.nima.2018.01.020} {\bibfield  {journal} {\bibinfo  {journal} {Nuclear Instruments and Methods in Physics Research Section A: Accelerators, Spectrometers, Detectors and Associated Equipment}\ }\textbf {\bibinfo {volume} {887}},\ \bibinfo {pages} {81} (\bibinfo {year} {2018})}\BibitemShut {NoStop}%
\bibitem [{\citenamefont {Mahajan}\ \emph {et~al.}(2022)\citenamefont {Mahajan}, \citenamefont {Adams}, \citenamefont {Allmond}, \citenamefont {Alvarez~Pol}, \citenamefont {{Argo, E.}}, \citenamefont {Ayyad}, \citenamefont {Bardayan}, \citenamefont {Bazin}, \citenamefont {Budner}, \citenamefont {Chen}, \citenamefont {Chipps}, \citenamefont {Davids}, \citenamefont {Dopfer}, \citenamefont {Friedman}, \citenamefont {Fynbo}, \citenamefont {Grzywacz}, \citenamefont {Jose}, \citenamefont {Liang}, \citenamefont {Pain}, \citenamefont {Perez-Loureiro}, \citenamefont {Pollacco}, \citenamefont {Psaltis}, \citenamefont {Ravishankar}, \citenamefont {Rogers}, \citenamefont {Schaedig}, \citenamefont {Sun}, \citenamefont {Surbrook}, \citenamefont {Wheeler}, \citenamefont {Weghorn},\ and\ \citenamefont {Wrede}}]{Mahajan}%
  \BibitemOpen
  \bibfield  {author} {\bibinfo {author} {\bibfnamefont {R.}~\bibnamefont {Mahajan}}, \bibinfo {author} {\bibfnamefont {A.}~\bibnamefont {Adams}}, \bibinfo {author} {\bibfnamefont {J.}~\bibnamefont {Allmond}}, \bibinfo {author} {\bibfnamefont {H.}~\bibnamefont {Alvarez~Pol}}, \bibinfo {author} {\bibnamefont {{Argo, E.}}}, \bibinfo {author} {\bibfnamefont {Y.}~\bibnamefont {Ayyad}}, \bibinfo {author} {\bibfnamefont {D.}~\bibnamefont {Bardayan}}, \bibinfo {author} {\bibfnamefont {D.}~\bibnamefont {Bazin}}, \bibinfo {author} {\bibfnamefont {T.}~\bibnamefont {Budner}}, \bibinfo {author} {\bibfnamefont {A.}~\bibnamefont {Chen}}, \bibinfo {author} {\bibfnamefont {K.}~\bibnamefont {Chipps}}, \bibinfo {author} {\bibfnamefont {B.}~\bibnamefont {Davids}}, \bibinfo {author} {\bibfnamefont {J.}~\bibnamefont {Dopfer}}, \bibinfo {author} {\bibfnamefont {M.}~\bibnamefont {Friedman}}, \bibinfo {author} {\bibfnamefont {H.}~\bibnamefont {Fynbo}}, \bibinfo {author} {\bibfnamefont {R.}~\bibnamefont {Grzywacz}}, \bibinfo {author}
  {\bibfnamefont {J.}~\bibnamefont {Jose}}, \bibinfo {author} {\bibfnamefont {J.}~\bibnamefont {Liang}}, \bibinfo {author} {\bibfnamefont {S.}~\bibnamefont {Pain}}, \bibinfo {author} {\bibfnamefont {D.}~\bibnamefont {Perez-Loureiro}}, \bibinfo {author} {\bibfnamefont {E.}~\bibnamefont {Pollacco}}, \bibinfo {author} {\bibfnamefont {A.}~\bibnamefont {Psaltis}}, \bibinfo {author} {\bibfnamefont {S.}~\bibnamefont {Ravishankar}}, \bibinfo {author} {\bibfnamefont {A.}~\bibnamefont {Rogers}}, \bibinfo {author} {\bibfnamefont {L.}~\bibnamefont {Schaedig}}, \bibinfo {author} {\bibfnamefont {L.~J.}\ \bibnamefont {Sun}}, \bibinfo {author} {\bibfnamefont {J.}~\bibnamefont {Surbrook}}, \bibinfo {author} {\bibfnamefont {T.}~\bibnamefont {Wheeler}}, \bibinfo {author} {\bibfnamefont {L.}~\bibnamefont {Weghorn}},\ and\ \bibinfo {author} {\bibfnamefont {C.}~\bibnamefont {Wrede}},\ }\bibfield  {title} {\bibinfo {title} {{Measuring the $^{15}$O($\alpha$, $\gamma$)$^{19}$Ne Reaction in Type I X-ray Bursts using the GADGET II TPC:
  Software}},\ }\href {https://doi.org/10.1051/epjconf/202226011034} {\bibfield  {journal} {\bibinfo  {journal} {EPJ Web Conf.}\ }\textbf {\bibinfo {volume} {260}},\ \bibinfo {pages} {11034} (\bibinfo {year} {2022})}\BibitemShut {NoStop}%
\bibitem [{\citenamefont {{Wheeler, Tyler}}\ \emph {et~al.}(2022)\citenamefont {{Wheeler, Tyler}}, \citenamefont {{Adams, A.}}, \citenamefont {{Allmond, J.}}, \citenamefont {{Alvarez Pol, H.}}, \citenamefont {{Argo, E.}}, \citenamefont {{Ayyad, Y.}}, \citenamefont {{Bardayan, D.}}, \citenamefont {{Bazin, D.}}, \citenamefont {{Budner, T.}}, \citenamefont {{Chen, A.}}, \citenamefont {{Chipps, K.}}, \citenamefont {{Davids, B.}}, \citenamefont {{Dopfer, J.}}, \citenamefont {{Friedman, M.}}, \citenamefont {{Fynbo, H.}}, \citenamefont {{Grzywacz, R.}}, \citenamefont {{Jose, J.}}, \citenamefont {{Liang, J.}}, \citenamefont {{Mahajan, R.}}, \citenamefont {{Pain, S.}}, \citenamefont {{P\'erez-Loureiro, D.}}, \citenamefont {{Pollacco, E.}}, \citenamefont {{Psaltis, A.}}, \citenamefont {{Ravishankar, S.}}, \citenamefont {{Rogers, A.}}, \citenamefont {{Schaedig, L.}}, \citenamefont {{Sun, L. J.}}, \citenamefont {{Surbrook, J.}}, \citenamefont {{Weghorn, L.}},\ and\ \citenamefont {{Wrede, C.}}}]{Wheeler}%
  \BibitemOpen
  \bibfield  {author} {\bibinfo {author} {\bibnamefont {{Wheeler, Tyler}}}, \bibinfo {author} {\bibnamefont {{Adams, A.}}}, \bibinfo {author} {\bibnamefont {{Allmond, J.}}}, \bibinfo {author} {\bibnamefont {{Alvarez Pol, H.}}}, \bibinfo {author} {\bibnamefont {{Argo, E.}}}, \bibinfo {author} {\bibnamefont {{Ayyad, Y.}}}, \bibinfo {author} {\bibnamefont {{Bardayan, D.}}}, \bibinfo {author} {\bibnamefont {{Bazin, D.}}}, \bibinfo {author} {\bibnamefont {{Budner, T.}}}, \bibinfo {author} {\bibnamefont {{Chen, A.}}}, \bibinfo {author} {\bibnamefont {{Chipps, K.}}}, \bibinfo {author} {\bibnamefont {{Davids, B.}}}, \bibinfo {author} {\bibnamefont {{Dopfer, J.}}}, \bibinfo {author} {\bibnamefont {{Friedman, M.}}}, \bibinfo {author} {\bibnamefont {{Fynbo, H.}}}, \bibinfo {author} {\bibnamefont {{Grzywacz, R.}}}, \bibinfo {author} {\bibnamefont {{Jose, J.}}}, \bibinfo {author} {\bibnamefont {{Liang, J.}}}, \bibinfo {author} {\bibnamefont {{Mahajan, R.}}}, \bibinfo {author} {\bibnamefont {{Pain, S.}}}, \bibinfo {author}
  {\bibnamefont {{P\'erez-Loureiro, D.}}}, \bibinfo {author} {\bibnamefont {{Pollacco, E.}}}, \bibinfo {author} {\bibnamefont {{Psaltis, A.}}}, \bibinfo {author} {\bibnamefont {{Ravishankar, S.}}}, \bibinfo {author} {\bibnamefont {{Rogers, A.}}}, \bibinfo {author} {\bibnamefont {{Schaedig, L.}}}, \bibinfo {author} {\bibnamefont {{Sun, L. J.}}}, \bibinfo {author} {\bibnamefont {{Surbrook, J.}}}, \bibinfo {author} {\bibnamefont {{Weghorn, L.}}},\ and\ \bibinfo {author} {\bibnamefont {{Wrede, C.}}},\ }\bibfield  {title} {\bibinfo {title} {{Measuring the $^{15}$O($\alpha$, $\gamma$)$^{19}$Ne reaction in Type I X-ray bursts using the GADGET II TPC: Hardware}},\ }\href {https://doi.org/10.1051/epjconf/202226011046} {\bibfield  {journal} {\bibinfo  {journal} {EPJ Web Conf.}\ }\textbf {\bibinfo {volume} {260}},\ \bibinfo {pages} {11046} (\bibinfo {year} {2022})}\BibitemShut {NoStop}%
\bibitem [{\citenamefont {Ayyad}\ \emph {et~al.}(2020)\citenamefont {Ayyad}, \citenamefont {Abgrall}, \citenamefont {Ahn}, \citenamefont {Álvarez Pol}, \citenamefont {Bazin}, \citenamefont {Beceiro-Novo}, \citenamefont {Carpenter}, \citenamefont {Cooper}, \citenamefont {Cortesi}, \citenamefont {Macchiavelli}, \citenamefont {Mittig}, \citenamefont {Olaizola}, \citenamefont {Randhawa}, \citenamefont {Santamaria}, \citenamefont {Watwood}, \citenamefont {Zamora},\ and\ \citenamefont {Zegers}}]{AYYAD2020161341}%
  \BibitemOpen
  \bibfield  {author} {\bibinfo {author} {\bibfnamefont {Y.}~\bibnamefont {Ayyad}}, \bibinfo {author} {\bibfnamefont {N.}~\bibnamefont {Abgrall}}, \bibinfo {author} {\bibfnamefont {T.}~\bibnamefont {Ahn}}, \bibinfo {author} {\bibfnamefont {H.}~\bibnamefont {Álvarez Pol}}, \bibinfo {author} {\bibfnamefont {D.}~\bibnamefont {Bazin}}, \bibinfo {author} {\bibfnamefont {S.}~\bibnamefont {Beceiro-Novo}}, \bibinfo {author} {\bibfnamefont {L.}~\bibnamefont {Carpenter}}, \bibinfo {author} {\bibfnamefont {R.}~\bibnamefont {Cooper}}, \bibinfo {author} {\bibfnamefont {M.}~\bibnamefont {Cortesi}}, \bibinfo {author} {\bibfnamefont {A.}~\bibnamefont {Macchiavelli}}, \bibinfo {author} {\bibfnamefont {W.}~\bibnamefont {Mittig}}, \bibinfo {author} {\bibfnamefont {B.}~\bibnamefont {Olaizola}}, \bibinfo {author} {\bibfnamefont {J.}~\bibnamefont {Randhawa}}, \bibinfo {author} {\bibfnamefont {C.}~\bibnamefont {Santamaria}}, \bibinfo {author} {\bibfnamefont {N.}~\bibnamefont {Watwood}}, \bibinfo {author} {\bibfnamefont
  {J.}~\bibnamefont {Zamora}},\ and\ \bibinfo {author} {\bibfnamefont {R.}~\bibnamefont {Zegers}},\ }\bibfield  {title} {\bibinfo {title} {{Next-generation experiments with the Active Target Time Projection Chamber (AT-TPC)}},\ }\href {https://doi.org/https://doi.org/10.1016/j.nima.2018.10.019} {\bibfield  {journal} {\bibinfo  {journal} {Nuclear Instruments and Methods in Physics Research Section A: Accelerators, Spectrometers, Detectors and Associated Equipment}\ }\textbf {\bibinfo {volume} {954}},\ \bibinfo {pages} {161341} (\bibinfo {year} {2020})},\ \bibinfo {note} {symposium on Radiation Measurements and Applications XVII}\BibitemShut {NoStop}%
\bibitem [{\citenamefont {Ayyad}\ \emph {et~al.}(2017)\citenamefont {Ayyad}, \citenamefont {Mittig}, \citenamefont {Bazin},\ and\ \citenamefont {Cortesi}}]{Ayyad_2017}%
  \BibitemOpen
  \bibfield  {author} {\bibinfo {author} {\bibfnamefont {Y.}~\bibnamefont {Ayyad}}, \bibinfo {author} {\bibfnamefont {W.}~\bibnamefont {Mittig}}, \bibinfo {author} {\bibfnamefont {D.}~\bibnamefont {Bazin}},\ and\ \bibinfo {author} {\bibfnamefont {M.}~\bibnamefont {Cortesi}},\ }\bibfield  {title} {\bibinfo {title} {Overview of the data analysis and new micro-pattern gas detector development for the active target time projection chamber (at-tpc) project.},\ }\href {https://doi.org/10.1088/1742-6596/876/1/012003} {\bibfield  {journal} {\bibinfo  {journal} {Journal of Physics: Conference Series}\ }\textbf {\bibinfo {volume} {876}},\ \bibinfo {pages} {012003} (\bibinfo {year} {2017})}\BibitemShut {NoStop}%
\bibitem [{\citenamefont {Ayyad}(2023)}]{ATTPCROOT_git}%
  \BibitemOpen
  \bibfield  {author} {\bibinfo {author} {\bibfnamefont {Y.}~\bibnamefont {Ayyad}},\ }\href@noop {} {\bibinfo {title} {Attpcroot}},\ \bibinfo {howpublished} {\url{https://github.com/ATTPC/ATTPCROOTv2}} (\bibinfo {year} {2023})\BibitemShut {NoStop}%
\bibitem [{\citenamefont {Anthony}\ \emph {et~al.}(2023)\citenamefont {Anthony}, \citenamefont {Ayyad}, \citenamefont {ACeulemans}, \citenamefont {freund17}, \citenamefont {lisacarpenter}, \citenamefont {cnhunt}, \citenamefont {Pol}, \citenamefont {Zamora}, \citenamefont {Rijal}, \citenamefont {Ramos}, \citenamefont {bolaizol}, \citenamefont {AriAtari}, \citenamefont {Ruchi-GADGETII}, \citenamefont {Simon}, \citenamefont {Wieskejo},\ and\ \citenamefont {diazcort}}]{anthony_2023_10027879}%
  \BibitemOpen
  \bibfield  {author} {\bibinfo {author} {\bibfnamefont {A.}~\bibnamefont {Anthony}}, \bibinfo {author} {\bibfnamefont {Y.}~\bibnamefont {Ayyad}}, \bibinfo {author} {\bibnamefont {ACeulemans}}, \bibinfo {author} {\bibnamefont {freund17}}, \bibinfo {author} {\bibnamefont {lisacarpenter}}, \bibinfo {author} {\bibnamefont {cnhunt}}, \bibinfo {author} {\bibfnamefont {H.~A.}\ \bibnamefont {Pol}}, \bibinfo {author} {\bibfnamefont {J.}~\bibnamefont {Zamora}}, \bibinfo {author} {\bibfnamefont {N.}~\bibnamefont {Rijal}}, \bibinfo {author} {\bibfnamefont {A.~M.}\ \bibnamefont {Ramos}}, \bibinfo {author} {\bibnamefont {bolaizol}}, \bibinfo {author} {\bibnamefont {AriAtari}}, \bibinfo {author} {\bibnamefont {Ruchi-GADGETII}}, \bibinfo {author} {\bibnamefont {Simon}}, \bibinfo {author} {\bibnamefont {Wieskejo}},\ and\ \bibinfo {author} {\bibnamefont {diazcort}},\ }\href {https://doi.org/10.5281/zenodo.10027879} {\bibinfo {title} {Attpcroot}} (\bibinfo {year} {2023})\BibitemShut {NoStop}%
\bibitem [{\citenamefont {Attié}\ \emph {et~al.}(2022)\citenamefont {Attié}, \citenamefont {Batkiewicz-Kwasniak}, \citenamefont {Billoir}, \citenamefont {Blanchet}, \citenamefont {Blondel}, \citenamefont {Bolognesi}, \citenamefont {Calvet}, \citenamefont {Catanesi}, \citenamefont {Cicerchia}, \citenamefont {Cogo}, \citenamefont {Colas}, \citenamefont {Collazuol}, \citenamefont {Delbart}, \citenamefont {Dumarchez}, \citenamefont {Emery-Schrenk}, \citenamefont {Feltre}, \citenamefont {Giganti}, \citenamefont {Gramegna}, \citenamefont {Grassi}, \citenamefont {Guigue}, \citenamefont {Hamacher-Baumann}, \citenamefont {Hassani}, \citenamefont {Iacob}, \citenamefont {Jesús-Valls}, \citenamefont {Kurjata}, \citenamefont {Lamoureux}, \citenamefont {Lehuraux}, \citenamefont {Longhin}, \citenamefont {Lux}, \citenamefont {Magaletti}, \citenamefont {Marchi}, \citenamefont {Maurel}, \citenamefont {Mellet}, \citenamefont {Mezzetto}, \citenamefont {Munteanu}, \citenamefont {Nguyen}, \citenamefont {Orain}, \citenamefont
  {Pari}, \citenamefont {Parraud}, \citenamefont {Pastore}, \citenamefont {Pepato}, \citenamefont {Pierre}, \citenamefont {Popov}, \citenamefont {Przybiliski}, \citenamefont {Radermacher}, \citenamefont {Radicioni}, \citenamefont {Riallot}, \citenamefont {Roth}, \citenamefont {Rychter}, \citenamefont {Scomparin}, \citenamefont {Steinmann}, \citenamefont {Suvorov}, \citenamefont {Swierblewski}, \citenamefont {Terront}, \citenamefont {Thamm}, \citenamefont {Toussenel}, \citenamefont {Valentino}, \citenamefont {Vasseur}, \citenamefont {Yevarouskaya}, \citenamefont {Ziembicki},\ and\ \citenamefont {Zito}}]{ATTIE2022166109}%
  \BibitemOpen
  \bibfield  {author} {\bibinfo {author} {\bibfnamefont {D.}~\bibnamefont {Attié}}, \bibinfo {author} {\bibfnamefont {M.}~\bibnamefont {Batkiewicz-Kwasniak}}, \bibinfo {author} {\bibfnamefont {P.}~\bibnamefont {Billoir}}, \bibinfo {author} {\bibfnamefont {A.}~\bibnamefont {Blanchet}}, \bibinfo {author} {\bibfnamefont {A.}~\bibnamefont {Blondel}}, \bibinfo {author} {\bibfnamefont {S.}~\bibnamefont {Bolognesi}}, \bibinfo {author} {\bibfnamefont {D.}~\bibnamefont {Calvet}}, \bibinfo {author} {\bibfnamefont {M.}~\bibnamefont {Catanesi}}, \bibinfo {author} {\bibfnamefont {M.}~\bibnamefont {Cicerchia}}, \bibinfo {author} {\bibfnamefont {G.}~\bibnamefont {Cogo}}, \bibinfo {author} {\bibfnamefont {P.}~\bibnamefont {Colas}}, \bibinfo {author} {\bibfnamefont {G.}~\bibnamefont {Collazuol}}, \bibinfo {author} {\bibfnamefont {A.}~\bibnamefont {Delbart}}, \bibinfo {author} {\bibfnamefont {J.}~\bibnamefont {Dumarchez}}, \bibinfo {author} {\bibfnamefont {S.}~\bibnamefont {Emery-Schrenk}}, \bibinfo {author} {\bibfnamefont
  {M.}~\bibnamefont {Feltre}}, \bibinfo {author} {\bibfnamefont {C.}~\bibnamefont {Giganti}}, \bibinfo {author} {\bibfnamefont {F.}~\bibnamefont {Gramegna}}, \bibinfo {author} {\bibfnamefont {M.}~\bibnamefont {Grassi}}, \bibinfo {author} {\bibfnamefont {M.}~\bibnamefont {Guigue}}, \bibinfo {author} {\bibfnamefont {P.}~\bibnamefont {Hamacher-Baumann}}, \bibinfo {author} {\bibfnamefont {S.}~\bibnamefont {Hassani}}, \bibinfo {author} {\bibfnamefont {F.}~\bibnamefont {Iacob}}, \bibinfo {author} {\bibfnamefont {C.}~\bibnamefont {Jesús-Valls}}, \bibinfo {author} {\bibfnamefont {R.}~\bibnamefont {Kurjata}}, \bibinfo {author} {\bibfnamefont {M.}~\bibnamefont {Lamoureux}}, \bibinfo {author} {\bibfnamefont {M.}~\bibnamefont {Lehuraux}}, \bibinfo {author} {\bibfnamefont {A.}~\bibnamefont {Longhin}}, \bibinfo {author} {\bibfnamefont {T.}~\bibnamefont {Lux}}, \bibinfo {author} {\bibfnamefont {L.}~\bibnamefont {Magaletti}}, \bibinfo {author} {\bibfnamefont {T.}~\bibnamefont {Marchi}}, \bibinfo {author} {\bibfnamefont
  {A.}~\bibnamefont {Maurel}}, \bibinfo {author} {\bibfnamefont {L.}~\bibnamefont {Mellet}}, \bibinfo {author} {\bibfnamefont {M.}~\bibnamefont {Mezzetto}}, \bibinfo {author} {\bibfnamefont {L.}~\bibnamefont {Munteanu}}, \bibinfo {author} {\bibfnamefont {Q.}~\bibnamefont {Nguyen}}, \bibinfo {author} {\bibfnamefont {Y.}~\bibnamefont {Orain}}, \bibinfo {author} {\bibfnamefont {M.}~\bibnamefont {Pari}}, \bibinfo {author} {\bibfnamefont {J.-M.}\ \bibnamefont {Parraud}}, \bibinfo {author} {\bibfnamefont {C.}~\bibnamefont {Pastore}}, \bibinfo {author} {\bibfnamefont {A.}~\bibnamefont {Pepato}}, \bibinfo {author} {\bibfnamefont {E.}~\bibnamefont {Pierre}}, \bibinfo {author} {\bibfnamefont {B.}~\bibnamefont {Popov}}, \bibinfo {author} {\bibfnamefont {H.}~\bibnamefont {Przybiliski}}, \bibinfo {author} {\bibfnamefont {T.}~\bibnamefont {Radermacher}}, \bibinfo {author} {\bibfnamefont {E.}~\bibnamefont {Radicioni}}, \bibinfo {author} {\bibfnamefont {M.}~\bibnamefont {Riallot}}, \bibinfo {author} {\bibfnamefont
  {S.}~\bibnamefont {Roth}}, \bibinfo {author} {\bibfnamefont {A.}~\bibnamefont {Rychter}}, \bibinfo {author} {\bibfnamefont {L.}~\bibnamefont {Scomparin}}, \bibinfo {author} {\bibfnamefont {J.}~\bibnamefont {Steinmann}}, \bibinfo {author} {\bibfnamefont {S.}~\bibnamefont {Suvorov}}, \bibinfo {author} {\bibfnamefont {J.}~\bibnamefont {Swierblewski}}, \bibinfo {author} {\bibfnamefont {D.}~\bibnamefont {Terront}}, \bibinfo {author} {\bibfnamefont {N.}~\bibnamefont {Thamm}}, \bibinfo {author} {\bibfnamefont {F.}~\bibnamefont {Toussenel}}, \bibinfo {author} {\bibfnamefont {V.}~\bibnamefont {Valentino}}, \bibinfo {author} {\bibfnamefont {G.}~\bibnamefont {Vasseur}}, \bibinfo {author} {\bibfnamefont {U.}~\bibnamefont {Yevarouskaya}}, \bibinfo {author} {\bibfnamefont {M.}~\bibnamefont {Ziembicki}},\ and\ \bibinfo {author} {\bibfnamefont {M.}~\bibnamefont {Zito}},\ }\bibfield  {title} {\bibinfo {title} {Characterization of resistive micromegas detectors for the upgrade of the t2k near detector time projection
  chambers},\ }\href {https://doi.org/https://doi.org/10.1016/j.nima.2021.166109} {\bibfield  {journal} {\bibinfo  {journal} {Nuclear Instruments and Methods in Physics Research Section A: Accelerators, Spectrometers, Detectors and Associated Equipment}\ }\textbf {\bibinfo {volume} {1025}},\ \bibinfo {pages} {166109} (\bibinfo {year} {2022})}\BibitemShut {NoStop}%
\bibitem [{\citenamefont {Chefdeville}\ \emph {et~al.}(2021)\citenamefont {Chefdeville}, \citenamefont {{de Oliveira}}, \citenamefont {Drancourt}, \citenamefont {Geffroy}, \citenamefont {Geralis}, \citenamefont {Gkountoumis}, \citenamefont {Kalamaris}, \citenamefont {Karyotakis}, \citenamefont {Nikas}, \citenamefont {Peltier}, \citenamefont {Pizzirusso}, \citenamefont {Psallidas}, \citenamefont {Teixeira}, \citenamefont {Titov},\ and\ \citenamefont {Vouters}}]{CHEFDEVILLE2021165268}%
  \BibitemOpen
  \bibfield  {author} {\bibinfo {author} {\bibfnamefont {M.}~\bibnamefont {Chefdeville}}, \bibinfo {author} {\bibfnamefont {R.}~\bibnamefont {{de Oliveira}}}, \bibinfo {author} {\bibfnamefont {C.}~\bibnamefont {Drancourt}}, \bibinfo {author} {\bibfnamefont {N.}~\bibnamefont {Geffroy}}, \bibinfo {author} {\bibfnamefont {T.}~\bibnamefont {Geralis}}, \bibinfo {author} {\bibfnamefont {P.}~\bibnamefont {Gkountoumis}}, \bibinfo {author} {\bibfnamefont {A.}~\bibnamefont {Kalamaris}}, \bibinfo {author} {\bibfnamefont {Y.}~\bibnamefont {Karyotakis}}, \bibinfo {author} {\bibfnamefont {D.}~\bibnamefont {Nikas}}, \bibinfo {author} {\bibfnamefont {F.}~\bibnamefont {Peltier}}, \bibinfo {author} {\bibfnamefont {O.}~\bibnamefont {Pizzirusso}}, \bibinfo {author} {\bibfnamefont {A.}~\bibnamefont {Psallidas}}, \bibinfo {author} {\bibfnamefont {A.}~\bibnamefont {Teixeira}}, \bibinfo {author} {\bibfnamefont {M.}~\bibnamefont {Titov}},\ and\ \bibinfo {author} {\bibfnamefont {G.}~\bibnamefont {Vouters}},\ }\bibfield  {title}
  {\bibinfo {title} {Development of micromegas detectors with resistive anode pads},\ }\href {https://doi.org/https://doi.org/10.1016/j.nima.2021.165268} {\bibfield  {journal} {\bibinfo  {journal} {Nuclear Instruments and Methods in Physics Research Section A: Accelerators, Spectrometers, Detectors and Associated Equipment}\ }\textbf {\bibinfo {volume} {1003}},\ \bibinfo {pages} {165268} (\bibinfo {year} {2021})}\BibitemShut {NoStop}%
\bibitem [{\citenamefont {Dixit}\ and\ \citenamefont {Rankin}(2006)}]{DIXIT2006281}%
  \BibitemOpen
  \bibfield  {author} {\bibinfo {author} {\bibfnamefont {M.}~\bibnamefont {Dixit}}\ and\ \bibinfo {author} {\bibfnamefont {A.}~\bibnamefont {Rankin}},\ }\bibfield  {title} {\bibinfo {title} {Simulating the charge dispersion phenomena in micro pattern gas detectors with a resistive anode},\ }\href {https://doi.org/https://doi.org/10.1016/j.nima.2006.06.050} {\bibfield  {journal} {\bibinfo  {journal} {Nuclear Instruments and Methods in Physics Research Section A: Accelerators, Spectrometers, Detectors and Associated Equipment}\ }\textbf {\bibinfo {volume} {566}},\ \bibinfo {pages} {281} (\bibinfo {year} {2006})}\BibitemShut {NoStop}%
\bibitem [{\citenamefont {Wu}(2007)}]{WU20071057}%
  \BibitemOpen
  \bibfield  {author} {\bibinfo {author} {\bibfnamefont {S.-C.}\ \bibnamefont {Wu}},\ }\bibfield  {title} {\bibinfo {title} {Nuclear data sheets for a = 216},\ }\href {https://doi.org/https://doi.org/10.1016/j.nds.2007.04.001} {\bibfield  {journal} {\bibinfo  {journal} {Nuclear Data Sheets}\ }\textbf {\bibinfo {volume} {108}},\ \bibinfo {pages} {1057} (\bibinfo {year} {2007})}\BibitemShut {NoStop}%
\bibitem [{\citenamefont {Auranen}\ and\ \citenamefont {McCutchan}(2020)}]{AURANEN2020117}%
  \BibitemOpen
  \bibfield  {author} {\bibinfo {author} {\bibfnamefont {K.}~\bibnamefont {Auranen}}\ and\ \bibinfo {author} {\bibfnamefont {E.}~\bibnamefont {McCutchan}},\ }\bibfield  {title} {\bibinfo {title} {Nuclear data sheets for a=212},\ }\href {https://doi.org/https://doi.org/10.1016/j.nds.2020.09.002} {\bibfield  {journal} {\bibinfo  {journal} {Nuclear Data Sheets}\ }\textbf {\bibinfo {volume} {168}},\ \bibinfo {pages} {117} (\bibinfo {year} {2020})}\BibitemShut {NoStop}%
\bibitem [{\citenamefont {Singh}\ and\ \citenamefont {Upadhyaya}(2012)}]{article}%
  \BibitemOpen
  \bibfield  {author} {\bibinfo {author} {\bibfnamefont {K.}~\bibnamefont {Singh}}\ and\ \bibinfo {author} {\bibfnamefont {S.}~\bibnamefont {Upadhyaya}},\ }\bibfield  {title} {\bibinfo {title} {Outlier detection: Applications and techniques},\ }\href {https://doi.org/http://dx.doi.org/10.1007/0-387-25465-X_7} {\bibfield  {journal} {\bibinfo  {journal} {International Journal of Computer Science Issues}\ }\textbf {\bibinfo {volume} {9}} (\bibinfo {year} {2012})}\BibitemShut {NoStop}%
\bibitem [{\citenamefont {Wold}\ \emph {et~al.}(1987)\citenamefont {Wold}, \citenamefont {Esbensen},\ and\ \citenamefont {Geladi}}]{WOLD198737}%
  \BibitemOpen
  \bibfield  {author} {\bibinfo {author} {\bibfnamefont {S.}~\bibnamefont {Wold}}, \bibinfo {author} {\bibfnamefont {K.}~\bibnamefont {Esbensen}},\ and\ \bibinfo {author} {\bibfnamefont {P.}~\bibnamefont {Geladi}},\ }\bibfield  {title} {\bibinfo {title} {Principal component analysis},\ }\href {https://doi.org/https://doi.org/10.1016/0169-7439(87)80084-9} {\bibfield  {journal} {\bibinfo  {journal} {Chemometrics and Intelligent Laboratory Systems}\ }\textbf {\bibinfo {volume} {2}},\ \bibinfo {pages} {37} (\bibinfo {year} {1987})},\ \bibinfo {note} {proceedings of the Multivariate Statistical Workshop for Geologists and Geochemists}\BibitemShut {NoStop}%
\bibitem [{\citenamefont {Giovinazzo}\ \emph {et~al.}(2018)\citenamefont {Giovinazzo}, \citenamefont {Pibernat}, \citenamefont {Goigoux}, \citenamefont {de Oliveira}, \citenamefont {Grinyer}, \citenamefont {Huss}, \citenamefont {Mauss}, \citenamefont {Pancin}, \citenamefont {Pedroza}, \citenamefont {Rebii}, \citenamefont {Roger}, \citenamefont {Rosier}, \citenamefont {Saillant},\ and\ \citenamefont {Wittwer}}]{ACTAR2018}%
  \BibitemOpen
  \bibfield  {author} {\bibinfo {author} {\bibfnamefont {J.}~\bibnamefont {Giovinazzo}}, \bibinfo {author} {\bibfnamefont {J.}~\bibnamefont {Pibernat}}, \bibinfo {author} {\bibfnamefont {T.}~\bibnamefont {Goigoux}}, \bibinfo {author} {\bibfnamefont {R.}~\bibnamefont {de Oliveira}}, \bibinfo {author} {\bibfnamefont {G.}~\bibnamefont {Grinyer}}, \bibinfo {author} {\bibfnamefont {C.}~\bibnamefont {Huss}}, \bibinfo {author} {\bibfnamefont {B.}~\bibnamefont {Mauss}}, \bibinfo {author} {\bibfnamefont {J.}~\bibnamefont {Pancin}}, \bibinfo {author} {\bibfnamefont {J.}~\bibnamefont {Pedroza}}, \bibinfo {author} {\bibfnamefont {A.}~\bibnamefont {Rebii}}, \bibinfo {author} {\bibfnamefont {T.}~\bibnamefont {Roger}}, \bibinfo {author} {\bibfnamefont {P.}~\bibnamefont {Rosier}}, \bibinfo {author} {\bibfnamefont {F.}~\bibnamefont {Saillant}},\ and\ \bibinfo {author} {\bibfnamefont {G.}~\bibnamefont {Wittwer}},\ }\bibfield  {title} {\bibinfo {title} {Metal-core pad-plane development for actar tpc},\ }\href
  {https://doi.org/https://doi.org/10.1016/j.nima.2018.03.007} {\bibfield  {journal} {\bibinfo  {journal} {Nuclear Instruments and Methods in Physics Research Section A: Accelerators, Spectrometers, Detectors and Associated Equipment}\ }\textbf {\bibinfo {volume} {892}},\ \bibinfo {pages} {114} (\bibinfo {year} {2018})}\BibitemShut {NoStop}%
\bibitem [{\citenamefont {Giovinazzo}\ \emph {et~al.}(2020)\citenamefont {Giovinazzo}, \citenamefont {Pancin}, \citenamefont {Pibernat},\ and\ \citenamefont {Roger}}]{ACTAR2020}%
  \BibitemOpen
  \bibfield  {author} {\bibinfo {author} {\bibfnamefont {J.}~\bibnamefont {Giovinazzo}}, \bibinfo {author} {\bibfnamefont {J.}~\bibnamefont {Pancin}}, \bibinfo {author} {\bibfnamefont {J.}~\bibnamefont {Pibernat}},\ and\ \bibinfo {author} {\bibfnamefont {T.}~\bibnamefont {Roger}},\ }\bibfield  {title} {\bibinfo {title} {Actar tpc performance with get electronics},\ }\href {https://doi.org/https://doi.org/10.1016/j.nima.2019.163184} {\bibfield  {journal} {\bibinfo  {journal} {Nuclear Instruments and Methods in Physics Research Section A: Accelerators, Spectrometers, Detectors and Associated Equipment}\ }\textbf {\bibinfo {volume} {953}},\ \bibinfo {pages} {163184} (\bibinfo {year} {2020})}\BibitemShut {NoStop}%
\bibitem [{\citenamefont {Heffner}\ \emph {et~al.}(2014)\citenamefont {Heffner}, \citenamefont {Asner}, \citenamefont {Baker}, \citenamefont {Baker}, \citenamefont {Barrett}, \citenamefont {Brune}, \citenamefont {Bundgaard}, \citenamefont {Burgett}, \citenamefont {Carter}, \citenamefont {Cunningham}, \citenamefont {Deaven}, \citenamefont {Duke}, \citenamefont {Greife}, \citenamefont {Grimes}, \citenamefont {Hager}, \citenamefont {Hertel}, \citenamefont {Hill}, \citenamefont {Isenhower}, \citenamefont {Jewell}, \citenamefont {King}, \citenamefont {Klay}, \citenamefont {Kleinrath}, \citenamefont {Kornilov}, \citenamefont {Kudo}, \citenamefont {Laptev}, \citenamefont {Leonard}, \citenamefont {Loveland}, \citenamefont {Massey}, \citenamefont {McGrath}, \citenamefont {Meharchand}, \citenamefont {Montoya}, \citenamefont {Pickle}, \citenamefont {Qu}, \citenamefont {Riot}, \citenamefont {Ruz}, \citenamefont {Sangiorgio}, \citenamefont {Seilhan}, \citenamefont {Sharma}, \citenamefont {Snyder}, \citenamefont {Stave},
  \citenamefont {Tatishvili}, \citenamefont {Thornton}, \citenamefont {Tovesson}, \citenamefont {Towell}, \citenamefont {Towell}, \citenamefont {Watson}, \citenamefont {Wendt}, \citenamefont {Wood},\ and\ \citenamefont {Yao}}]{FissionTPC}%
  \BibitemOpen
  \bibfield  {author} {\bibinfo {author} {\bibfnamefont {M.}~\bibnamefont {Heffner}}, \bibinfo {author} {\bibfnamefont {D.}~\bibnamefont {Asner}}, \bibinfo {author} {\bibfnamefont {R.}~\bibnamefont {Baker}}, \bibinfo {author} {\bibfnamefont {J.}~\bibnamefont {Baker}}, \bibinfo {author} {\bibfnamefont {S.}~\bibnamefont {Barrett}}, \bibinfo {author} {\bibfnamefont {C.}~\bibnamefont {Brune}}, \bibinfo {author} {\bibfnamefont {J.}~\bibnamefont {Bundgaard}}, \bibinfo {author} {\bibfnamefont {E.}~\bibnamefont {Burgett}}, \bibinfo {author} {\bibfnamefont {D.}~\bibnamefont {Carter}}, \bibinfo {author} {\bibfnamefont {M.}~\bibnamefont {Cunningham}}, \bibinfo {author} {\bibfnamefont {J.}~\bibnamefont {Deaven}}, \bibinfo {author} {\bibfnamefont {D.}~\bibnamefont {Duke}}, \bibinfo {author} {\bibfnamefont {U.}~\bibnamefont {Greife}}, \bibinfo {author} {\bibfnamefont {S.}~\bibnamefont {Grimes}}, \bibinfo {author} {\bibfnamefont {U.}~\bibnamefont {Hager}}, \bibinfo {author} {\bibfnamefont {N.}~\bibnamefont {Hertel}},
  \bibinfo {author} {\bibfnamefont {T.}~\bibnamefont {Hill}}, \bibinfo {author} {\bibfnamefont {D.}~\bibnamefont {Isenhower}}, \bibinfo {author} {\bibfnamefont {K.}~\bibnamefont {Jewell}}, \bibinfo {author} {\bibfnamefont {J.}~\bibnamefont {King}}, \bibinfo {author} {\bibfnamefont {J.}~\bibnamefont {Klay}}, \bibinfo {author} {\bibfnamefont {V.}~\bibnamefont {Kleinrath}}, \bibinfo {author} {\bibfnamefont {N.}~\bibnamefont {Kornilov}}, \bibinfo {author} {\bibfnamefont {R.}~\bibnamefont {Kudo}}, \bibinfo {author} {\bibfnamefont {A.}~\bibnamefont {Laptev}}, \bibinfo {author} {\bibfnamefont {M.}~\bibnamefont {Leonard}}, \bibinfo {author} {\bibfnamefont {W.}~\bibnamefont {Loveland}}, \bibinfo {author} {\bibfnamefont {T.}~\bibnamefont {Massey}}, \bibinfo {author} {\bibfnamefont {C.}~\bibnamefont {McGrath}}, \bibinfo {author} {\bibfnamefont {R.}~\bibnamefont {Meharchand}}, \bibinfo {author} {\bibfnamefont {L.}~\bibnamefont {Montoya}}, \bibinfo {author} {\bibfnamefont {N.}~\bibnamefont {Pickle}}, \bibinfo {author}
  {\bibfnamefont {H.}~\bibnamefont {Qu}}, \bibinfo {author} {\bibfnamefont {V.}~\bibnamefont {Riot}}, \bibinfo {author} {\bibfnamefont {J.}~\bibnamefont {Ruz}}, \bibinfo {author} {\bibfnamefont {S.}~\bibnamefont {Sangiorgio}}, \bibinfo {author} {\bibfnamefont {B.}~\bibnamefont {Seilhan}}, \bibinfo {author} {\bibfnamefont {S.}~\bibnamefont {Sharma}}, \bibinfo {author} {\bibfnamefont {L.}~\bibnamefont {Snyder}}, \bibinfo {author} {\bibfnamefont {S.}~\bibnamefont {Stave}}, \bibinfo {author} {\bibfnamefont {G.}~\bibnamefont {Tatishvili}}, \bibinfo {author} {\bibfnamefont {R.}~\bibnamefont {Thornton}}, \bibinfo {author} {\bibfnamefont {F.}~\bibnamefont {Tovesson}}, \bibinfo {author} {\bibfnamefont {D.}~\bibnamefont {Towell}}, \bibinfo {author} {\bibfnamefont {R.}~\bibnamefont {Towell}}, \bibinfo {author} {\bibfnamefont {S.}~\bibnamefont {Watson}}, \bibinfo {author} {\bibfnamefont {B.}~\bibnamefont {Wendt}}, \bibinfo {author} {\bibfnamefont {L.}~\bibnamefont {Wood}},\ and\ \bibinfo {author} {\bibfnamefont
  {L.}~\bibnamefont {Yao}},\ }\bibfield  {title} {\bibinfo {title} {A time projection chamber for high accuracy and precision fission cross-section measurements},\ }\href {https://doi.org/https://doi.org/10.1016/j.nima.2014.05.057} {\bibfield  {journal} {\bibinfo  {journal} {Nuclear Instruments and Methods in Physics Research Section A: Accelerators, Spectrometers, Detectors and Associated Equipment}\ }\textbf {\bibinfo {volume} {759}},\ \bibinfo {pages} {50} (\bibinfo {year} {2014})}\BibitemShut {NoStop}%
\bibitem [{\citenamefont {Konczykowski}\ \emph {et~al.}(2019)\citenamefont {Konczykowski}, \citenamefont {Fernández-Dominguez}, \citenamefont {Alvarez-Pol}, \citenamefont {Caamaño}, \citenamefont {Grinyer}, \citenamefont {Laffoley}, \citenamefont {Mauss}, \citenamefont {Pancin}, \citenamefont {Pérez-Loureiro},\ and\ \citenamefont {Roger}}]{KONCZYKOWSKI2019125}%
  \BibitemOpen
  \bibfield  {author} {\bibinfo {author} {\bibfnamefont {P.}~\bibnamefont {Konczykowski}}, \bibinfo {author} {\bibfnamefont {B.}~\bibnamefont {Fernández-Dominguez}}, \bibinfo {author} {\bibfnamefont {H.}~\bibnamefont {Alvarez-Pol}}, \bibinfo {author} {\bibfnamefont {M.}~\bibnamefont {Caamaño}}, \bibinfo {author} {\bibfnamefont {G.}~\bibnamefont {Grinyer}}, \bibinfo {author} {\bibfnamefont {A.}~\bibnamefont {Laffoley}}, \bibinfo {author} {\bibfnamefont {B.}~\bibnamefont {Mauss}}, \bibinfo {author} {\bibfnamefont {J.}~\bibnamefont {Pancin}}, \bibinfo {author} {\bibfnamefont {D.}~\bibnamefont {Pérez-Loureiro}},\ and\ \bibinfo {author} {\bibfnamefont {T.}~\bibnamefont {Roger}},\ }\bibfield  {title} {\bibinfo {title} {Validation of the energy-loss response of $\alpha$ particles in $ic_4h_{10}$ with actarsim},\ }\href {https://doi.org/https://doi.org/10.1016/j.nima.2019.02.013} {\bibfield  {journal} {\bibinfo  {journal} {Nuclear Instruments and Methods in Physics Research Section A: Accelerators, Spectrometers,
  Detectors and Associated Equipment}\ }\textbf {\bibinfo {volume} {927}},\ \bibinfo {pages} {125} (\bibinfo {year} {2019})}\BibitemShut {NoStop}%
\bibitem [{\citenamefont {{Gutarra-Leon}}\ \emph {et~al.}(2017)\citenamefont {{Gutarra-Leon}}, \citenamefont {{Barazandeh}},\ and\ \citenamefont {{Majewski}}}]{2017APSG}%
  \BibitemOpen
  \bibfield  {author} {\bibinfo {author} {\bibfnamefont {A.}~\bibnamefont {{Gutarra-Leon}}}, \bibinfo {author} {\bibfnamefont {C.}~\bibnamefont {{Barazandeh}}},\ and\ \bibinfo {author} {\bibfnamefont {W.}~\bibnamefont {{Majewski}}},\ }\bibfield  {title} {\bibinfo {title} {{Atmospheric Muon Lifetime, Standard Model of Particles and the Lead Stopping Power for Muons}},\ }in\ \href@noop {} {\emph {\bibinfo {booktitle} {APS April Meeting Abstracts}}},\ \bibinfo {series} {APS Meeting Abstracts}, Vol.\ \bibinfo {volume} {2017}\ (\bibinfo {year} {2017})\ p.\ \bibinfo {pages} {E2.008}\BibitemShut {NoStop}%
\bibitem [{\citenamefont {Barney}\ \emph {et~al.}(2021)\citenamefont {Barney} \emph {et~al.}}]{spirit}%
  \BibitemOpen
  \bibfield  {author} {\bibinfo {author} {\bibfnamefont {J.}~\bibnamefont {Barney}} \emph {et~al.},\ }\bibfield  {title} {\bibinfo {title} {{The S$\pi$RIT time projection chamber}},\ }\href {https://doi.org/10.1063/5.0041191} {\bibfield  {journal} {\bibinfo  {journal} {Review of Scientific Instruments}\ }\textbf {\bibinfo {volume} {92}},\ \bibinfo {pages} {063302} (\bibinfo {year} {2021})}\BibitemShut {NoStop}%
\bibitem [{\citenamefont {John}\ \emph {et~al.}(2005)\citenamefont {John}, \citenamefont {Mary},\ and\ \citenamefont {Robert}}]{diff}%
  \BibitemOpen
  \bibfield  {author} {\bibinfo {author} {\bibfnamefont {S.}~\bibnamefont {John}}, \bibinfo {author} {\bibfnamefont {J.}~\bibnamefont {Mary}},\ and\ \bibinfo {author} {\bibfnamefont {B.}~\bibnamefont {Robert}},\ }\bibfield  {title} {\bibinfo {title} {Diffusion of gases: The concentration dependence of the diffusivity of carbon dioxide and argon in nitrogen},\ }\href {https://doi.org/10.1021/jp0455430} {\bibfield  {journal} {\bibinfo  {journal} {Journal of Physical Chemistry}\ }\textbf {\bibinfo {volume} {109}},\ \bibinfo {pages} {3654} (\bibinfo {year} {2005})}\BibitemShut {NoStop}%
\bibitem [{\citenamefont {Grinyer}(2018)}]{GF}%
  \BibitemOpen
  \bibfield  {author} {\bibinfo {author} {\bibfnamefont {G.}~\bibnamefont {Grinyer}},\ }\href@noop {} {\bibinfo {title} {Private communication}} (\bibinfo {year} {2018})\BibitemShut {NoStop}%
\bibitem [{\citenamefont {Agostinelli}\ \emph {et~al.}(2003)\citenamefont {Agostinelli}, \citenamefont {Allison}, \citenamefont {Amako}, \citenamefont {Apostolakis}, \citenamefont {Araujo}, \citenamefont {Arce}, \citenamefont {Asai}, \citenamefont {Axen}, \citenamefont {Banerjee}, \citenamefont {Barrand}, \citenamefont {Behner}, \citenamefont {Bellagamba}, \citenamefont {Boudreau}, \citenamefont {Broglia}, \citenamefont {Brunengo}, \citenamefont {Burkhardt}, \citenamefont {Chauvie}, \citenamefont {Chuma}, \citenamefont {Chytracek}, \citenamefont {Cooperman}, \citenamefont {Cosmo}, \citenamefont {Degtyarenko}, \citenamefont {Dell'Acqua}, \citenamefont {Depaola}, \citenamefont {Dietrich}, \citenamefont {Enami}, \citenamefont {Feliciello}, \citenamefont {Ferguson}, \citenamefont {Fesefeldt}, \citenamefont {Folger}, \citenamefont {Foppiano}, \citenamefont {Forti}, \citenamefont {Garelli}, \citenamefont {Giani}, \citenamefont {Giannitrapani}, \citenamefont {Gibin}, \citenamefont {{Gómez Cadenas}}, \citenamefont
  {González}, \citenamefont {{Gracia Abril}}, \citenamefont {Greeniaus}, \citenamefont {Greiner}, \citenamefont {Grichine}, \citenamefont {Grossheim}, \citenamefont {Guatelli}, \citenamefont {Gumplinger}, \citenamefont {Hamatsu}, \citenamefont {Hashimoto}, \citenamefont {Hasui}, \citenamefont {Heikkinen}, \citenamefont {Howard}, \citenamefont {Ivanchenko}, \citenamefont {Johnson}, \citenamefont {Jones}, \citenamefont {Kallenbach}, \citenamefont {Kanaya}, \citenamefont {Kawabata}, \citenamefont {Kawabata}, \citenamefont {Kawaguti}, \citenamefont {Kelner}, \citenamefont {Kent}, \citenamefont {Kimura}, \citenamefont {Kodama}, \citenamefont {Kokoulin}, \citenamefont {Kossov}, \citenamefont {Kurashige}, \citenamefont {Lamanna}, \citenamefont {Lampén}, \citenamefont {Lara}, \citenamefont {Lefebure}, \citenamefont {Lei}, \citenamefont {Liendl}, \citenamefont {Lockman}, \citenamefont {Longo}, \citenamefont {Magni}, \citenamefont {Maire}, \citenamefont {Medernach}, \citenamefont {Minamimoto}, \citenamefont {{Mora de
  Freitas}}, \citenamefont {Morita}, \citenamefont {Murakami}, \citenamefont {Nagamatu}, \citenamefont {Nartallo}, \citenamefont {Nieminen}, \citenamefont {Nishimura}, \citenamefont {Ohtsubo}, \citenamefont {Okamura}, \citenamefont {O'Neale}, \citenamefont {Oohata}, \citenamefont {Paech}, \citenamefont {Perl}, \citenamefont {Pfeiffer}, \citenamefont {Pia}, \citenamefont {Ranjard}, \citenamefont {Rybin}, \citenamefont {Sadilov}, \citenamefont {{Di Salvo}}, \citenamefont {Santin}, \citenamefont {Sasaki}, \citenamefont {Savvas}, \citenamefont {Sawada}, \citenamefont {Scherer}, \citenamefont {Sei}, \citenamefont {Sirotenko}, \citenamefont {Smith}, \citenamefont {Starkov}, \citenamefont {Stoecker}, \citenamefont {Sulkimo}, \citenamefont {Takahata}, \citenamefont {Tanaka}, \citenamefont {Tcherniaev}, \citenamefont {{Safai Tehrani}}, \citenamefont {Tropeano}, \citenamefont {Truscott}, \citenamefont {Uno}, \citenamefont {Urban}, \citenamefont {Urban}, \citenamefont {Verderi}, \citenamefont {Walkden}, \citenamefont
  {Wander}, \citenamefont {Weber}, \citenamefont {Wellisch}, \citenamefont {Wenaus}, \citenamefont {Williams}, \citenamefont {Wright}, \citenamefont {Yamada}, \citenamefont {Yoshida},\ and\ \citenamefont {Zschiesche}}]{geant_paper}%
  \BibitemOpen
  \bibfield  {author} {\bibinfo {author} {\bibfnamefont {S.}~\bibnamefont {Agostinelli}}, \bibinfo {author} {\bibfnamefont {J.}~\bibnamefont {Allison}}, \bibinfo {author} {\bibfnamefont {K.}~\bibnamefont {Amako}}, \bibinfo {author} {\bibfnamefont {J.}~\bibnamefont {Apostolakis}}, \bibinfo {author} {\bibfnamefont {H.}~\bibnamefont {Araujo}}, \bibinfo {author} {\bibfnamefont {P.}~\bibnamefont {Arce}}, \bibinfo {author} {\bibfnamefont {M.}~\bibnamefont {Asai}}, \bibinfo {author} {\bibfnamefont {D.}~\bibnamefont {Axen}}, \bibinfo {author} {\bibfnamefont {S.}~\bibnamefont {Banerjee}}, \bibinfo {author} {\bibfnamefont {G.}~\bibnamefont {Barrand}}, \bibinfo {author} {\bibfnamefont {F.}~\bibnamefont {Behner}}, \bibinfo {author} {\bibfnamefont {L.}~\bibnamefont {Bellagamba}}, \bibinfo {author} {\bibfnamefont {J.}~\bibnamefont {Boudreau}}, \bibinfo {author} {\bibfnamefont {L.}~\bibnamefont {Broglia}}, \bibinfo {author} {\bibfnamefont {A.}~\bibnamefont {Brunengo}}, \bibinfo {author} {\bibfnamefont {H.}~\bibnamefont
  {Burkhardt}}, \bibinfo {author} {\bibfnamefont {S.}~\bibnamefont {Chauvie}}, \bibinfo {author} {\bibfnamefont {J.}~\bibnamefont {Chuma}}, \bibinfo {author} {\bibfnamefont {R.}~\bibnamefont {Chytracek}}, \bibinfo {author} {\bibfnamefont {G.}~\bibnamefont {Cooperman}}, \bibinfo {author} {\bibfnamefont {G.}~\bibnamefont {Cosmo}}, \bibinfo {author} {\bibfnamefont {P.}~\bibnamefont {Degtyarenko}}, \bibinfo {author} {\bibfnamefont {A.}~\bibnamefont {Dell'Acqua}}, \bibinfo {author} {\bibfnamefont {G.}~\bibnamefont {Depaola}}, \bibinfo {author} {\bibfnamefont {D.}~\bibnamefont {Dietrich}}, \bibinfo {author} {\bibfnamefont {R.}~\bibnamefont {Enami}}, \bibinfo {author} {\bibfnamefont {A.}~\bibnamefont {Feliciello}}, \bibinfo {author} {\bibfnamefont {C.}~\bibnamefont {Ferguson}}, \bibinfo {author} {\bibfnamefont {H.}~\bibnamefont {Fesefeldt}}, \bibinfo {author} {\bibfnamefont {G.}~\bibnamefont {Folger}}, \bibinfo {author} {\bibfnamefont {F.}~\bibnamefont {Foppiano}}, \bibinfo {author} {\bibfnamefont {A.}~\bibnamefont
  {Forti}}, \bibinfo {author} {\bibfnamefont {S.}~\bibnamefont {Garelli}}, \bibinfo {author} {\bibfnamefont {S.}~\bibnamefont {Giani}}, \bibinfo {author} {\bibfnamefont {R.}~\bibnamefont {Giannitrapani}}, \bibinfo {author} {\bibfnamefont {D.}~\bibnamefont {Gibin}}, \bibinfo {author} {\bibfnamefont {J.}~\bibnamefont {{Gómez Cadenas}}}, \bibinfo {author} {\bibfnamefont {I.}~\bibnamefont {González}}, \bibinfo {author} {\bibfnamefont {G.}~\bibnamefont {{Gracia Abril}}}, \bibinfo {author} {\bibfnamefont {G.}~\bibnamefont {Greeniaus}}, \bibinfo {author} {\bibfnamefont {W.}~\bibnamefont {Greiner}}, \bibinfo {author} {\bibfnamefont {V.}~\bibnamefont {Grichine}}, \bibinfo {author} {\bibfnamefont {A.}~\bibnamefont {Grossheim}}, \bibinfo {author} {\bibfnamefont {S.}~\bibnamefont {Guatelli}}, \bibinfo {author} {\bibfnamefont {P.}~\bibnamefont {Gumplinger}}, \bibinfo {author} {\bibfnamefont {R.}~\bibnamefont {Hamatsu}}, \bibinfo {author} {\bibfnamefont {K.}~\bibnamefont {Hashimoto}}, \bibinfo {author} {\bibfnamefont
  {H.}~\bibnamefont {Hasui}}, \bibinfo {author} {\bibfnamefont {A.}~\bibnamefont {Heikkinen}}, \bibinfo {author} {\bibfnamefont {A.}~\bibnamefont {Howard}}, \bibinfo {author} {\bibfnamefont {V.}~\bibnamefont {Ivanchenko}}, \bibinfo {author} {\bibfnamefont {A.}~\bibnamefont {Johnson}}, \bibinfo {author} {\bibfnamefont {F.}~\bibnamefont {Jones}}, \bibinfo {author} {\bibfnamefont {J.}~\bibnamefont {Kallenbach}}, \bibinfo {author} {\bibfnamefont {N.}~\bibnamefont {Kanaya}}, \bibinfo {author} {\bibfnamefont {M.}~\bibnamefont {Kawabata}}, \bibinfo {author} {\bibfnamefont {Y.}~\bibnamefont {Kawabata}}, \bibinfo {author} {\bibfnamefont {M.}~\bibnamefont {Kawaguti}}, \bibinfo {author} {\bibfnamefont {S.}~\bibnamefont {Kelner}}, \bibinfo {author} {\bibfnamefont {P.}~\bibnamefont {Kent}}, \bibinfo {author} {\bibfnamefont {A.}~\bibnamefont {Kimura}}, \bibinfo {author} {\bibfnamefont {T.}~\bibnamefont {Kodama}}, \bibinfo {author} {\bibfnamefont {R.}~\bibnamefont {Kokoulin}}, \bibinfo {author} {\bibfnamefont
  {M.}~\bibnamefont {Kossov}}, \bibinfo {author} {\bibfnamefont {H.}~\bibnamefont {Kurashige}}, \bibinfo {author} {\bibfnamefont {E.}~\bibnamefont {Lamanna}}, \bibinfo {author} {\bibfnamefont {T.}~\bibnamefont {Lampén}}, \bibinfo {author} {\bibfnamefont {V.}~\bibnamefont {Lara}}, \bibinfo {author} {\bibfnamefont {V.}~\bibnamefont {Lefebure}}, \bibinfo {author} {\bibfnamefont {F.}~\bibnamefont {Lei}}, \bibinfo {author} {\bibfnamefont {M.}~\bibnamefont {Liendl}}, \bibinfo {author} {\bibfnamefont {W.}~\bibnamefont {Lockman}}, \bibinfo {author} {\bibfnamefont {F.}~\bibnamefont {Longo}}, \bibinfo {author} {\bibfnamefont {S.}~\bibnamefont {Magni}}, \bibinfo {author} {\bibfnamefont {M.}~\bibnamefont {Maire}}, \bibinfo {author} {\bibfnamefont {E.}~\bibnamefont {Medernach}}, \bibinfo {author} {\bibfnamefont {K.}~\bibnamefont {Minamimoto}}, \bibinfo {author} {\bibfnamefont {P.}~\bibnamefont {{Mora de Freitas}}}, \bibinfo {author} {\bibfnamefont {Y.}~\bibnamefont {Morita}}, \bibinfo {author} {\bibfnamefont
  {K.}~\bibnamefont {Murakami}}, \bibinfo {author} {\bibfnamefont {M.}~\bibnamefont {Nagamatu}}, \bibinfo {author} {\bibfnamefont {R.}~\bibnamefont {Nartallo}}, \bibinfo {author} {\bibfnamefont {P.}~\bibnamefont {Nieminen}}, \bibinfo {author} {\bibfnamefont {T.}~\bibnamefont {Nishimura}}, \bibinfo {author} {\bibfnamefont {K.}~\bibnamefont {Ohtsubo}}, \bibinfo {author} {\bibfnamefont {M.}~\bibnamefont {Okamura}}, \bibinfo {author} {\bibfnamefont {S.}~\bibnamefont {O'Neale}}, \bibinfo {author} {\bibfnamefont {Y.}~\bibnamefont {Oohata}}, \bibinfo {author} {\bibfnamefont {K.}~\bibnamefont {Paech}}, \bibinfo {author} {\bibfnamefont {J.}~\bibnamefont {Perl}}, \bibinfo {author} {\bibfnamefont {A.}~\bibnamefont {Pfeiffer}}, \bibinfo {author} {\bibfnamefont {M.}~\bibnamefont {Pia}}, \bibinfo {author} {\bibfnamefont {F.}~\bibnamefont {Ranjard}}, \bibinfo {author} {\bibfnamefont {A.}~\bibnamefont {Rybin}}, \bibinfo {author} {\bibfnamefont {S.}~\bibnamefont {Sadilov}}, \bibinfo {author} {\bibfnamefont {E.}~\bibnamefont
  {{Di Salvo}}}, \bibinfo {author} {\bibfnamefont {G.}~\bibnamefont {Santin}}, \bibinfo {author} {\bibfnamefont {T.}~\bibnamefont {Sasaki}}, \bibinfo {author} {\bibfnamefont {N.}~\bibnamefont {Savvas}}, \bibinfo {author} {\bibfnamefont {Y.}~\bibnamefont {Sawada}}, \bibinfo {author} {\bibfnamefont {S.}~\bibnamefont {Scherer}}, \bibinfo {author} {\bibfnamefont {S.}~\bibnamefont {Sei}}, \bibinfo {author} {\bibfnamefont {V.}~\bibnamefont {Sirotenko}}, \bibinfo {author} {\bibfnamefont {D.}~\bibnamefont {Smith}}, \bibinfo {author} {\bibfnamefont {N.}~\bibnamefont {Starkov}}, \bibinfo {author} {\bibfnamefont {H.}~\bibnamefont {Stoecker}}, \bibinfo {author} {\bibfnamefont {J.}~\bibnamefont {Sulkimo}}, \bibinfo {author} {\bibfnamefont {M.}~\bibnamefont {Takahata}}, \bibinfo {author} {\bibfnamefont {S.}~\bibnamefont {Tanaka}}, \bibinfo {author} {\bibfnamefont {E.}~\bibnamefont {Tcherniaev}}, \bibinfo {author} {\bibfnamefont {E.}~\bibnamefont {{Safai Tehrani}}}, \bibinfo {author} {\bibfnamefont {M.}~\bibnamefont
  {Tropeano}}, \bibinfo {author} {\bibfnamefont {P.}~\bibnamefont {Truscott}}, \bibinfo {author} {\bibfnamefont {H.}~\bibnamefont {Uno}}, \bibinfo {author} {\bibfnamefont {L.}~\bibnamefont {Urban}}, \bibinfo {author} {\bibfnamefont {P.}~\bibnamefont {Urban}}, \bibinfo {author} {\bibfnamefont {M.}~\bibnamefont {Verderi}}, \bibinfo {author} {\bibfnamefont {A.}~\bibnamefont {Walkden}}, \bibinfo {author} {\bibfnamefont {W.}~\bibnamefont {Wander}}, \bibinfo {author} {\bibfnamefont {H.}~\bibnamefont {Weber}}, \bibinfo {author} {\bibfnamefont {J.}~\bibnamefont {Wellisch}}, \bibinfo {author} {\bibfnamefont {T.}~\bibnamefont {Wenaus}}, \bibinfo {author} {\bibfnamefont {D.}~\bibnamefont {Williams}}, \bibinfo {author} {\bibfnamefont {D.}~\bibnamefont {Wright}}, \bibinfo {author} {\bibfnamefont {T.}~\bibnamefont {Yamada}}, \bibinfo {author} {\bibfnamefont {H.}~\bibnamefont {Yoshida}},\ and\ \bibinfo {author} {\bibfnamefont {D.}~\bibnamefont {Zschiesche}},\ }\bibfield  {title} {\bibinfo {title} {Geant4—a simulation
  toolkit},\ }\href {https://doi.org/https://doi.org/10.1016/S0168-9002(03)01368-8} {\bibfield  {journal} {\bibinfo  {journal} {Nuclear Instruments and Methods in Physics Research Section A: Accelerators, Spectrometers, Detectors and Associated Equipment}\ }\textbf {\bibinfo {volume} {506}},\ \bibinfo {pages} {250} (\bibinfo {year} {2003})}\BibitemShut {NoStop}%
\bibitem [{\citenamefont {Anthony}(2023)}]{Anyhony2023}%
  \BibitemOpen
  \bibfield  {author} {\bibinfo {author} {\bibfnamefont {A.~K.}\ \bibnamefont {Anthony}},\ }\emph {\bibinfo {title} {Fission in the LEAD region}},\ \href@noop {} {Ph.D. thesis},\ \bibinfo  {school} {MSU} (\bibinfo {year} {2023}),\ \bibinfo {note} {published thesis}\BibitemShut {NoStop}%
\bibitem [{\citenamefont {Tarasov}\ and\ \citenamefont {Bazin}(2016)}]{TARASOV2016185}%
  \BibitemOpen
  \bibfield  {author} {\bibinfo {author} {\bibfnamefont {O.}~\bibnamefont {Tarasov}}\ and\ \bibinfo {author} {\bibfnamefont {D.}~\bibnamefont {Bazin}},\ }\bibfield  {title} {\bibinfo {title} {Lise++: Exotic beam production with fragment separators and their design},\ }\href {https://doi.org/https://doi.org/10.1016/j.nimb.2016.03.021} {\bibfield  {journal} {\bibinfo  {journal} {Nuclear Instruments and Methods in Physics Research Section B: Beam Interactions with Materials and Atoms}\ }\textbf {\bibinfo {volume} {376}},\ \bibinfo {pages} {185} (\bibinfo {year} {2016})}\BibitemShut {NoStop}%
\bibitem [{\citenamefont {Hausmann}\ \emph {et~al.}(2013)\citenamefont {Hausmann}, \citenamefont {Aaron}, \citenamefont {Amthor}, \citenamefont {Avilov}, \citenamefont {Bandura}, \citenamefont {Bennett}, \citenamefont {Bollen}, \citenamefont {Borden}, \citenamefont {Burgess}, \citenamefont {Chouhan}, \citenamefont {Graves}, \citenamefont {Mittig}, \citenamefont {Morrissey}, \citenamefont {Pellemoine}, \citenamefont {Portillo}, \citenamefont {Ronningen}, \citenamefont {Schein}, \citenamefont {Sherrill},\ and\ \citenamefont {Zeller}}]{HAUSMANN2013349}%
  \BibitemOpen
  \bibfield  {author} {\bibinfo {author} {\bibfnamefont {M.}~\bibnamefont {Hausmann}}, \bibinfo {author} {\bibfnamefont {A.}~\bibnamefont {Aaron}}, \bibinfo {author} {\bibfnamefont {A.}~\bibnamefont {Amthor}}, \bibinfo {author} {\bibfnamefont {M.}~\bibnamefont {Avilov}}, \bibinfo {author} {\bibfnamefont {L.}~\bibnamefont {Bandura}}, \bibinfo {author} {\bibfnamefont {R.}~\bibnamefont {Bennett}}, \bibinfo {author} {\bibfnamefont {G.}~\bibnamefont {Bollen}}, \bibinfo {author} {\bibfnamefont {T.}~\bibnamefont {Borden}}, \bibinfo {author} {\bibfnamefont {T.}~\bibnamefont {Burgess}}, \bibinfo {author} {\bibfnamefont {S.}~\bibnamefont {Chouhan}}, \bibinfo {author} {\bibfnamefont {V.}~\bibnamefont {Graves}}, \bibinfo {author} {\bibfnamefont {W.}~\bibnamefont {Mittig}}, \bibinfo {author} {\bibfnamefont {D.}~\bibnamefont {Morrissey}}, \bibinfo {author} {\bibfnamefont {F.}~\bibnamefont {Pellemoine}}, \bibinfo {author} {\bibfnamefont {M.}~\bibnamefont {Portillo}}, \bibinfo {author} {\bibfnamefont {R.}~\bibnamefont
  {Ronningen}}, \bibinfo {author} {\bibfnamefont {M.}~\bibnamefont {Schein}}, \bibinfo {author} {\bibfnamefont {B.}~\bibnamefont {Sherrill}},\ and\ \bibinfo {author} {\bibfnamefont {A.}~\bibnamefont {Zeller}},\ }\bibfield  {title} {\bibinfo {title} {Design of the advanced rare isotope separator aris at frib},\ }\href {https://doi.org/https://doi.org/10.1016/j.nimb.2013.06.042} {\bibfield  {journal} {\bibinfo  {journal} {Nuclear Instruments and Methods in Physics Research Section B: Beam Interactions with Materials and Atoms}\ }\textbf {\bibinfo {volume} {317}},\ \bibinfo {pages} {349} (\bibinfo {year} {2013})},\ \bibinfo {note} {xVIth International Conference on ElectroMagnetic Isotope Separators and Techniques Related to their Applications, December 2–7, 2012 at Matsue, Japan}\BibitemShut {NoStop}%
\bibitem [{\citenamefont {Tammina}(2019)}]{tamira}%
  \BibitemOpen
  \bibfield  {author} {\bibinfo {author} {\bibfnamefont {S.}~\bibnamefont {Tammina}},\ }\bibfield  {title} {\bibinfo {title} {Transfer learning using vgg-16 with deep convolutional neural network for classifying images},\ }\href {https://doi.org/10.29322/IJSRP.9.10.2019.p9420} {\bibfield  {journal} {\bibinfo  {journal} {International Journal of Scientific and Research Publications (IJSRP)}\ }\textbf {\bibinfo {volume} {9}},\ \bibinfo {pages} {p9420} (\bibinfo {year} {2019})}\BibitemShut {NoStop}%
\bibitem [{\citenamefont {Simonyan}\ and\ \citenamefont {Zisserman}(2014)}]{Simonyan2014VeryDC}%
  \BibitemOpen
  \bibfield  {author} {\bibinfo {author} {\bibfnamefont {K.}~\bibnamefont {Simonyan}}\ and\ \bibinfo {author} {\bibfnamefont {A.}~\bibnamefont {Zisserman}},\ }\bibfield  {title} {\bibinfo {title} {Very deep convolutional networks for large-scale image recognition},\ }\href@noop {} {\bibfield  {journal} {\bibinfo  {journal} {CoRR}\ }\textbf {\bibinfo {volume} {abs/1409.1556}} (\bibinfo {year} {2014})}\BibitemShut {NoStop}%
\bibitem [{\citenamefont {Kuchera}\ \emph {et~al.}(2019)\citenamefont {Kuchera}, \citenamefont {Ramanujan}, \citenamefont {Taylor}, \citenamefont {Strauss}, \citenamefont {Bazin}, \citenamefont {Bradt},\ and\ \citenamefont {Chen}}]{KUCHERA2019156}%
  \BibitemOpen
  \bibfield  {author} {\bibinfo {author} {\bibfnamefont {M.}~\bibnamefont {Kuchera}}, \bibinfo {author} {\bibfnamefont {R.}~\bibnamefont {Ramanujan}}, \bibinfo {author} {\bibfnamefont {J.}~\bibnamefont {Taylor}}, \bibinfo {author} {\bibfnamefont {R.}~\bibnamefont {Strauss}}, \bibinfo {author} {\bibfnamefont {D.}~\bibnamefont {Bazin}}, \bibinfo {author} {\bibfnamefont {J.}~\bibnamefont {Bradt}},\ and\ \bibinfo {author} {\bibfnamefont {R.}~\bibnamefont {Chen}},\ }\bibfield  {title} {\bibinfo {title} {Machine learning methods for track classification in the at-tpc},\ }\href {https://doi.org/https://doi.org/10.1016/j.nima.2019.05.097} {\bibfield  {journal} {\bibinfo  {journal} {Nuclear Instruments and Methods in Physics Research Section A: Accelerators, Spectrometers, Detectors and Associated Equipment}\ }\textbf {\bibinfo {volume} {940}},\ \bibinfo {pages} {156} (\bibinfo {year} {2019})}\BibitemShut {NoStop}%
\bibitem [{\citenamefont {Mueller}\ \emph {et~al.}(2001)\citenamefont {Mueller}, \citenamefont {Church}, \citenamefont {Glasmacher}, \citenamefont {Gutknecht}, \citenamefont {Hackman}, \citenamefont {Hansen}, \citenamefont {Hu}, \citenamefont {Miller},\ and\ \citenamefont {Quirin}}]{MUELLER2001492}%
  \BibitemOpen
  \bibfield  {author} {\bibinfo {author} {\bibfnamefont {W.}~\bibnamefont {Mueller}}, \bibinfo {author} {\bibfnamefont {J.}~\bibnamefont {Church}}, \bibinfo {author} {\bibfnamefont {T.}~\bibnamefont {Glasmacher}}, \bibinfo {author} {\bibfnamefont {D.}~\bibnamefont {Gutknecht}}, \bibinfo {author} {\bibfnamefont {G.}~\bibnamefont {Hackman}}, \bibinfo {author} {\bibfnamefont {P.}~\bibnamefont {Hansen}}, \bibinfo {author} {\bibfnamefont {Z.}~\bibnamefont {Hu}}, \bibinfo {author} {\bibfnamefont {K.}~\bibnamefont {Miller}},\ and\ \bibinfo {author} {\bibfnamefont {P.}~\bibnamefont {Quirin}},\ }\bibfield  {title} {\bibinfo {title} {Thirty-two-fold segmented germanium detectors to identify $\gamma$-rays from intermediate-energy exotic beams},\ }\href {https://doi.org/https://doi.org/10.1016/S0168-9002(01)00257-1} {\bibfield  {journal} {\bibinfo  {journal} {Nuclear Instruments and Methods in Physics Research Section A: Accelerators, Spectrometers, Detectors and Associated Equipment}\ }\textbf {\bibinfo {volume}
  {466}},\ \bibinfo {pages} {492} (\bibinfo {year} {2001})}\BibitemShut {NoStop}%
\bibitem [{deg(2020)}]{dega}%
  \BibitemOpen
  \href {http://(site fds.ornl.gov/wp-content/uploads/2020/09/FDS-WP} {} (\bibinfo {year} {2020})\BibitemShut {NoStop}%
\bibitem [{\citenamefont {Prokop}\ \emph {et~al.}(2014)\citenamefont {Prokop}, \citenamefont {Liddick}, \citenamefont {Abromeit}, \citenamefont {Chemey}, \citenamefont {Larson}, \citenamefont {Suchyta},\ and\ \citenamefont {Tompkins}}]{PROKOP2014163}%
  \BibitemOpen
  \bibfield  {author} {\bibinfo {author} {\bibfnamefont {C.}~\bibnamefont {Prokop}}, \bibinfo {author} {\bibfnamefont {S.}~\bibnamefont {Liddick}}, \bibinfo {author} {\bibfnamefont {B.}~\bibnamefont {Abromeit}}, \bibinfo {author} {\bibfnamefont {A.}~\bibnamefont {Chemey}}, \bibinfo {author} {\bibfnamefont {N.}~\bibnamefont {Larson}}, \bibinfo {author} {\bibfnamefont {S.}~\bibnamefont {Suchyta}},\ and\ \bibinfo {author} {\bibfnamefont {J.}~\bibnamefont {Tompkins}},\ }\bibfield  {title} {\bibinfo {title} {Digital data acquisition system implementation at the national superconducting cyclotron laboratory},\ }\href {https://doi.org/https://doi.org/10.1016/j.nima.2013.12.044} {\bibfield  {journal} {\bibinfo  {journal} {Nuclear Instruments and Methods in Physics Research Section A: Accelerators, Spectrometers, Detectors and Associated Equipment}\ }\textbf {\bibinfo {volume} {741}},\ \bibinfo {pages} {163} (\bibinfo {year} {2014})}\BibitemShut {NoStop}%
\end{thebibliography}%
\end{document}